\documentclass{mnras}

\usepackage{newtxtext,newtxmath}

\usepackage[T1]{fontenc}
\usepackage{ae,aecompl}
\usepackage{graphicx}	
\usepackage{amsmath}	
\usepackage{amssymb}	
\usepackage{ltxtable}
\usepackage{pdflscape}
\usepackage{placeins}
\usepackage{longtable}
\usepackage{tabularx} 
\usepackage{tablefootnote}

\newcommand{\ovi}{O\,{\sc vi}}

\newcommand{\ha}{H\,$\alpha$} 
\newcommand{\hbeta}{H\,$\beta$}
\newcommand{\hc}{H\,$\gamma$}

\newcommand{\heliumb}{He\,{\sc ii}}

\newcommand{\oxygeniii}{[O\,{\sc iii}]}

\newcommand{\DD}{D$^{\prime}$}

\def\vhel{\ifmmode{V_{{\rm HEL}}}\else{$V_{{\rm HEL}}$}\fi}
\def\vsys{\ifmmode{V_{\rm sys}}\else{$V_{\rm sys}$}\fi}
\def\kms{\ifmmode{~{\rm km\,s}^{-1}}\else{~km~s$^{-1}$}\fi}
\def\vlsr{\ifmmode{v_{\rm lsr}}\else{$v_{\rm lsr}$}\fi}

\title[A machine learning approach to identify symbiotic stars]{\center A machine learning approach for identification and classification of symbiotic stars using 2MASS and WISE}

\author[S. Akras et al.]{Stavros Akras$^{1,2}$\thanks{CNPq Fellow (PDI-DA 300336/2016-0)},\thanks{e-mail: stavrosakras@on.br}, 
Marcelo L. Leal-Ferreira$^{3,4}$\thanks{CNPq Fellow (248503/2013-8)}, Lizette Guzman-Ramirez$^{3,5}$, 
\newauthor Gerardo Ramos-Larios$^{6}$ \\
$^{1}$ Observat\'orio Nacional/MCTI, Rua Gen. Jos\'{e} Cristino, 77, 20921-400, Rio de Janeiro, Brazil\\
$^{2}$ Observat\'orio do Valongo, Universidade Federal do Rio de Janeiro, Ladeira Pedro Antonio 43, 20080-090, Rio de Janeiro, Brazil\\
$^{3}$ Leiden Observatory, Leiden University, Niels Bohrweg 2, 2333 CA Leiden, Netherlands\\
$^{4}$ Argelander-Institut f\"{u}r Astronomie, Universit\"{u}t Bonn, Auf dem H\"{u}gel 71, 53121, Bonn, Germany\\
$^{5}$ European Southern Observatory, Alonso de C\'ordova 3107, Casilla 19001, Santiago, Chile\\
$^{6}$ Instituto de Astronom\'ia y Meteorolog\'ia, Av. Vallarta No. 2602, Col. Arcos Vallarta, C.P. 44130 Guadalajara, Jalisco, Mexico\\
}

\date{Accepted XXX. Received YYY; in original form ZZZ}

\pubyear{2016}

\begin{document}
\label{firstpage}
\pagerange{\pageref{firstpage}--\pageref{lastpage}}
\maketitle

\begin{abstract}

In this second paper in a series of papers based on the most-up-to-date catalogue of symbiotic stars (SySts), we present a new  
approach for identifying and distinguishing SySts from other \ha\ emitters in photometric surveys using machine learning algorithms 
such as classification tree, linear discriminant analysis, and K-nearest neighbour. The motivation behind of this work is to seek for possible 
colour indices in the regime of near- and mid-infrared covered by the 2MASS and WISE surveys.
A number of diagnostic colour-colour diagrams are generated for all the known Galactic SySts and several classes of stellar objects that mimic 
SySts such as planetary nebulae, post-AGB, Mira, single K and M giants, cataclysmic variables, Be, AeBe, YSO, weak and classical 
T Tauri stars, and Wolf-Rayet. The classification tree algorithm unveils that primarily {\it J--H}, {\it W1--W4} and {\it K$_{\rm s}$--W3}  and secondarily {\it H--W2}, {\it W1--W2} and {\it W3--W4} are ideal colour indices to identify SySts. Linear discriminant analysis method is also applied 
to determine the linear combination of 2MASS and AllWISE magnitudes that better distinguish SySts. The probability of a source being a SySt 
is determined using the K-nearest neighbour method on the LDA components. By applying our classification tree model to the list of 
candidate SySts (Paper~I), the IPHAS list of candidate SySts, and the DR2 VPHAS+ catalogue, we find 125 (72 new candidates) sources that 
pass our criteria while we also recover 90 per cent of the known Galactic SySts.  

\end{abstract}

\begin{keywords}
general: catalogues - stars: binaries: symbiotic - stars: fundamental parameters - methods: statistical - methods: data analysis
\end{keywords}

\section{Introduction}

This is the second in a series of papers based on the new catalogue of symbiotic stars (SySts). In Paper~I (Akras et al. 2019, accepted for 
publication in ApJS) the compilation of known (323) and candidate (87) SySts as well as an atlas of 348 spectral energy distributions (SED) from 1 to 
22$\mu$m, using the Two Micron All Sky Survey (2MASS, Skrutskie et al. 2006) and the Wide-field Infrared Survey Explorer (WISE, 
Wright et al. 2010) data are presented. The classification of all known SySts in the S-D-\DD scheme, based on their SED profiles, is revised. 
Seventy-four per cent are classified as S-type (stellar), 13 per cent as D-type (dusty), 8 per cent as S$+$IR-type (stellar $+$ infrared excess)
and 3.5 per cent as \DD-type. 

SySts are ideal astrophysical laboratories for investigating and studying the formation of aspherical circumstellar envelopes, mass transfer 
accretion disks processes, formation of soft and hard-X rays emission, dust forming regions, colliding winds among others (e.g. 
Jordan et al. 1996; Totov 2003; Sokoloski 2003; Leedj\"{a}rv 2004, Mikolajewska 2012; Luna et al. 2013, Skopal \& Carikov\'{i}a 
2015; Mukai et al. 2016). Beside all these phenomena and processes, they are also considered as candidates for the 
progenitors of type Ia supernova (SN Ia, Munari \& Renzini 1992; Han \& Podsiadlowski 2004; Di Stefano 2010;  Wang et al. 2010; 
Dilday et al. 2012).

Yet, the numbers of known SySts in the Milky Way (257, Paper~I) and nearby galaxies (66, Paper~I) are still far from being 
consistent with the expected number derived from population models (e.g. 3$\times$10$^5$, Munari \& Renzini 1992; 4$\times$10$^5$, 
Magrini, Corradi \& Munari 2003; 1.2-15$\times$10$^3$, L\"{u}, Yungelson \& Han 2016).

Many attempts have been made to discover new members by developing diagnostic colour-colour diagrams (DCCD) in the optical 
(\oxygeniii\ 4363/\hc\ vs. \oxygeniii\ 5007/\hbeta, Gutierrez-Moreno, Moreno \& Cort\'{e}s 1995; various combinations of emission line ratios, Ilkiewicz \& Mikolajewska 2017; {\it r}--\ha\ vs. {\it r--i}, Corradi et al. 2008, 2010; Rodr\'{i}guez-Flores et al. 2014), 
near-IR  ({\it J--H} vs. {\it H--K$_s$}, Allen \& Glass 1974; Phillips 2007; Corradi et al. 2008; Clyne et al. 2015, {\it I--J} vs. {\it J--K$_s$},  Schmeja \& Kimeswenger 2001) and mid-IR regime ({\it K}-[12] vs. [12]-[25], Luud \& Tuvikene 1987;  Leedj\"{a}rv 1992).

The motivation of this work is to find new colour criteria in the regime of near and mid-IR that will identify SySts using machine learning 
algorithms. Recall that SySts display SED peaks in the wavelength range between $\sim$1 and $\sim$25$\mu$m (Ivison et al. 1995, Paper~I). 
Therefore, the 2MASS/WISE surveys are very helpful to distinguish SySts from other strong \ha\ emitters (e.g. genuine planetary nebulae (PNe), 
Wolf-Rayet stars (WR), Be stars, AeBe stars, cataclysmic variables (CV), Mira stars, weak and classical T Tauri stars (WTT,ClTT), 
young stellar objects (YSO)).

The paper is organized as follows: new 2MASS/AllWISE DCCDs are generated and presented in Section~2. The results from a machine learning 
approach, classification tree, linear discriminant analysis (LDA) and K-nearest neighbours (KNN), are presented in Sections 3 and 4. 
In Section 5, we apply our classification criteria to a compilation of candidate SySts. This compilation includes candidates from the list 
of candidates (Paper~I), the IPHAS (Corradi et al. 2008; Drew et al. 2005) and the VPHAS+ (DR2; Drew et al. 2014) surveys. A number of new 
very likely SySts candidates are presented. We finish with our conclusions in Sect. 6.

\begin{table}
\caption[]{List of references for all the classes of objects.}
\label{table4}
\begin{tabular}{lllllllllll}
\hline 
Class of Object & Sample & References \\     			
\hline 
PNe            & 188  & Ramos-Larios \& Phillips 2005\\ 
Post-AGB       & 180  & Vickers et al.2015,Akras et al.2017\\
			   &      & Suarez et al. 2006, Yoon et al. 2014\\  
Wolf-Rayet     & 162  & van der Hucht 2001\\
Be             & 185  & Chojnowski et al. 2015\\ 
AeBe           & 173  & Vieira et al. 2003,\\
			   &      & Herbst \& Shevchenko 1999 \\
			   &      & Rodrigues et al. 2009\\
CV			   & 191  & Hoard et al. 2002\\
Mira           & 316  & Huemmerich \& Bernhard 2012,\\
               &      & Whitelock et al. 2008 \\
K giants 	   & 240  & Carlberg et al. 2011, Gray et al. 2016 \\
M giants       &  210  & Tabur et al. 2009, Gray et al. 2016 \\
Classical T Tauri& 183 & Galli et al. 2015, France et al. 2014\\
               &      & Grankin et al. 2007, \\
               &      & Herbst \& Shevchenko 1999\\             
Weak T Tauri   & 213  & Grankin et al. 2008, Galli et al. 2015\\
			   &      & Cieza et al. 2007, \\
			   &      & Herbst \& Shevchenko 1999\\
YSO			   & 260  & Rebull et al. 2011, Harvey et al. 2007\\\			    			   
SySts		   & 220$^{\dag}$  & Paper~I and references therein\\
\hline 
\end{tabular}
\medskip{}
\begin{flushleft}
$^{\dag}$ Galactic SySts
\end{flushleft}
\end{table}

\section{Diagnostic colour-colour diagrams (DCCD)}

\begin{figure*}
\includegraphics[scale=0.52]{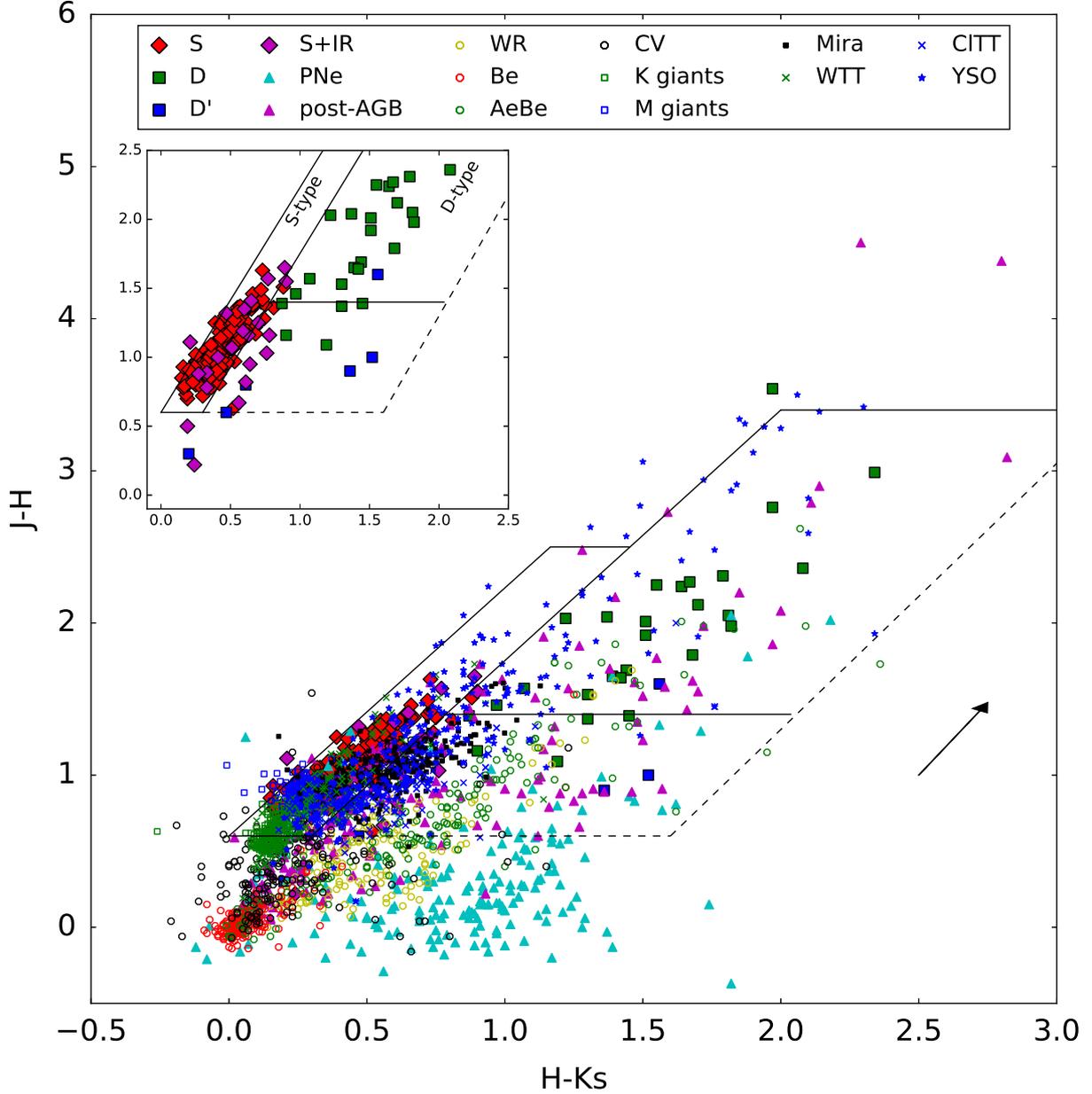}
\caption[]{The 2MASS {\it J--H} vs. {\it H--K$_s$} DCCD for different classes of objects. The same DCCD for the four type of 
SySts is presented in the inset plot. The dashed and solid boxes define the regimes of S- and D-type SySts from Corradi et al. (2008) 
and Rodr\'{i}gues-Flores et al. (2014). The black arrow corresponds to 4~mag extinction in the {\it V} band.
The names in the box correspond to S-, D-, \DD- and S$+$IR-type SySts, planetary nebulae (PNe), post-AGB stars (post-AGB), Wolf-Rayet stars 
(WR), Be stars (Be), AeBe stars (AeBe), cataclysmic variables (CV), K/M giants, weak/classical 
T Tauri stars (WTT/ClTT) and YSO.} 
\label{fig9}
\end{figure*}

\begin{figure*}
\includegraphics[scale=0.52]{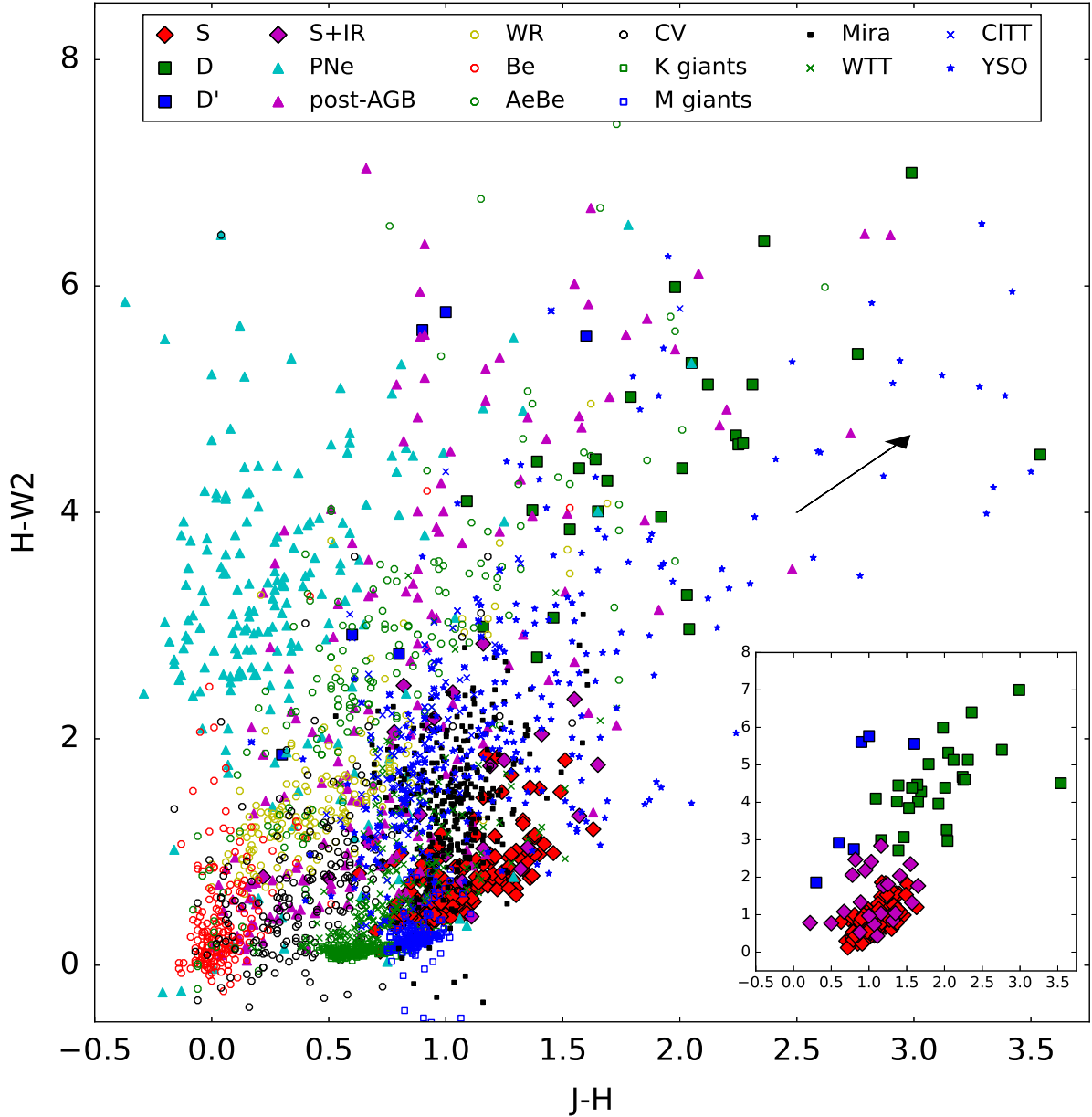}
\caption[]{The 2MASS/AllWISE {\it H--W2} vs. {\it J-H} DCCD for different classes of objects as well as  for the four type of 
SySts presented in the inset plot. The black arrow corresponds to 4~mag extinction in the {\it V} band.}
\label{fig10}
\end{figure*}

\begin{figure*}
\includegraphics[scale=0.52]{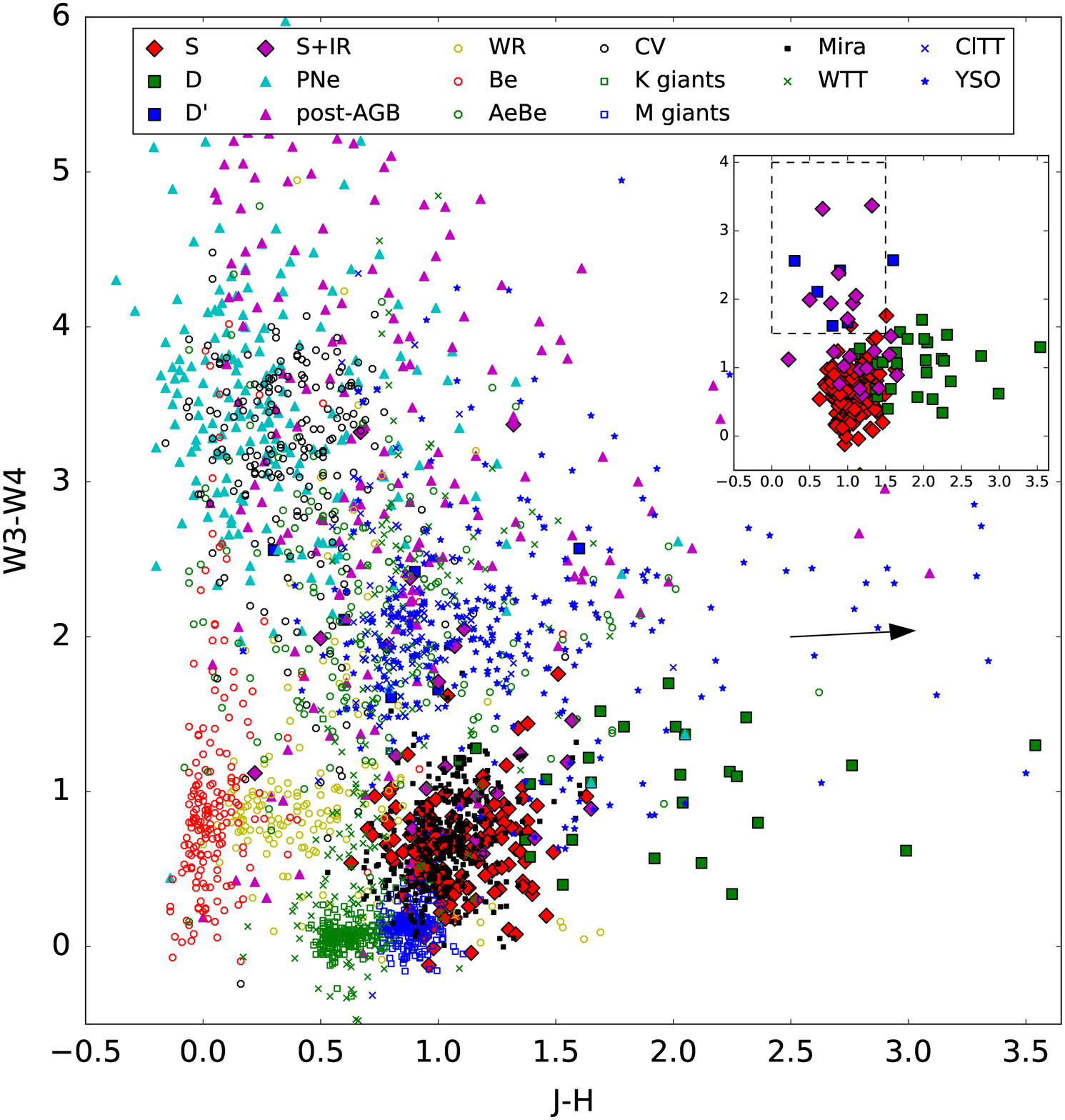}
\caption[]{The 2MASS/AllWISE {\it W3--W4} vs. {\it J--H} DCCD for different classes of objects as well as  for the four type of 
SySts presented in the inset plot. The black arrow corresponds to 4~mag extinction in the {\it V} band. The dashed box in the 
inset plot indicates the vertical branched region of SySts discussed in the text. The {\it W3} and {\it W4} magnitudes of the majority of 
CVs correspond to upper limit values.}
\label{fig11}
\end{figure*}

\begin{figure*}
\includegraphics[scale=0.52]{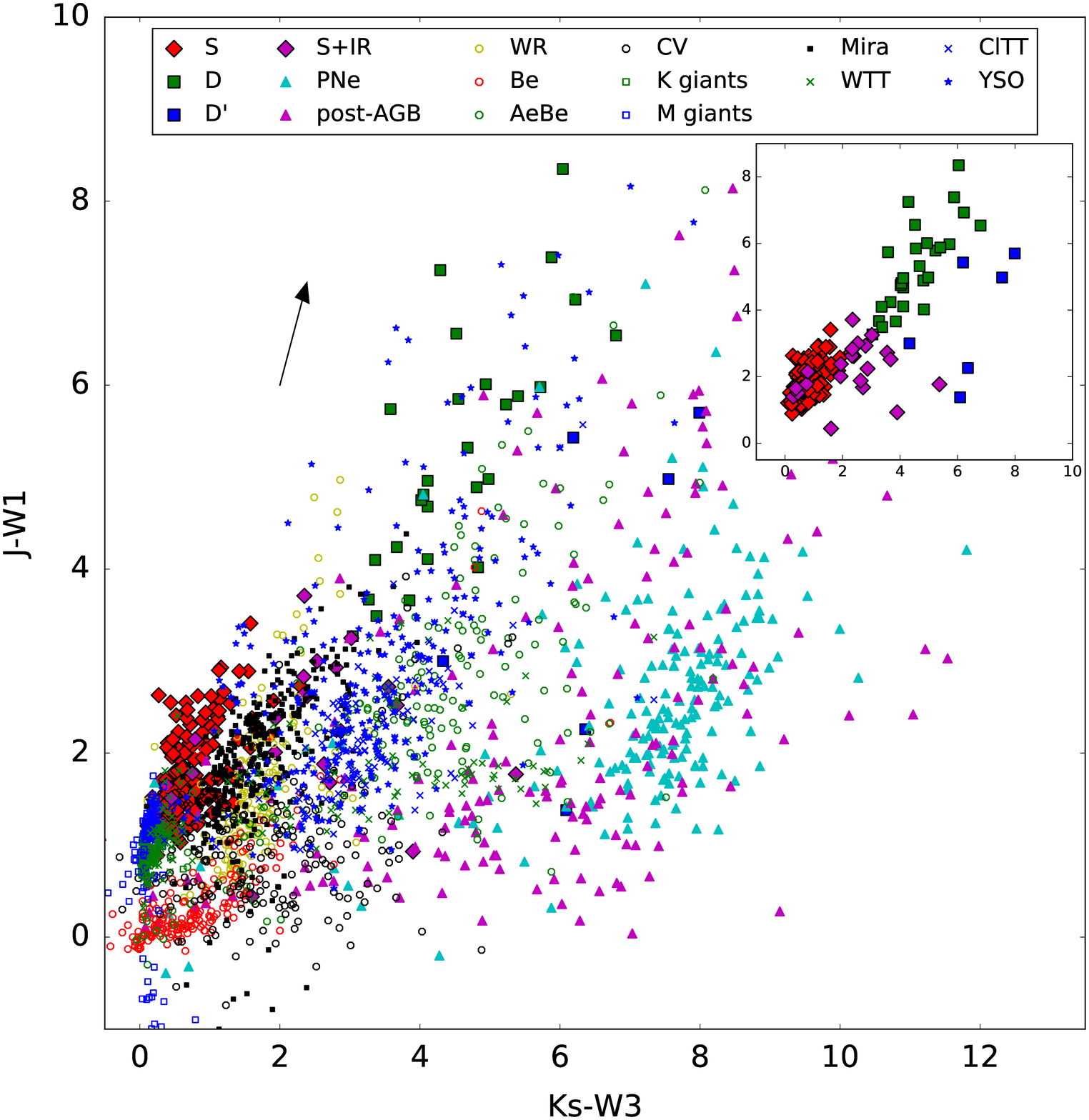}
\caption[]{The 2MASS/AllWISE {\it J--W1} vs. {\it K$_s$--W3} DCCD for different classes of objects as well as  for the four type of 
SySts presented in the inset plot. The black arrow corresponds to 4~mag extinction in the {\it V} band. The {\it W3} magnitude of the 
majority of CVs corresponds to upper limit values.}
\label{fig12}
\end{figure*}

\begin{figure*}
\includegraphics[scale=0.52]{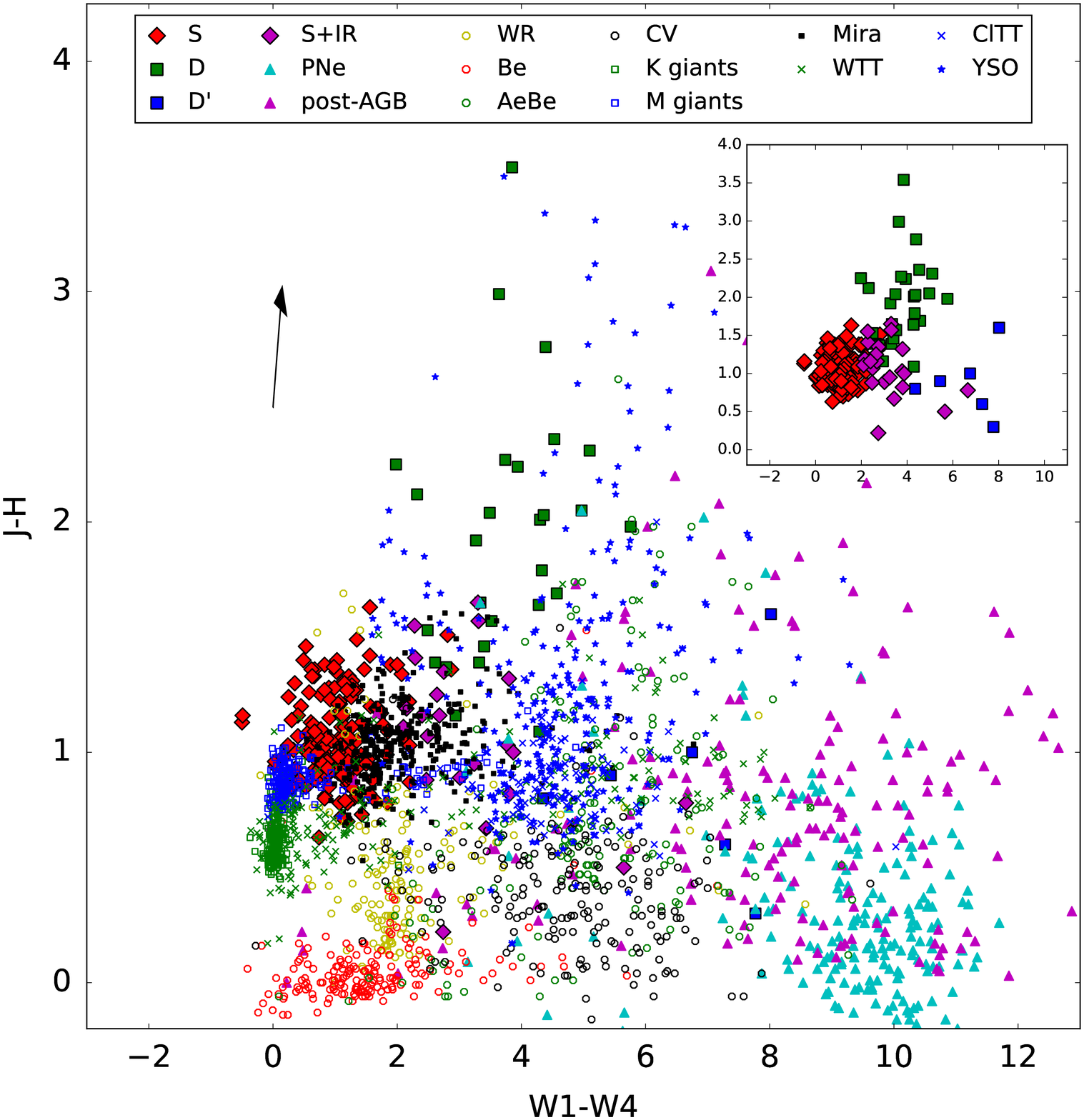}
\caption[]{The 2MASS/AllWISE {\it J--H} vs. {\it W1--W4} DCCD for different classes of objects as well as  for the four type of 
SySts presented in the inset plot. The black arrow corresponds to 4~mag extinction in the {\it V} band. The {\it W4} magnitude of the 
majority of CVs corresponds to upper limit values.}
\label{fig13}
\end{figure*}

\begin{figure*}
\includegraphics[scale=0.52]{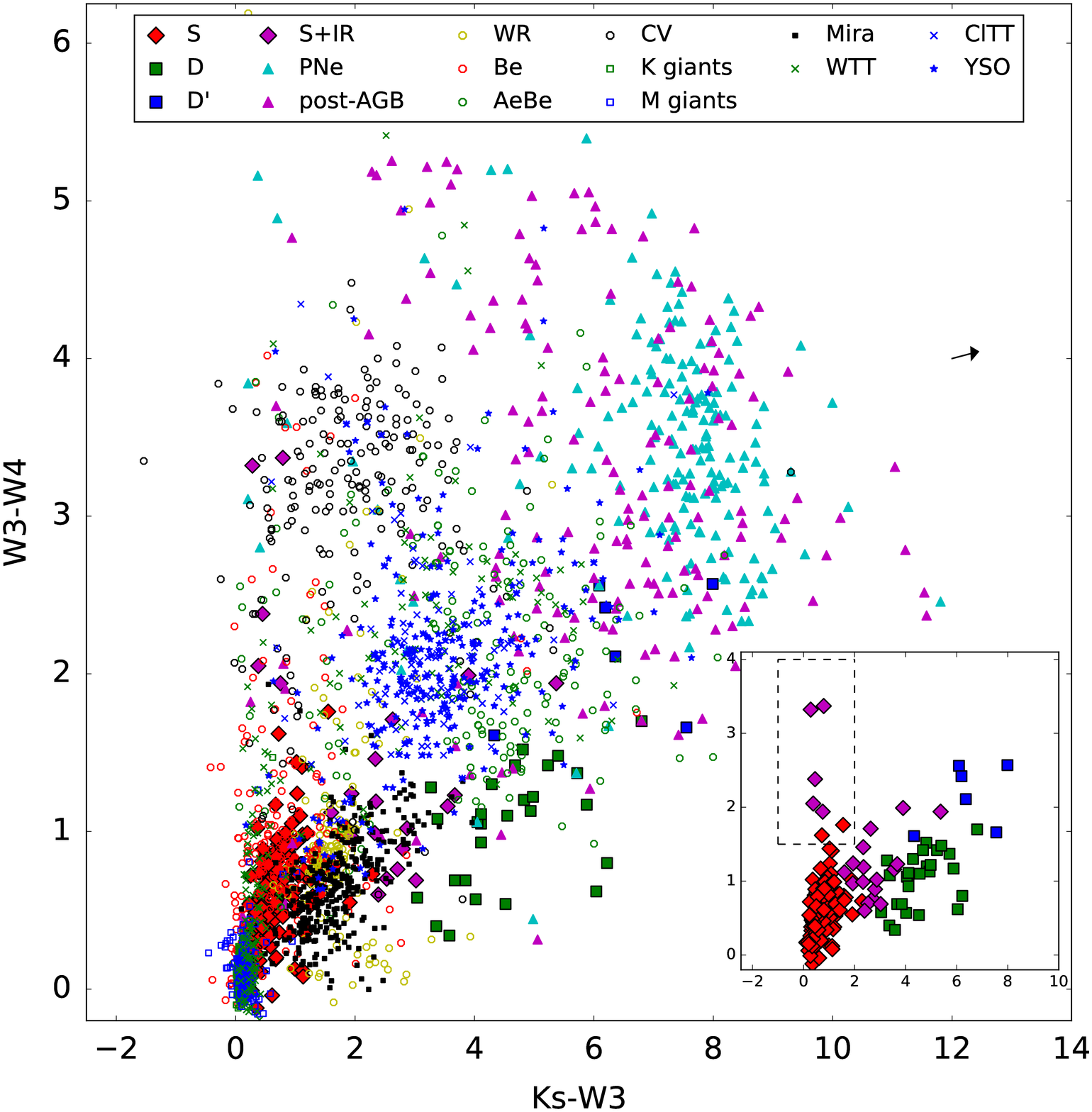}
\caption[]{The 2MASS/AllWISE {\it W3--W4} vs. {\it K$_s$--W3} DCCD for different classes of objects as well as  for the four type of 
SySts presented in the inset plot. The black arrow corresponds to 4~mag extinction in the {\it V} band. The dashed box in the inset plot 
indicates the vertical branched region of SySts discussed in the text. The {\it W3} and {\it W4} magnitudes of the majority of CVs 
correspond to upper limit values.}
\label{fig13b}
\end{figure*}

The 2MASS {\it J--H} vs. {\it H--K$_s$} DCCD has extensively been used to study the near-IR 
properties of SySts, to classify them into S- and D-types or to identify new candidates (Allen \& Glass 1974; Rodriguez-Flores 2006; Phillips 2007; Corradi et al. 2008, 2010; Baella, Pereira \& Miranda 2013; 
Baella et al. 2016; Clyne et al. 2015). 

Corradi et al. (2008) propose two specific regions in which the majority of the S- and D-type are placed. Few years later, these
regions were redefined by Rodriguez-Flores et al. (2014) being more restricted. In the {\it J--H} vs. {\it H--K$_s$} DCCD from
Corradi et al. (2008), one can see that there is a small overlap between the S- and D-types probably because of some mis-classifications. 
The same overlap is not observed in the DCCD from Rodriguez-Flores et al. (2014) due to the preliminary selection of 
SySts from the IPHAS {\it r}--\ha\ vs. {\it r--i} DCCD and the likely better classification by the authors (see Fig.~1 in Rodriguez-Flores et al. 2014). 

In Figure~\ref{fig9}, we present the 2MASS {\it J--H} vs. {\it H--K$_s$} DCCD for all the known Galactic SySts. For the vast majority of them, the classification is based on the SED profiles (Paper~I), whereas for those without an SED profile and thus a new classification, 
the old one has been considered. Besides SySts, various classes of objects that show \ha\ emission such as PNe, WR stars, Be and AeBe stars, 
CVs, Mira stars, CTT and WTT stars as well as post-AGB stars and single K-M giants are also included. 

\begin{figure*}
\includegraphics[scale=0.75]{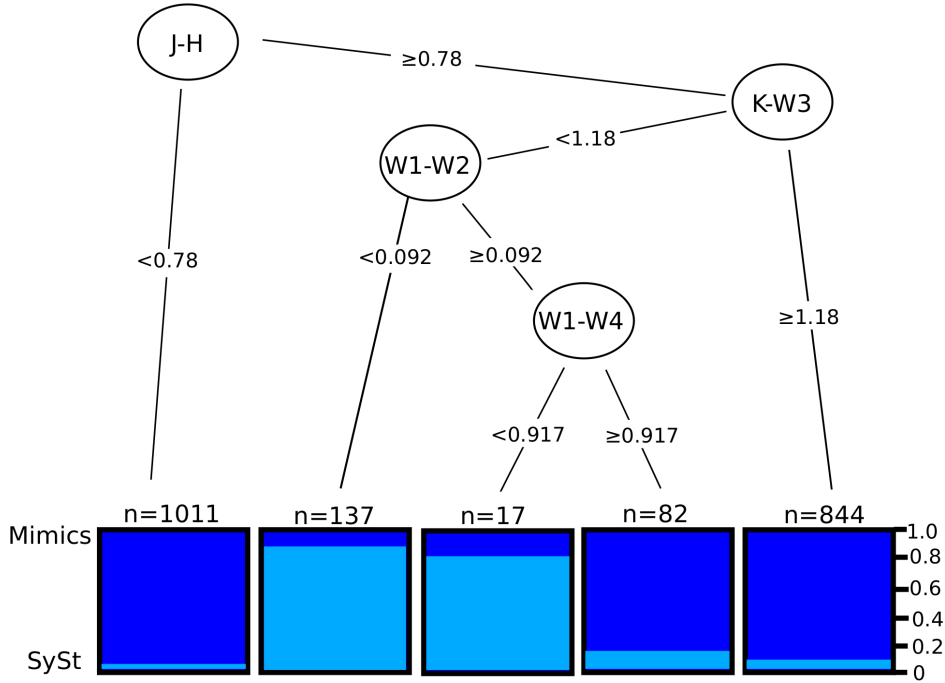}
\caption[]{Classification tree plot using as training sample the groups of all known Galactic SySts and the mimics. Each column represents
the population of the sources (normalised to one) that satisfy the colour criteria and the total numbers of these objects is given at the top 
of each column. The sum of the population in each column or criterion corresponds to the total population of the training sample. 
The colour criteria are given inside the ellipses. The dark and light blue colours correspond to the sample of the mimics and SySts, 
respectively. }
\label{figB1}
\end{figure*}

The sample size of all these classes of object as well as the references are given in Table~\ref{table4}.
The photometric magnitudes were obtained from the AllWISE (Cutri et al. 2014) and 2MASS (Cutri et al. 2003) catalogues 
using a searching radius of 6~arcsec due to the resolution of the {\it W3} and {\it W4} bands. For approximately 90 per cent of the sources, 
the cross-matching of the 2MASS and AllWISE was made in a radius less than 1~arcsec.
Only sources with actual measurements and no upper limit values were selected for all the classes of objects except 
CV\footnote{By verifying various catalogues of CVs, we found that the vast majority have only upper limit magnitudes in {\it W3} and 
{\it W4}. Moreover, more than 95 per cent of CVs have {\it J--H} colour index lower than 0.75 which means that the upper limit values 
magnitudes in {\it W3} and {\it W4} do not affect our classification tree model (see \S~3, Figure~\ref{figB1} and Figure~\ref{figB7}). 
Therefore, we decided not to excluded the CVs with upper limit {\it W3} and {\it W4} magnitudes in order to keep their sample size 
comparable with the rest of the mimics and the sample size of SySts.}.

Genuine PNe often mimic SySts mainly in the optical regime. In the 2MASS {\it J--H} vs. {\it H--K$_s$} DCCD, the majority of PNe are 
found to be bluer in the {\it H--J} colour index ($<$0.9) compared to SySts ($>$0.8) occupying the lower part of the DCCD. 
However, there is small number of PNe with {\it H--J}$>$0.9 that are mixed up with S- and D-type SySts. These are likely denser 
and younger members. On the other hand, post-AGB, YSO and AeBe stars are found to be well mixed with the D-type SySts. 
{\it This clearly illustrates than the dusty SySts cannot easily be distinguished from other dusty sources in the near-infrared wavelength regime.}
In the left part of the plot with  0$<${\it H--K$_s$}$<$1 (Fig.~\ref{fig9}), we find the locus of S-type SySts as well as a number 
of other sources such as single M and K type giants, WTT and ClTT stars and Mira stars. All these sources are well 
mixed making hard to distinguish them based only on the 2MASS colours. The single K and M type red giants are found to be  bluer in the 
{\it J--H} colour index ($<$1) compared to the S-type SySts with a cool companion of the same spectral type (see also Catchpole \& Glass 1974). 
A similar behaviour is also found between the single Mira stars and D-type SySts. This may be associated with a higher dust formation rate in 
SySts than in single giants. Evidence of fast rotation in some S-type SySts and the majority of \DD-types\ compared to 
single giants may indicate a substantial increase in mass-loss rate by a factor of 10 (Zamanov et al. 2006, 2008) or the higher mass-loss rate 
of symbiotic Mira stars compared to normal ones (Gromadzki et al. 2009). WTT and ClTT stars appear to occupy different areas in this DCCD. 
WTT are well mixed with S-type SySts while ClTT show a redder {\it H--K$_s$} colour index.	 
The bulk of WR stars is found to occupy a region between SySts and PNe. However there is a small number of WR stars which exhibit similar 
colour indices with the D- and \DD-type SySts. 
 
In the inset plot, we display for clarity only the four types of SySts. There is a clear separation between the S and D types 
which agrees with the regions defined by Corradi et al. (2008) and Rodriguez-Flores et al. (2014). \DD-type SySts are found to be highly 
dispersed in this DCCD without occupying any specific region with the {\it J--H} and {\it H--K$_s$} colour indices range from 0 to 1.75. 
The new S$+$IR-type SySts (see definition in Paper~I) are found to lie in the same region as the S-type. This is not 
surprising since the only difference between the S- and S$+$IR-types is, by definition, an infrared excess at longer wavelengths 
(11.6 and 22.1~$\mu$m), which explains why they have not been recovered before. 

The {\it J--K$_s$} vs. {\it Y--J} DCCD has also been reported for distinguishing SySts from PNe (Miszalski et al. 2011).
D-type lie in different region from S-type based on the {\it J--K$_s$} colour index (S-type:$\leq$2.20; D-type:$\geq$2.20) 
equivalently to the {\it J--H} colour index.  

Besides the common {\it J--H} vs. {\it H--K$_s$} DCCD, we explored all the possible DCCDs using all the different combinations between
2MASS and AllWISE data. We present, here, the most representatives DCCDs that provide a good separation among the different classes of objects: 
{\it H--W2} vs. {\it J--H} (Fig.~\ref{fig10}), {\it W3--W4} vs. {\it J--H} (Fig.~\ref{fig11}), {\it J--W1} vs. {\it K$_s$--W3} (Fig.~\ref{fig12}), 
{\it J--H} vs. {\it W1--W4} (Fig.~\ref{fig13}) and {\it W3--W4} vs. {\it K$_s$--W3} (Fig.~\ref{fig13b}).

The {\it H--W2} vs. {\it J--H} DCCD (Fig.~\ref{fig10}) provides a better separation among the different type of objects than the previous one. 
Mira and WTT stars, which are well mixed with the S-type SySts in the previous DCCD, are found to be redder in the {\it H--W2} 
colour index compared to the bulk of S-type and bluer compared to the D and \DD-ype SySts. PNe, post-AGB, YSO, D- and \DD-type SySts 
have the same range of {\it H--W2} colour index (from 2 to 7) but different  {\it J--H} colour index -- PNe are bluer and occupy the 
upper-left part, YSO are redder occupying the upper-right corner, while post-AGB, D and \DD-type SySts are well mixed and occupy the region
between PNe and YSO. Previous studies have shown that \DD-type SySts have SEDs that resemble those of post-AGB stars. 
Be, WR, CV and M/K giants are located in the lower-left corner of the DCCD.

Miszalski et al. (2011) argue that {\it J} --[4.5] colour index is ideal for separating PNe and H~{\sc ii} regions from SySts with the 
former having {\it J} -- [4.5]$<$4 and the latter $>$5 (see Fig. 7 in Miszalski et al. 2011). At least for the Galactic SySts, we 
find that the {\it J -- W2} colour index (not presented here, or equivalently {\it J} -- [4.5]) alone is not a good indicator for identify SySts as a significant 
number of PNe also exhibit {\it J} -- [4.5]$>$5. Reid (2014) also find the same result for PNe in the LMC.

The {\it W3--W4} vs. {\it J--H} DCCD (Fig.~\ref{fig11}) provides a good separation of SySts from other types of objects. In particular, 
D-type SySts are found to be located in the bottom-right corner of the DCCD while PNe/post-AGB stars are found in the upper-left corner and 
YSO in the centre of the plot. S-type are also well separated from sources like WTT, ClTT, K and M giant, Be and CV but not 
from Mira stars.

From this DCCD, we conclude that {\it W3--W4} colour index is a good indicator for SySts with the vast majority of them displaying
0$<${\it W3--W4}$<$1.5. Although, there is a small number of SySts with values between 1.5 and 4 (dashed-line box).
These objects deserve a further study in order to reveal their true nature and understand why they display higher 
{\it W3--W4} colour index while the {\it J--H} is nearly constant (see also Fig.~\ref{fig13b}). The possibility of unreliable {\it W3} 
and {\it W4} photometric magnitudes cannot be ruled out given that one SySt has photometric errors higher than 0.1~dex (UKS Ce-1).

The {\it J--W1} vs. {\it K$_s$--W3} DCCD (Fig.~\ref{fig12}) also separates the different classes of 
objects as well as the four types of SySts. In particular, the majority of S$+$IR-type SySts are found to occupy a specific 
region between the S- and D-types. This may suggest that the S$+$IR-type are a transition type between the S- and D-type SySts.
The intriguing \DD-type SySts have lower {\it J--W1} and higher {\it K$_s$--W3} colour indices compared to D-types but similar to 
those of post-AGB and PNe. Again, D-type SySts and YSO are found to occupy the same regime.
Moreover, Mira stars and D-types seem to form a continuous branch in which D-type are redder in both colours than single 
Mira stars due to the dusty shells around the binary systems in SySts. This agrees with the hypothesis that Mira stars in SySts have higher 
mass-loss rate compared to normal Mira stars (Gromadzki et al. 2009). S-type SySts (bluer in {\it K$_s$--W3}) are found to be well 
separated from Mira stars (redder in {\it K$_s$--W3}) in this DCCD. 

In the following DCCD ({\it J--H} vs. {\it W1--W4}, Fig.~\ref{fig13})  S-, S$+$IR-types and Mira are located in a region with 
{\it J--H}$\sim$1 and 0$<${\it W1--W4}$<$4. D-type, on the other hand, are clearly distinguished from all other objects having 
{\it J--H}$>$1.25 and 3$<${\it W1--W4}$<$6. The common mimics of D-type SySts, PNe and post-AGB, are located in the bottom-right 
corner of the DCCD. \DD-type SySts have colour indices similar to those of PNe and post-AGB and occupy the same region. 
The systematically low {\it W1--W4} colour index of D-type is attributed to a weaker emission in the 22$\mu$m relative to \DD-type, PNe 
and post-AGB stars.

The last DCCD, {\it W3--W4} vs. {\it K$_s$--W3}, (Fig.~\ref{fig13b}), provides the best separation among the four types of SySts 
covering different values ranging {\it K$_s$--W3}. In particular, S-, S$+$IR-, D- and \DD-type SySts exhibit {\it K$_s$--W3}$<$2, 
2$<${\it K$_s$--W3}$<$3, 3$<${\it K$_s$--W3}$<$6 and {\it K$_s$--W3}$>$6, respectively. {\it K$_s$--W3} index is, thus, a good indicator 
for an infrared excess or the presence of a dusty shell. Regarding the other classes of objects, D-type SySts are very well separated 
from all the dusty sources like PNe, post-AGB, YSO and AeBe with very little contamination. \DD-type are still hard to be separated from 
post-AGB and PNe.

S-types are well distinguished from WR and Mira stars which are found to be redder in the {\it K$_s$--W3} colour index by approximately 
1~dex. However, S-type are strongly contaminated with Be, WTT and K/M giants. The vertical branch of S- and S$+$IR-type SySts becomes 
apparent in this DCCD similar to Figure~\ref{fig13b} (bashed box in the inset plot). 
SySts in that region display {\it W3--W4}$>$2.0 occupying the same locus with CVs. 
From Figure~\ref{fig13b}, one can see that {\it W3--W4} increases with the increase of {\it K$_s$--W3} (e.g. blue square symbols for 
\DD-type SySts). We conclude that SySts with {\it W3--W4}$>$2.0 in the vertical branch of Figure~\ref{fig13b} are likely more dusty or 
the photometric data are uncertain. We argue that a more careful study of these specific S- and S$+$IR-type SySts is 
necessary. It is also worth mentioning that this DCCD is equivalent to the IRAS K-[12] vs. [12]-[25] DCCD from Luud \& Tuvikene (1987) 
who argued that it can identify the S-, D- and \DD-type SySts.

By cross-matching the WISE and {\it Kepler-Isaac Newton Telescope Survey} (KIS, Greiss et al.2012) catalogues, Scaringi et al. (2013) 
demonstrate that CVs occupy a specific region in the {\it W1--W2} vs {\it W2--W3} and {\it W1--W2} vs {\it W3--W4} DCCDs well 
separated from quasi-stellar objects (QSOs). Despite the low number of CVs, it seems that they exhibit {\it W2--W3}$>$1 and 
{\it W3--W4}$>$2.75. Our {\it W3--W4} vs. {\it K$_s$--W3} DCCD also provides the same result with the CVs lying in the region of 2.5$<${\it W3--W4}$<$4. But, most of the CVs have only upper limit 
{\it W3} and {\it W4} magnitudes which make their position quite uncertain. 

We also conclude that CVs, S- and S$+$IR SySts have {\it W1--W2} colour index between 0 and 0.4 in agreement with Debes et al. (2011). 
In particular, CVs have an average {\it W1--W2} colour index equal to 0.16 with a standard deviation of 0.24, S-type have 
an average value of 0.02 (SD=0.16) and S$+$IR-type have an average value of 0.37 (SD=0.32). Given that CVs, S- and S$+$IR SySts are 
composed of a white dwarf (WD), their {\it W1--W2} values are consistent with Debes et al. (2011) for single white dwarfs.
Regarding D-, \DD-type SySts and PNe, which are also composed of a WD, we find systematically higher {\it W1--W2} colour index 
of 1.12 (SD=0.43), 1.16 (SD=0.45) and 0.91 (SD=0.46), respectively. Dust emission in these specific classes of objects is strong enough to 
overwhelm the emission from the WDs resulting in higher colour index compared to the stellar SySts and CVs.


Overall, these new 2MASS/AllWISE DCCDs provide essential information for studying SySts as well as 
distinguish them from other stellar objects.  {\it J--H} vs. {\it H--K$_s$}, {\it H--W2} vs. {\it J--H} and 
{\it J--W1} vs. {\it K$_s$--W3} DCCDs provide a good separation among  SySts, PNe, YSO and post-AGB stars. The last two DCCDs also  
separate S-type SySts from Mira stars. The {\it W3--W4} vs. {\it J--H} and {\it J--H} vs. {\it W1--W4} DCCDs 
can distinguish SySts from mimics like CVs, WR, WTT, ClTT star, Be stars and single K and M giants. 

The {\it K$_s$--W3} vs. {\it W1--W4} DCCD has been proposed to separate very well SySts from PNe and ClTT stars, but only once sources like Mira, CVs, Be stars have already been discarded from the samples based on their \ha\ emission i.e. {\it r--i} vs. {\it r}--\ha\ IPHAS DCCD (Corradi private communication). 

It should be noted that none of the previous DCCDs provide an adequate separation between the \DD-type SySts and PNe occupying 
the same regions on these IR DCCDs. On the other hand, D-types are better distinguished (e.g. {\it J--H} vs. {\it H--K$_s$} or 
{\it J--H} vs. {\it W1--W4}). 

This difference between the \DD- and D-type SySts is likely associated with the progenitor of the circumstellar nebula around these systems: 
(i) the hot WD companion when entered in the AGB phase (\DD-type) or (ii) the cold giant companion (D-type) (Schwarz \& Corradi 1992; 
Munari \& Patat 1993; Pereira, Smith, Cunha 2005). According to the first scenario, a \DD-type SySt can also be consider as a genuine 
PNe since the circumstellar envelope has been expelled by the same star than ionizes it, while the second scenario implies that the 
circumstellar evelope, ionized by the WD, is the material lost by the cold giant and transferred to the WD. Therefore, 
there may exist objects with a dual nature, classified either as SySts or PNe with a binary central system.

\section{Classification trees}
DCCDs are widely used to distinguish different classes of objects as well as to find new candidates.
However, in order to perform a more quantitative analysis and derive the criteria that can easily be used, 
the machine learning algorithm of classification tree is used (Moret 1982, Buntine 1993).

In astrophysics, the classification tree method has been applied to set of observable parameters in order to derive the 
criteria that provide the lowest contaminated groups (e.g. da Silva, Milone \& Rocha-Pinto 2015). For this analysis, 
the evtree function (Grubinger, Zeileis \& Pfeiffer 2011) in R software R was used with a set of 10 
representative colour indices ({\it J--H}, {\it H--K$_s$}, {\it K$_s$--W1}, {\it W1--W2}, {\it W2--W3}, {\it W3--W4}, 
{\it W1--W4}, {\it J--W1}, {\it H--W2} and {\it K$_s$--W3}).
 
As a training sample for our classification tree model, we used all possible sources that mimics SySts in the optical regime i.e. \ha\ emitters, 
as the criteria to identify SySts are based on information in the optical wavelengths (see Kenyon 1986; Mikolajewska, Acker,
Stenholm 1997; Belczynski et al. 2000). There are several classes of objects that show \ha\ emission like SySts, but we 
selected only the bright ones such as PNe, WR, Mira, CVs, YSO, ClTT, WTT, AeBe and Be sources (see also Witham et al. 2006, 
Corradi et al. 2008) which satisfy the IPHAS criterion ({\it (r-\ha)$\geq$ 0.25$\times$(r-i)+0.65}, Corradi et al. 2008) and 
it is applied to the IPHAS and VPHAS+ catalogues to get the final samples (see Section~5).

Late K or M dwarf stars as well as main sequence stars and red giant stars also show \ha\ emission. However, 
it is generally very weak and all these sources can be easily distinguished from SySts (see Corradi et al. 2008). Supergiants are also 
\ha\ emitters but according to their synthetic {\it r-\ha} and {\it r-i} colour indices (Drew et al. 2005), they occupy a totally 
different in the IPHAS diagnostic diagram. Therefore, we decided not to include them in our training sample, since the IPHAS criterion 
will automatically excluded them from our validate samples.
 
We also excluded H~II regions and QSOs, two known \ha\ emitters for two reasons: (i) our analysis is focused on the Milky way and H~II 
regions can be easily distinguished from compact SySts by looking at the observed images and (ii) the prior star/galaxy separation in 
photometric surveys minimises the contamination from QSOs. We should note here that in extragalactic surveys of SySts, the 
contamination of H~II regions (or diffuse ionized gas, Mikolajewska et al. 2017) is significant and it has to be taken into consideration.

The main goal of this work is to reveal the hidden SySts population in \ha\ photometric surveys like IPHAS, VPHAS+, {\it the Javalambre Physics 
of the Accelerating Universe Astrophysical Survey} (J-PAS, Benitez et al. 2014), {\it the Javalambre Photometric Local Universe Survey} 
(J-PLUS, Cenarro et al. 2018, submitted) and {\it the Southern Photometric Local Universe Survey} (S-PLUS, Mendes de Oliveira et al. submitted), 
among others, taking into considerations, apart from the \ha\ excess, the 2MASS and AllWISE data. 

Table~\ref{table4} lists the most common mimics of SySts that may occupy the same area in the {\it (r-\ha)} vs {\it (r-i)} DCCD 
(Corradi et al. 2008) as well as their sample sizes and the references. All the mimics have sample sizes approximately equal to the 
population of known Galactic SySts with no upper limit values (220). As we are interested in searching for the criteria that separate 
better SySts from all the mimics in Table~\ref{table4}, we merged their samples into one sample, namely "Mimics". 
This yields to a training sample of 220 SySts and 1871 mimics or in other words imbalanced training samples. This is a 
well known problem in data mining (e.g. He \& Garcia 2009). Given that our goal is to identify the minority class (SySts), 
it may impose a bias to the resulting model toward to the majority class. A few methods 
have been developed to overcome the imbalanced learning problem such as oversampling, undersampling or synthetic sampling among others 
(e.g. Weiss \& Provost 2003; He \& Garcia 2009; Longadge, Dongre, Malik 2013 and references therein). 

In our case, the {\it between-class imbalance} of our training sample is of the order of 8.5:1, which is not such high but at the same 
time enough to be considered as imbalanced. If a training sample that represents the real population of SySts and mimics in the Milky Way is constructed, it may result in a significantly higher {\it between-class imbalance} and presumably to a poorer classification model biased 
towards the mimics. In addition to that, the true Galactic population of SySts (between 2000 and 400000) as well as of mimics are not 
very accurate and may provide less representative and more problematic training samples. It has been demonstrated that training 
samples with small sizes can also provide good classification models as training samples with bigger sizes (Weiss \& Provost 2003). 

By keeping constant the sample size of mimics and randomly reducing the sample size of SySts to 50, 100 and 150, or in other words 
increasing the {\it between-class imbalance} to 38:1, 19:1 and 13:1, respectively, we found that the colour criteria change no more than 
8-9 per cent relative to the values in Figure~\ref{figB1}. The lower the size of SySts the high the difference. For instance, 
the low prevalence of S+IR-type SySts in the training samples results in lower {\it W1--W2} colour.  This is a characteristic example of 
training samples that suffer from lack of information (Visa \& Ralescu 2005). Recall that we are interested in the minority class of SySts, 
their whole sample size of 220 sources provides all the available information of this class of objects. 
 
By replicating the sample of SySts a few times (oversampling method), the distribution of SySts in various colour indices becomes significantly different compared to the distribution of mimics. For instance, the numbers of D-types and \DD-type substantially increase relative to the number of dusty mimics like YSO and PNe. Equivalently, if we randomly reduce the number of mimics (undersampling method), we may get significantly less Miras, WTT or ClTT star relative to S-type SySts. Because of the small number of known SySts, these two methods of re-sampling the training samples 
are not ideal.

Despite our models are eventually trained using imbalanced samples, the construction of the training samples with equal populations for all classes 
of mimics and SySts assures that they are unbiased towards any of these sources.

Classification tree was also applied to several training samples with three different classes of objects each 
in order to derive those colour criteria that identify SySts among various classes of objects. A training sample with the four types 
of SySts (S-type/S$+$IR-type/D-type/\DD-type) was also used to train our model in order to seek for the colour criteria that can 
separate these four types.

\begin{figure*}
\includegraphics[scale=0.75]{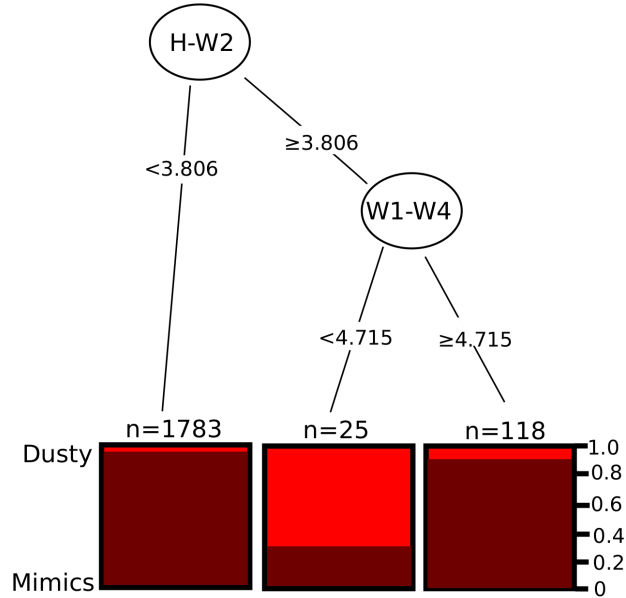}
\caption[]{Classification tree plot using as training sample the groups of all dusty (S+IR, D and \DD) known Galactic SySts and 
all the mimics.The dark and light red colours correspond to the sample of the mimics and dusty SySts, respectively.}
\label{figB3}
\end{figure*}

Figure~\ref{figB1} displays the classification tree plot using as a training sample the group of all the known Galactic SySts 
(light blue) and mimics (dark blue). The majority of SySts can be distinguished from their mimics using the {\it J--H}, 
{\it K$_s$--W3}, {\it W1--W2} and {\it W1--W4} colour indices. Almost 50 per cent of the population of mimics (1011 sources) 
exhibit {\it J--H}$<$0.78 while SySts appear to have {\it J--H}$>$0.78. This first group contains mainly Be, CV, PNe WR, and 
WTT as well as a small number of S-type SySts with {\it J--H}$<$0.78 (10 sources).
The second criterion {\it K$_s$--W3}$<$1.18 separates SySts from the remaining mimics. However, there is a number of 
SySts (62 sources) that exhibit {\it K$_s$--W3}$>$1.18 and they are misclassified. These SySts are mainly the dusty ones like 
S+IR-, D- or \DD-type. From the DCCDs above, we have shown that the dusty SySts are well mixed with YSO, PNe, and AeBe stars
and they are very hard to be distinguished. The third and forth criteria give us all the remaining S-type SySts. 
In particular, 137 S-type SySts or 83 per cent satisfy the criterion {\it W1--W2}$<$0.092 with 10 per cent contamination\footnote{ 
the contamination levels are given relative to the total number of objects that satisfy a specific criterion},
whereas 15 S-type pass the fourth criterion {\it W1--W4}$<$0.917 with 15 per cent contamination. 
In total, these two criterion give us 93 per cent of the S-type SySts and they are mainly contaminated with K/M giants, 
WTT and Mira stars. Overall, we argue that these four colour criteria can be used to distinguish and identify S-type 
SySts with an accuracy up to 90 per cent.

In order to find the right colour criteria that distinguish the dusty SySts (S+IR, D and \DD), we used as a training sample the subgroup 
of the dusty SySts and the sample of mimics (Figure~\ref{figB3}). Two criteria {\it H--W2}$>$3.806 and {\it W1--W4}$<$4.715 are found to 
provide the best combination for identifying dusty SySts. However, these colour criteria are not as good as the previous ones of S-type 
for two reasons: (i) only 25 SySts or 45 per cent satisfy both criteria and (ii) the high contamination of 25 per cent with other classes of objects. 
Examining all the dusty SySts one by one, we conclude that the criteria works only for the D-type SySts. The S$+$IR have {\it H--W2}$<$3.806 
and the \DD-type {\it W1--W4}$>$4.715 being misclassified (see also Fig.~\ref{figB10}).

The overall accuracy of the algorithm was verified by randomly selected 80\% of the Galactic SySts sample as 
training set and 20\% as testing set and repeated it for a few times. We find an accuracy range from 71\% to 77\% while the values of 
the criteria vary from 5 to 8\%. The low accuracy of our model is attributed to the dusty SySts which cannot 
be distinguished from other dusty sources (e.g. PNe, YSO). Hence, we repeated the same procedure but this time the testing 
set was generated by randomly selected 20\% from the S-type SySts. In this case, the accuracy of the method becomes higher from 82\% to 88\%.
Moreover, the false identifications were found to be around 1\% for the mimics and 13\% for the S-type SySts.

\begin{figure*}
\includegraphics[scale=0.75]{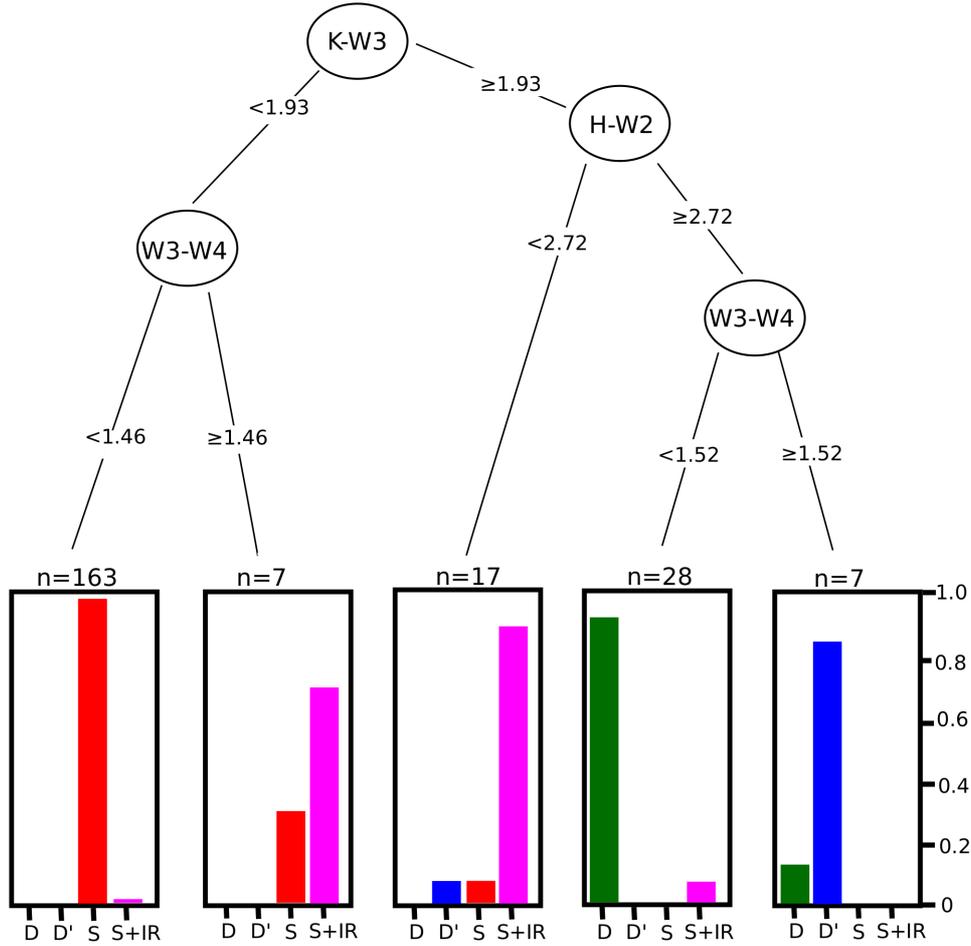}
\caption[]{Classification tree plot using as the training sample the four types of SySts.}
\label{figB2}
\end{figure*}

Figure~\ref{figB2} displays the classification tree plot for the four types of SySts (S, S$+$IR, D and \DD). The {\it K$_s$--W3}
is the principal criterion that distinguishes the stellar from the dusty SySts. The left branch includes those SySts that are bluer in the 
{\it K$_s$--W3} ($<$1.93) and they are divided into two subgroups: the S-type with {\it W3--W4}$<$1.46 (163 sources or 99 per cent) suffering of only 0.5 percent contamination from S$+$IR and the S$+$IR-type with {\it W3--W4}$>$1.46 (5 sources or only 22 per cent) suffering of 29 per cent 
contamination from S-type. The right branch ({\it K$_s$--W3}$>$1.93) contains mainly all the dusty SySts. The S$+$IR-type have {\it H--W2}$<$2.72 
with 11 per cent contamination (one S-type and one D-type) whereas the more dusty SySts (D and \DD) have {\it H--W2}$>$2.72 (i.e. strong 
infrared excess) and they are further separated by the {\it W3--W4} colour (D-type: {\it W3--W4}$<$1.52 (97 per cent);
\DD-type: {\it W3--W4}$>$1.52 (86 per cent)).

Evidently, the {\it H--W2} and {\it W3--W4} indices are strong indicatives for the presence of a dusty shell in SySts and likely in any 
other dusty sources (see Fig.~\ref{figB3}). Moreover, if we take the S$+$IR-type SySts out from this analysis as well as the {\it H--W2} 
colour index criterion, the classification tree yields as the best indicators the {\it K$_s$--W3} and {\it W3--W4} colour indices. 
This is the result in which  Luud \& Tuvikene (1987) concluded using the IRAS colours -- K-[12] and [12]-[25] -- which are equivalent to 
our colour criteria.

\begin{figure}
\includegraphics[scale=0.55]{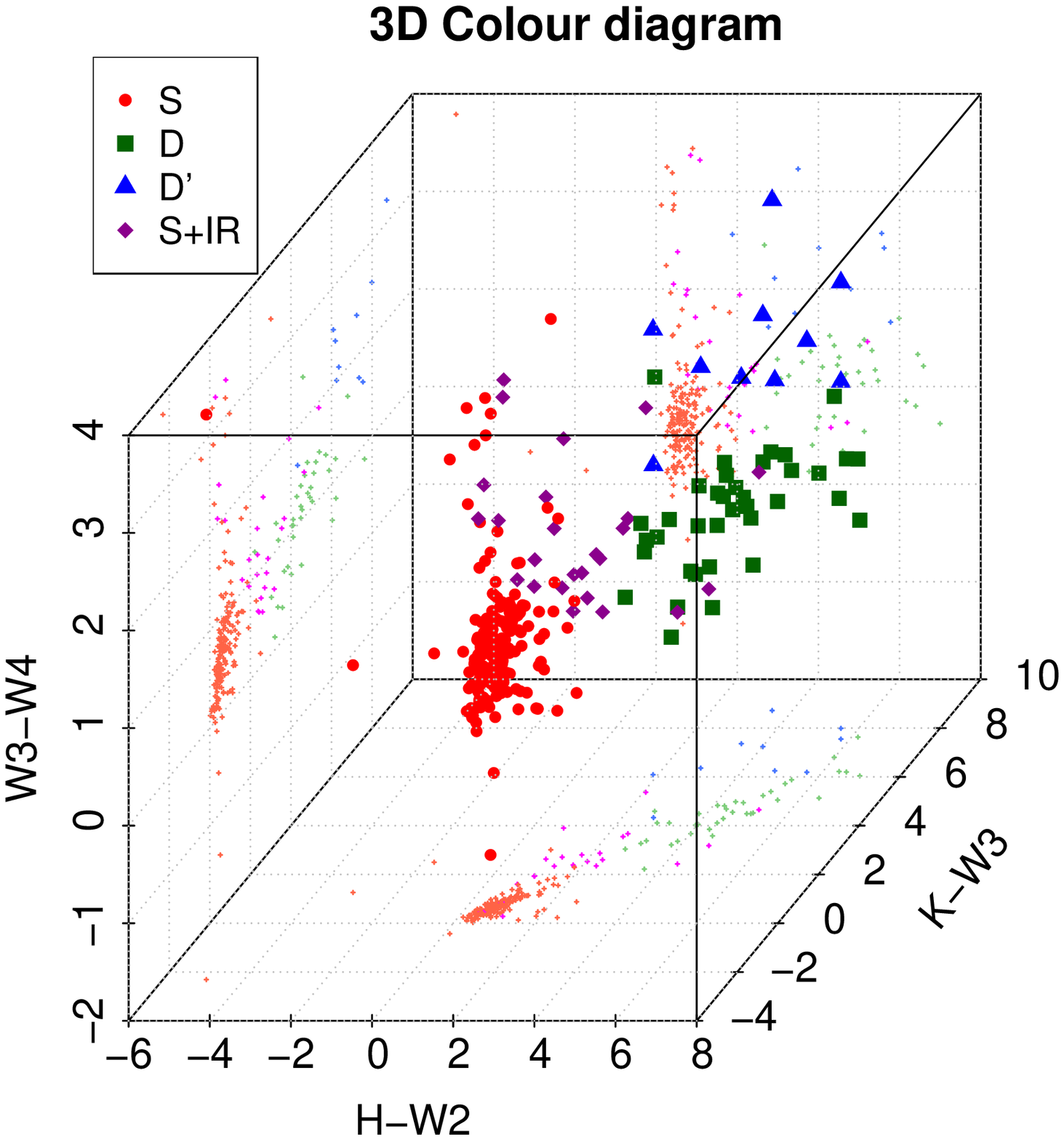}
\caption[]{3D Colour diagram of the four types of SySts. The colour indices have been obtained from the classification 
tree algorithm.}
\label{fig15a}
\end{figure}

Figure~\ref{fig15a} illustrates a 3D colour plot among the most relevant colour indices according to the results from the classification tree 
algorithm. The four SySt types can be better illustrated in this 3D colour diagram than the previous 2D DCCD. S$+$IR-type SySts 
are vividly occuping a different region between the S- and D-type SySts, whereas two distinct locus are also defined for the D- and 
\DD-types SySts.

We then used the following training samples with different classes of source in order to find those criteria that distinguish SySts from 
specific classes of sources: (i) SySts/PNe/Be, (ii) SySts/CV/Mira, (iii) SySts/CV/YSO, (iv) SySts/WR/post-AGB, (v) SySts/K-giants/M-giants, 
(vi) SySts/WTT/ClTT, (vii) SySts/Be/AeBe. The first training sample includes SySts, PNe and Be (see Fig.~\ref{figB4}), two of the most 
common mimic of SySts due to the emission of several common lines. The first {\it W1--W4} colour index criterion discriminates SySts 
and Be stars from PNe. Almost all PNe (166 sources or 88 per cent) satisfy the criterion {\it W1--W4}$>$7.285 while they suffer by a 3 per 
cent contamination from Be and SySts. Interestingly, all SySts are \DD-type, a further proof that \DD-type SySts do resemble PNe. 
It is worth mentioning that except one \DD-type SySt (K~5-33) none of them emit the \ovi\ $\lambda$6830 Raman-scattered line, 
which is a strong indicator of the symbiotic activity (Paper~I). Therefore, an additional confirmation of this line in K~5-33 is necessary due to 
the low signal-to-noise ratio of its detection (Miszalski, Mikolajewska \& Udalski 2013). 
Be stars and SySts exhibit lower infrared excess {\it W1--W4}$<$7.285 compared to PNe. 98 per cent of Be stars show {\it J--H}$<$0.541.
On the other hand, 96 per cent of the known Galactic SySts show {\it J--H}$>$0.541 and {\it W3--W4}$<$2.56 suffering of only 3 per cent 
contamination. Finally, those nine sources (seven PNe, two SySts and one Be star) with {\it W3--W4}$>$2.56 deserve further study in order to 
explore possible link among them. 

Figure~\ref{figB5} shows the classification tree plot between SySts, CV and Mira stars. At this point, we have to clarify 
that only a set of 6 colour indices were used in our classification tree model. Due to the upper limit magnitudes of CVs in 
{\it W3} and {\it W4}, we did not use the colours {\it W2--W3}, {\it W3--W4}, {\it W1--W4} and {\it K$_s$--W3}).
SySts are separated into two groups depending on the {\it W1--W2} colour index. SySts with {\it W1--W2}$<$0.151 are classified as S-type 
and they are systematically redder in the {\it J--H} colour than CVs. On the other side, the dusty SySts exhibit {\it W1--W2}$>$0.151 but 
they are the minority compared with Mira stars and CVs which show clearly different {\it J--H} colours (CVs: {\it J--H}$<$0.684, Mira: 
{\it J--H}$>$0.684)).

The next training sample contains SySts, CV and YSO (Figure~\ref{figB7}). In this case, CVs are easily separated from SySts and YSO 
based on the criterion {\it J--H}$<$0.663 and the upper limit magnitudes in {\it W3} and {\it W4} do not affect our model. SySts and YSO 
exhibit both {\it J--H}$>$0.663 and they are distinguished based on the {\it K$_{\rm s}$--W3} colour. SySts with {\it K$_{\rm s}$-W3}$<$1.344 correspond to S-type whereas those with {\it W3--W4}$>$1.344 correspond to the dusty SySts and are mixed with YSO. 

SySts, post-AGB and WR stars can also be separated very well using the {\it J--H} and {\it W1--W4} colour indices (Fig.~\ref{figB8}). 
The vast majority of post-AGB stars show {\it W1--W4}$>$4.735 and they are contaminated by only few WR and SySts.
SySts (92 per cent) are found to exhibit {\it W1--W4}$<$4.735 and {\it J--H}$>$0.774 while WR stars have  {\it J--H}$<$0.774 

In Figure~\ref{figB9}, we present the classification tree plot for the training sample among SySts and M/K giants. Given that S-type SySts 
have a M or K giant companion, it is coherent to explore the colour indices that discriminate SySts from single M/K giants. 
Almost all SySts (95 per cent) have {\it H--W2}$>$0.206 and {\it K$_{\rm s}$--W3}$>$0.27 with a very small contamination mostly from 
M giants and few K giants (3 per cent), which should be further investigated. Approximately 50 per cent of M giants show the same {\it H--W2} 
colour with SySts but are bluer in the {\it K$_{\rm s}$--W3} colour index ($<$0.27). Regarding the K giants, the majority of them shows 
{\it H--W2} colour index $<$0.206 and they are separated from M giants based on the {\it W2--W3} colour index. A similar work using the 
{\it i}, {\it Y} or {\it Z} bands may be very useful for discriminating SySts and red giants.

The resultant classification tree plot among the SySts, WTT and ClTT is displayed in Figure~\ref{figB10}. Almost all SySts have
{\it W3--W4}$<$1.483 and {\it J--H}$>$0.78 and they suffer of only 5 per cent contamination. Half of WTT stars are found to be 
bluer in the {\it J--H} colour index ($<$0.78) compared to SySts. The remaining of WTT have {\it W3--W4}$>$1.483 and they are mixed 
with ClTT and few SySts. The {\it W1--W2} colour index separates further the WTT and the ClTT stars with the former being bluer and the latter redder.

AeBe stars also emit strong Balmer lines and mimic SySts in the optical regime. It is thus consistent to train our model 
with a training sample among SySts, Be and AeBe stars (Figure~\ref{figB11}). The {\it W1--W4} colour index is the first 
criterion that strongly discriminated SySts and Be stars from AeBe stars.
SySts and Be are found to be bluer in the {\it W1--W4} colour ($<$3.949) compared to the AeBe stars ($>$3.949). SySts and Be stars 
are further separated based on the {\it J--H} colour index (Be$<$0.63, SySts$>$0.63). The contamination of these two groups is small of the order of 
5.9 and 1.7 percent, respectively. On the other hand, AeBe stars exhibit {\it W1--W2}$>$0.03 with a very small contamination of SySts and Be.

Overall, we conclude that primarily {\it J--H}, {\it W1--W4} and {\it K$_s$--W3} and secondarily {\it H--W2}, {\it W1--W2} and {\it W3--W4} 
colour indices provide the best combinations of colours for distinguishing SySts from their mimics. Classification tree is evidently a 
powerful statistical tool to separate/classify different classes of objects, especially the current epoch with so many ongoing and upcoming 
photometric surveys. 

In this work, we have used only 2MASS and WISE photometry, but the classification may be even expand to other wavelengths 
ranges as well.

\section{Linear discriminant analysis and K-nearest neighbour method}

\subsection{LDA}

For a more robust identification/discrimination of SySts, we also explore the linear discriminant analysis (LDA) method or the Fisher 
discrimination analysis (Fisher 1936, Rao 1948). The main idea of this technique is to find the discriminant components, the linear correlation of 
a set of observed variables (such as the 2MASS and WISE photometry), which better distinguish two known groups 
of object (e.g. da Silva, Milone \& Rocha-Pinto 2015). For this analysis the lda and predict functions in R software 
were used R as well as the training samples SySts/PNe/Be, SySts/CV/Mira, SySts/CV/YSO, SySts/WR/post-AGB, SySts/K-giants/M-giants,
SySts/WTT/ClTT and SySts/Be/AeBe. The LDA algorithm is not applied to the training sample of SySts and mimics because of their imbalanced 
samples which would result to a poor classification. LDA is more depended on the sample sizes of training samples than classification tree. 

First of all, we applied the LDA method to the training sample of the different types of SySts in order to examine how the different type 
of SySts are separated. The resulting discriminant components LD1 and LD2 that provide the best separation of the four type of SySts are given 
below. 

\begin{eqnarray}
  \label{ldaeq}
LD1&=&1.947+0.314{\it J}-0.663{\it H}-1.426{\it K}\\&&-0.373{\it W1}+
1.385{\it W2}+0.100{\it W3}+0.742{\it W4} \nonumber \\
LD2&=&-1.236+1.187{\it J}+1.967{\it H}-4.022{\it K}\\&&+1.235{\it W1}-
1.417{\it W2}-0.033{\it W3}+1.081{\it W4} \nonumber 
\end{eqnarray}
 
\begin{figure}
\includegraphics[scale=0.75]{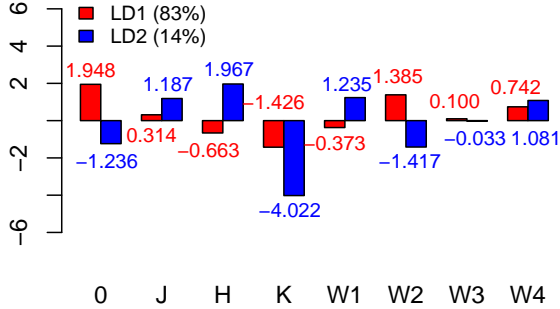}
\caption[]{Coefficient spectrum plot among the four types of SySts. The \rq\rq proportion of trace" or discriminability  of each component 
is given in percentage (see text for more details). The third linear discriminant component is not presented since provides only 1\% of discriminability. The \lq\lq 0" parameter corresponds to the zero point of the linear discriminant components due to the scaling so 
that the variables have mean value zero.}
\label{fig16}
\end{figure}

Figure~\ref{fig16} displays the coefficient spectrum plot of the discriminant components. In all the plots of Table~\ref{table5}, 
the red colour corresponds to the first discriminant component (LD1) and the blue to the second component (LD2)\footnote{For the cases in which 
CVs are used the {\it W3} and {\it W4} magnitudes were not used due to the upper limit values of CVs in these two bands.}. Moreover, 
the so-called \lq\lq proportion of trace" or discriminability of each component -- the proportion of each component that explains the 
between-groups variance in a given data set -- is given in percentage for each component. For the case of the four types of SySts, 
the discriminability is found 0.84 and 0.14 per cent for the to components, respectively.

\subsection{KNN}

In addition to the LDA algorithm, we also apply the K-nearest neighbour (KNN) method on the LDA components in order to explore the locus 
of each type of SySts or among the different class of sources in the training samples.

To perform this analysis, as describe below, we used the following external R packages: {\it class} to apply the KNN method, {\it dplyr} 
for data manipulation and {\it ggplot2} for graphical purpose. 
First, we randomly mixed our sample of 220 known Galactic SySts (with available 2MASS and AllWISE data) and selected 80$\%$ of them as training 
set and 20$\%$ as testing set. Secondly, the LD1 and LD2 entries were normalized (equations in Appendix B) in order to avoid different 
weights between the parameters. Then, we performed the KNN calculation in the training sample and used the testing sample to verify the 
accuracy of the results. This procedure was repeated a few times to examine how the accuracy would change when different random training/testing 
sample are chosen. Assuming the accuracy increases with the growth of the sample, we applied the KNN technique one last time, 
but now to the whole sample of 220 Galactic SySts. We find that the accuracy is in the range between 85 and 96 per cent.

\begin{figure}
\includegraphics[scale=0.50]{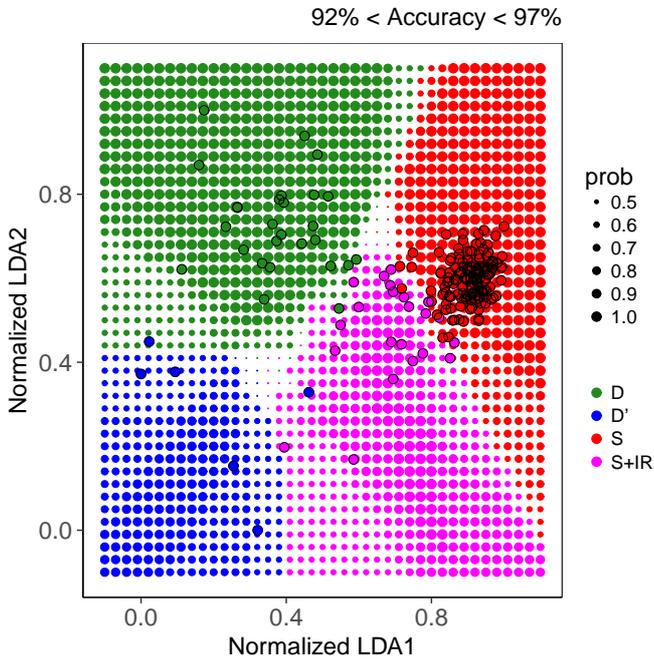}
\caption[]{Plot of normalised LDA components overlaid the four regions defined by the KNN algorithm of each SySt type (LDA/KNN plot). 
Different colour corresponds to different type of SySts. The size of the background circles corresponds to the probability of being 
classified as a specific type.}
\label{fig17}
\end{figure}

In Figure~\ref{fig17}, we present the normalised LD1 versus LD2 plot and the results obtained from the KNN analysis. The dots with a black contour 
represent the observed data and the background dots (with no contours) identify the regions expected for each one of the SySts types to occur. 
The size of the background dots is proportional to the probability of a given source of belonging to that class. We notice some clear overfitting 
in a few regions of the border between the different types (e.g. isolated, tiny background red dots appear around x = 0.4, y = 0.4). 
This occur in regions where probabilities are anyway low, and are similar for two or more types. Thus, the classification of an unknown 
source is uncertain on those areas. In conclusion, one should always have the probabilities in mind when studying this plot.

It is important to clarify that for the LDA and KNN analysis we conclude that the "a priori" classification of each Galactic SySt 
(taken from Paper~I) is correct except from some cases. These classifications were used as input in the LDA and KNN calculations. 
The outliers that one can identify in Figure~\ref{fig17} do not imply that they are wrongly classified before, as we do expect that some mixture naturally exists between the different type clusters. Thus, as stated above, a random position in the S+IR region (magenta colour) of the plot can still has a non-zero probability to shelter D-type (green) sources for example. In conclusion, the KNN result should not be used to reclassify the objects originally used to calculate the KNN (however, it would be interesting to further investigate the nature of such outliers).

Figure~\ref{fig17} can be used to help classifying any newly observed source. To transform any set of 2MASS and AllWISE magnitudes into 
the coordinates of the figure, one should apply the relations:
  \begin{eqnarray}
    Normalized LD1 = \frac{(LD1 + 10.20)}{13.21} \\
    Normalized LD2 = \frac{(LD2 + 8.12)}{13.44},
  \end{eqnarray}

where LD1 and LD2 are given by the equations 1 and 2.
  
The advantage of applying KNN to the LDA components is that the locus for each type of SySts, or different class of objects can be 
defined with a more robust technique whereas the probabilities of being a specific type of objects are also determined.

\subsection{Characterising SySts with LDA and KNN}

We then applied the LDA and KNN methods to the training samples of different classes of objects -- also applied to the 
classification tree -- in order to find those models that provide the best discrimination. For several training samples, 
we find an LDA model that separates SySts from other stellar objects very well. The coefficients of the first (LD1) and the second 
(LD2) discriminant components as well as the LDA/KNN plots of the normalised LDA components are given in Table~\ref{table5}. 

The LDA method provides a very important framework for discriminating objects. For most of the training samples examined in this work, 
SySts are found to fill a clearly distinguished locus. However, the discrimination is not the best for the case of the SySts/Mira/CV and 
SySts/CV/YSO training samples due to the limitation of not using the {\it W3} and {\it W4} magnitudes. 

\section{Application to real data}
Having already developed the classification criteria, the next logical step is to look for new candidate SySts in publicly 
available catalogues. Our classification tree criteria derived from the training sample of SySts and mimics were applied 
to the list of candidate SySts (Paper~I, Belczy\'{n}ski et al. 2000), to the IPHAS list of candidate SySts (Corradi et al. 2008) and 
finally to the second data release of the VPHAS+ catalogue (Drew et al. 2014) in order to search for a hidden SySts population.

We found 13 strong candidate SySts in the list of candidates from Paper~I, 9 new candidates in the IPHAS list 
of candidates SySts (2 S-type and 7 D-type), and 63 new candidates (34 S-type and 29 D-type) in the DR2 VPHAS+ catalogue (Table~\ref{tableC1}). 
The classification tree criteria were applied directly to the first two list of candidates whereas for the VPHAS+ catalogue, we should 
first apply the IPHAS \ha\ excess criterion in order to get only the sample of strong \ha-emitters (Corradi et al. 2008, Rodr\'{i}guez-Flores et al. 2014). Additional criteria regarding the quality of the measurements (photometric errors) were also applied. In particular, we accepted as candidates 
only the sources with errors in the \ha, r and i bands less than 0.1 (or signal to noise higher than 10) for both catalogues, IPHAS 
and VPHAS+. Moreover, we restricted the selection of the candidates based on their error in the 2MASS (less than 0.2) and AllWISE 
data (less than 0.3). The AllWISE images of the candidates were also visually inspected to ensure that a compact source is present in 
all the bands.

The particular case of W16-294 in the list of candidates SySts (Paper~I), a sources with strong \ha\ and \heliumb\ lines as well as a 
red continuum spectrum of a K giant star (Mikolajewska, Acker \& Stenholm 1997), satisfies all the classification tree 
criteria. We thus argue that it is a genuine SySts.

Certainly, there are several known, spectroscopically confirmed, SySts in the IPHAS candidate list and the VPHAS+ cataloque 
(column nine in Table~\ref{tableC1}). In particularly, there are 10 known SySts in the IPHAS list and all of them were recovered 
(100 per cent success). By cross-matching the whole DR2 IPHAS catalogue with the list of known SySts (Paper~I), we found
that there are in total 27 known SySts. Many of them are not included in the IPHAS list of candidates because of no available 
{\it r}, {\it i}, and/or \ha\ magnitudes. According to our classification criteria, only 18 out of 27 SySts or 66 per cent are recovered. 
But, 6 of them are classified as S$+$IR, D and \DD-type and it is well known that they are not recovered with the current colour criteria 
(see \S 3, Fig.~\ref{figB1}). This signifies that the current criteria recover 18 out of 21 known S-types SySts or 86\% success.

As for the DR2 VPHAS+ catalogue, there are 40 known SySts and 27 of them are classified as S-type and 13 as S$+$IR, D and \DD-type. 
We recovered 23 out of 27 known S-type or 85 per cent. Note that only 13 are presented in Table~\ref{tableC1}. The rest of them do not pass 
the IPHAS criterion (not available {\it r}, {\it i} and \ha\ magnitudes). The four missing S-type SySts that do not satisfy the classification criteria are: (i) SS73~17: {\it J--H}=0.67 with a photometric {\it H} error of 0.21~mag , (ii) Hen 3-1410: {\it K$_s$--W3}=1.29, 
(iii) K~6-6: {\it K$_s$--W3}=2.29 and (iv) AS~289: {\it W1--W4}=1.05. In conclusion, three of them could be recovered but they show 
slightly different colour indices while K~6-6 exhibits strong {\it K$_s$--W3} excess indicative of an S$+$IR-type SySt. 

Recall that the classification criteria derived from the SySts and mimics training samples do not recover the dusty ones (S$+$IR, D and 
\DD-type, Fig.~\ref{figB1}). Therefore, we also applied the classification criteria derived from the training samples of dusty SySts and mimics 
(Figure~\ref{figB3}). These criteria revealed three more candidate SySts in the list of candidates, seven in the IPHAS list of candidates
and 29 in the DR2 VPHAS+ catalogue. Regarding the known SySts in the IPHAS and VPHAS+ catalogues, there are 14 D-type, 6 S+IR-type and 1 \DD-type 
SySts. 93 per cent of the D-type are recovered but none of the S$+$IR or \DD-type since all S$+$IR have 
{\it H--W2}$<$3.806 and all \DD\-type have {\it W1--W4}$>$4.715.

In order to verify the feasibility of our method, we also examined whether the non-SySt sources (spectroscopically classified) in 
Rodr\'{i}guez-Flores et al. (2014) satisfy our criteria. After analysing all of the 13 sources, we found out that only one of them (IPHASJ201550.96+373004.2) satisfies all the criteria of being SySts. This candidate emerged from the dusty SySts/mimics model 
which suffers by a 25 per cent of contamination (see \S~3). All the 13 sources were later observed and none of them was found to be a SySt. 
The IPHASJ201550.96+373004.2 candidate that satisfies out criteria was classified as a Be star or YSO (Rodr\'{i}guez-Flores et al. 2014). 
This is a strong proof that our classification criteria works very well, and it can indicate very likely SySts candidates or reject sources 
from follow-up observations.

Corradi et al. (2010) have also obtained spectroscopic data of two sources in our list of candidate SySts (IPHASJ194907.23+211742.0 and IPHASJ202947.93+355926.5). The first one is a young PN but according to Viironen et al. (2009) its spectrum resembles those of 
D-type SySts and it may belong to the rare group of objects whose its nature is still not clear like M 2-9. The second one is 
classified as YSO (see also Krause et al. 2003). Our classification criteria also indicate a likely YSO or AeBe star. Note that the last 
two objects were derived from the dusty/mimic model for which the group of SySts suffers by a 25 per cent contamination. 
Therefore, the possibility of finding other dusty sources like YSO is not small.

Baella et al. (2016) also searched for new yellow SySts by observing five candidate SySts and they ended up discovering one new SySt (StHa~63) 
while the remaining sources were classified as K giants. From our classification criteria, we conclude that all of them are good candidate 
SySts and deserved to be observed. Nevertheless, by using the criteria from the SySts$+$K/M giants training sample (Fig.~\ref{figB9}), 
only two objects StHa~63 (the confirmed) and SS~360 (classified as M3 III, see Baella et al. 2016) pass the criteria of 
being SySts while the remaining not.

The interesting point here is that SySts with low luminosity WDs produce very weak optical emission line and they can be 
misclassified. SU Lyn belongs to this specific group of SySts. Despite its optical spectrum resembling that of an M6~III star having a 
very weak \ha\ emission line, its UV-excess indicates the presence of a hot white dwarf (Mukai et al. 2016). According to our classification 
tree criteria, SU Lyn is indeed a SySt and not an isolate red giant. Therefore, we claim that SS~360 may also be a member of SySts with a low 
luminosity WD. Notice that SS~360 is the only object, besides StHa~63, with an \ha\ emission (see Fig.~4 in Baella et al. 2016).

The final step is to apply our classification tree and LDA/KNN criteria derived from the training samples of different classes of sources 
to our list of candidate SySts. In Tables~\ref{tableC1} and \ref{tableC2}, we present a probable classification of each source in 
columns 2 to 8, for the classification tree and LDA/KNN methods, respectively. All the known SySts in this list are recovered from both
methods. Therefore, at least an 80-90 per cent of the sources are very likely SySts. A spectroscopic study of these sources will
be presented in a forthcoming paper.

\section{Discussion and Conclusions}
 
We carried out and presented a machine learning approach to find new SySts in publicly available \ha\ photometric catalogues using \ha-excess, 
2MASS and WISE photometric data. First, we explored a number of different combinations of colour indices that can provide a good separation of SySts 
from other classes of objects that mimic SySts such as PNe, post-AGB stars, CVs, WR stars, WTT and ClTT stars, single K and M giants, 
and Be stars. We shown that the widely used {\it J--H} vs. {\it H--K$_{\rm s}$} is not an adequate DCCD for identifying 
SySts. S-type SySts, Mira, YSO and WTT stars occupy the same regions making very hard to distinguish them. The {\it W3--W4} vs. {\it K$_s$--W3} 
and {\it J--H} vs. {\it W1--W4} DCCDs were found to be better DCCDs.

Machine learning methods such as classification tree, linear discriminant analysis and K-nearest neighbours were also used to 
derive new criteria that distinguish SySts from other stellar objects. Classification tree revealed that the {\it K$_{\rm s}$--W3}, 
{\it H--W2} and {\it W3--W4} colour indices are the best observable parameters for classifying SySts into the S, D, \DD and S$+$IR scheme.
The {\it J--H}, {\it K$_{\rm s}$--W3},{\it W1--W2} colour indices were found to provide the best combination for separating 
S-type SySts from mimics, whereas the {\it H--W2} and {\it W1--W4} colour indices are better for identifying the dusty SySts. 
By training the classification tree using samples with different combinations of classes of objects, 
we deduced that primarily {\it J--H}, {\it W1--W4} and {\it K$_{\rm s}$--W3} and secondarily {\it H--W2}, {\it W1--W2} and {\it W3--W4} provide ideal 
colour indices to distinguish SySts.

Linear discrimination analysis and K-nearest neighbour were also used in order to find the linear combination of 2MASS and AllWISE data, 
that better discriminate SySts. SySts were found to define, in most of the cases, distinct regions. Diagnostic diagrams obtained from 
the LDA+KNN analysis were also provided. The accuracy of these diagrams vary between 80 and 98 per cent. Mira stars were found to be the 
objects which cannot be easily distinguished from the SySts, especially the S-type, as they have very similar colour indices. 

Finally, we applied our classification tree model derived from the SySts and mimics training samples to the list of candidate SySts from Paper~I, 
the IPHAS list of candidate SySts, and the DR2 VPHAS+ catalogues. We ended up with 125 sources that pass the criteria. 72 of them 
(36 S-type and 36 D-type) are new candidate SySts. All the criteria derived from the training samples with different combinations of sources 
were also applied to our final list of candidates and the most likely identification was provided for each source. Our models recovered 
up to 90 per cent of the known SySts in these three lists. Around 80-90 per cent of the sources in our list are very likely SySts but a 
spectroscopic confirmation is required.

\section*{Acknowledgments} 
All the authors thank the anonymous referee for very insightful comments and for helping us to significantly improve our paper.
S.A. and M.L.L.-F. acknowledge support of CNPq, Conselho Nacional de Desenvolvimento Cient\'ifico e Tecnol\'ogico -
Brazil (grant 300336/2016-0 and 248503/2013-8 respectively). GRL acknowledges support from Universidad de
Guadalajara, CONACyT, PRODEP and SEP (Mexico). LGR is supported by NWO funding towards the Allegro group at Leiden University. 
Authors would like to thank Vaselina Kalinova, Dario Colombo, Jens Kauffmann and Helio Jaques Rocha-Pinto for the fruitful 
discussion on machine learning. M.L.L.-F. would also like to thank Xander Tielens and Richard Stancliffe, for host him in their 
respective research groups at the Leiden Observatory and Argelander Instit\"{u}t fur Astronomie. This publication makes use of data 
from the Two Micron All-Sky Survey which is a joint project of the University of Massachusetts and the Infrared Processing and 
Analysis Center/California Institute of Technology, funded by the NASA and the National Science Foundation, data products from 
the Wide-field Infrared Survey Explorer, which is a joint project of the University of California, Los Angeles, and the Jet 
Propulsion Laboratory/California Institute of Technology, funded by the National Aeronautics and Space Administration. 
This paper also makes use of data obtained as part of the INT Photometric \ha\ Survey of the Northern Galactic Plane (IPHAS,
www.iphas.org) carried out at the Isaac Newton Telescope (INT). The INT is operated on the island of La Palma by the Isaac Newton 
Group in the Spanish Observatorio del Roque de los Muchachos of the Instituto de Astrofisica de Canarias. All IPHAS data are processed 
by the Cambridge Astronomical Survey Unit, at the Institute of Astronomy in Cambridge. The band merged DR2 catalogue was assembled at the 
Centre for Astrophysics Research, University of Hertfordshire, supported by STFC grant ST/J001333/1.
Based on data products from observations made with ESO Telescopes at the La Silla Paranal Observatory under programme ID 177.D-3023, 
as part of the VST Photometric \ha\ Survey of the Southern Galactic Plane and Bulge (VPHAS+, www.vphas.eu). 
Finally, this publication makes use of many software packages in {\sc Python}, including: {\sc Matplotlib} (Hunter 2007), {\sc NumPy} 
(van der Walt et al. 2011), {\sc SciPy} (Jones et al. 2001) and {\sc AstroPy Python} (Astropy Collaboration et al. 2013; 
Muna et al. 2016).

\bibliographystyle{mnras}

\appendix

\section{ Classification tree}
The classification tree plots derived from the training samples of the following groups: SySts/PNe/Be, SySts/CV/Mira, 
SySts/CV/YSO, SySts/WR/post-AGB, SySts/K-giants/M-giants, SySts/WTT/ClTT and SySts/Be/AeBe are presented here.\\

\begin{figure*}
\includegraphics[scale=0.65]{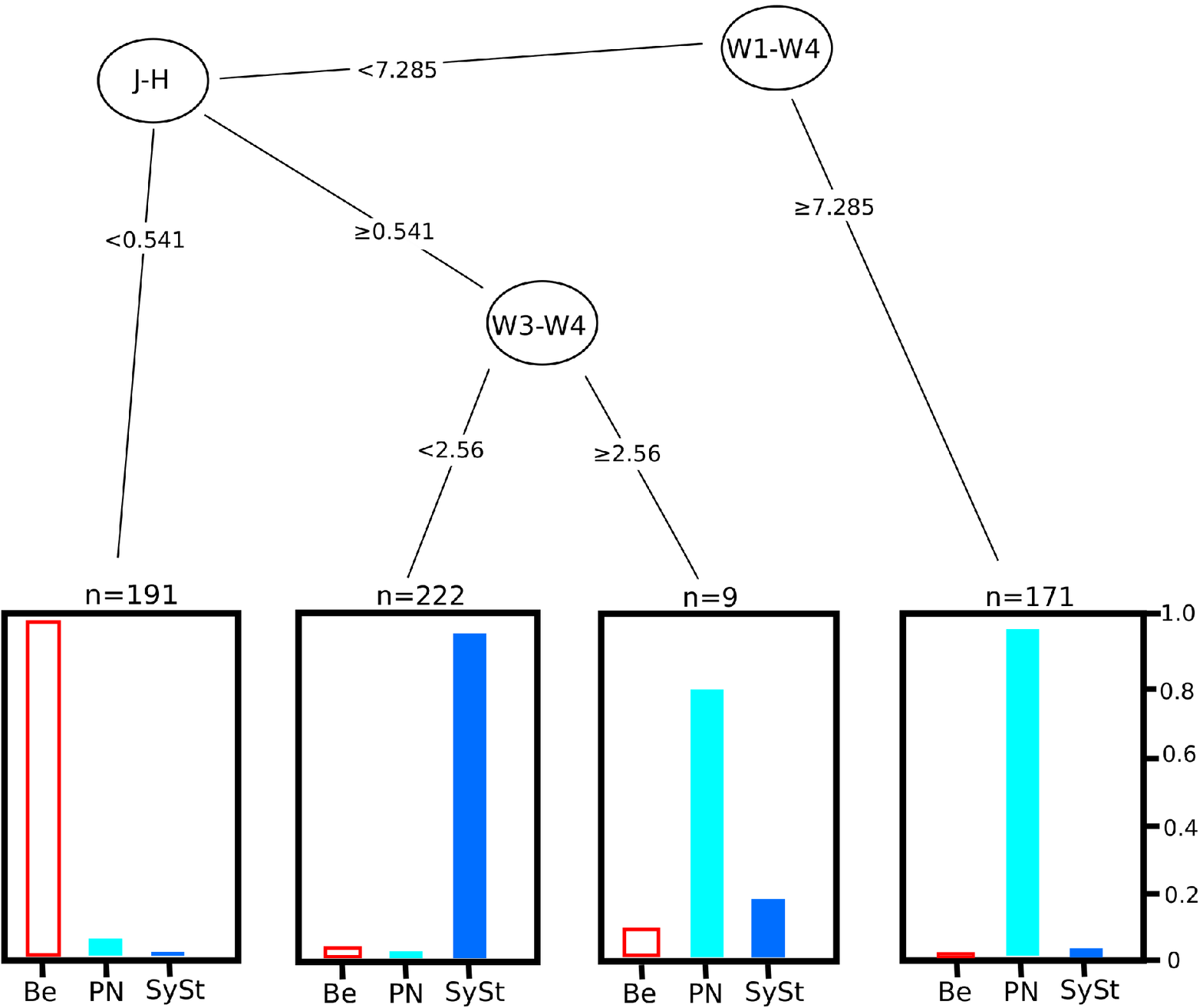}
\caption[]{Classification tree plot from the SySts/PNe/Be training sample. }
\label{figB4}
\end{figure*}

\begin{figure*}
\includegraphics[scale=0.65]{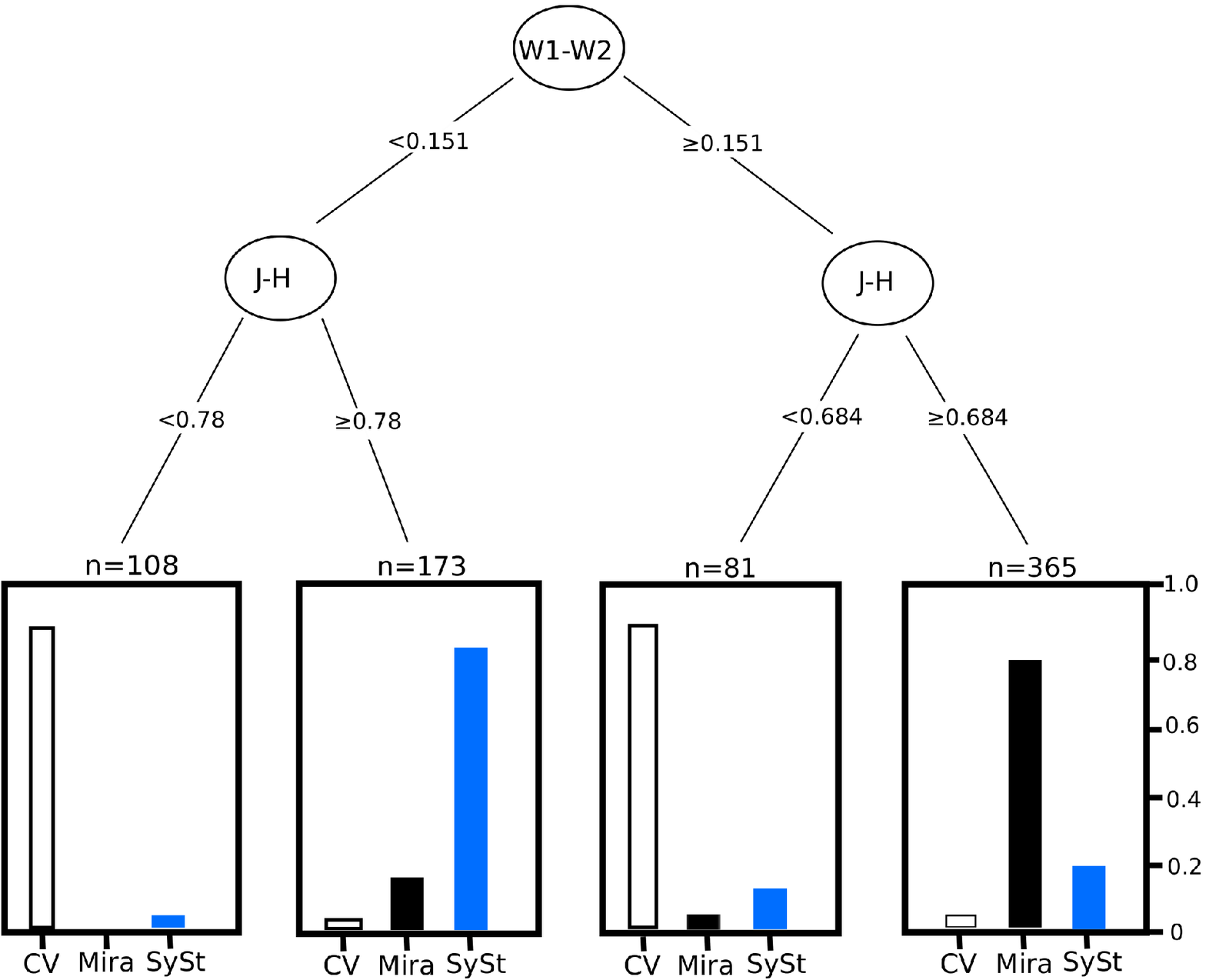}
\caption[]{Classification tree plot from the SySts/CV/Mira training samples.}
\label{figB5}
\end{figure*}

\begin{figure*}
\includegraphics[scale=0.65]{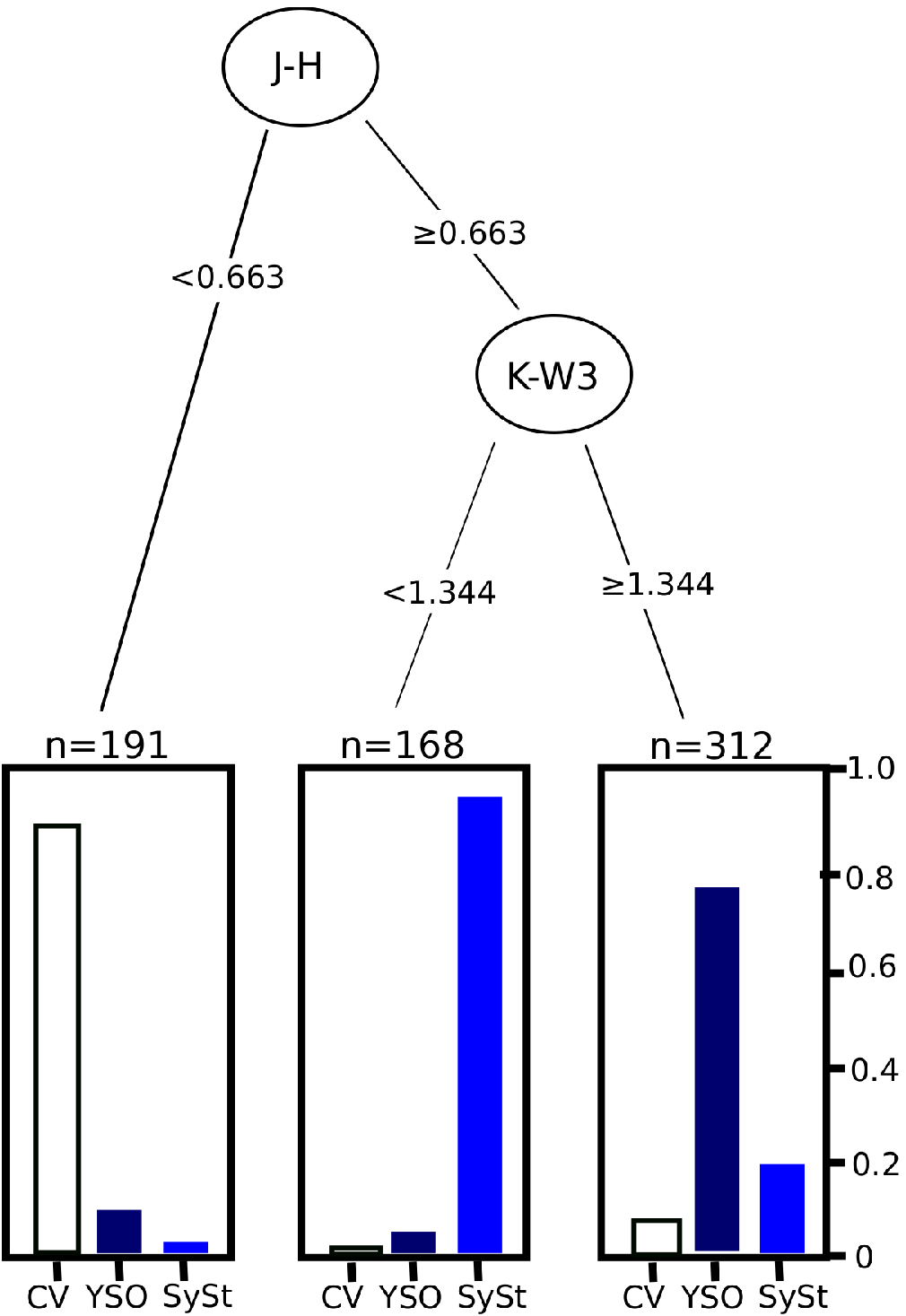}
\caption[]{Classification tree plot from the SySts/CV/YSO training sample.}
\label{figB7}
\end{figure*}

\begin{figure*}
\includegraphics[scale=0.65]{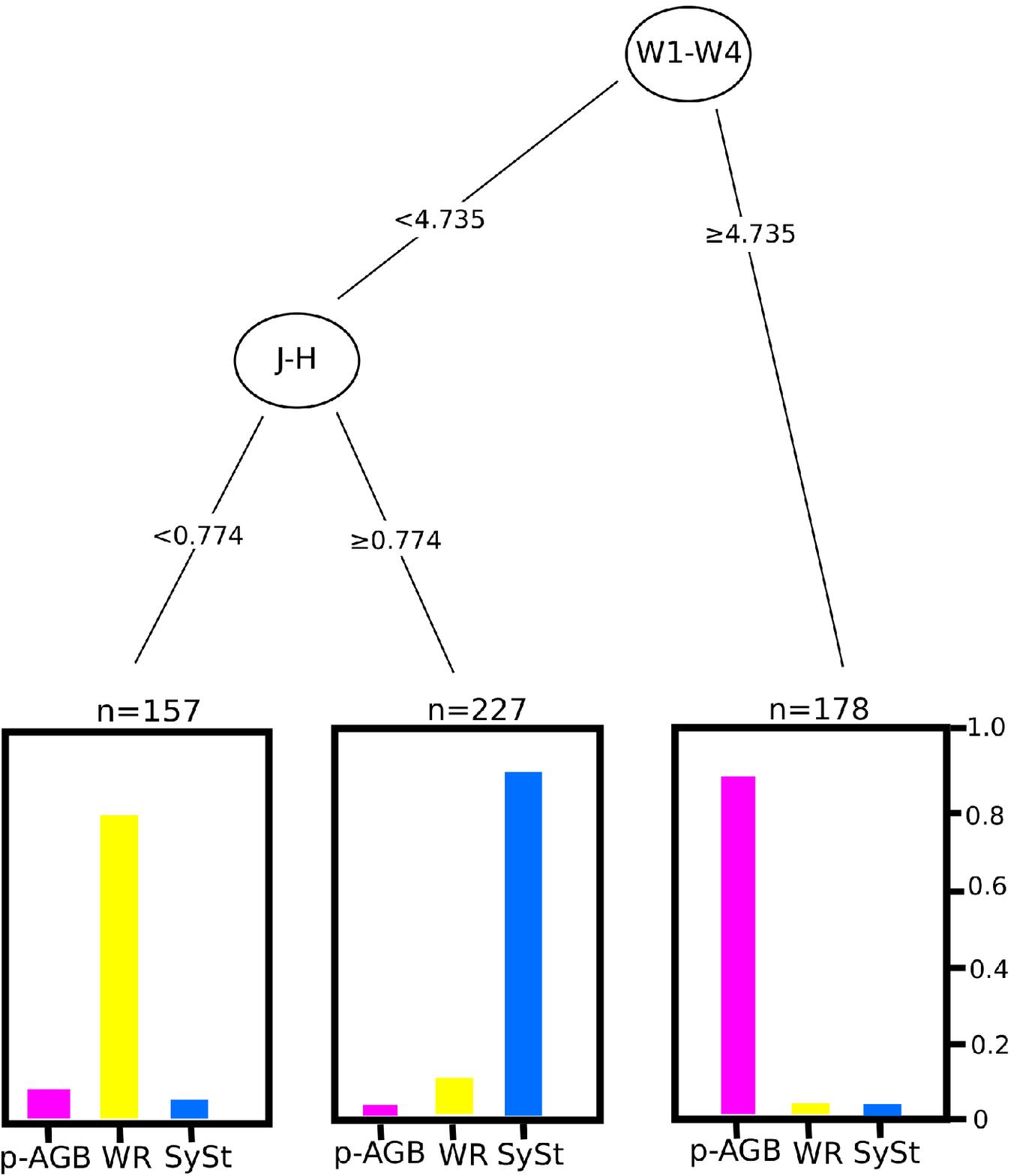}
\caption[]{Classification tree plot from the SySts/WR/post-AGB training sample.}
\label{figB8}
\end{figure*}

\begin{figure*}
\includegraphics[scale=0.65]{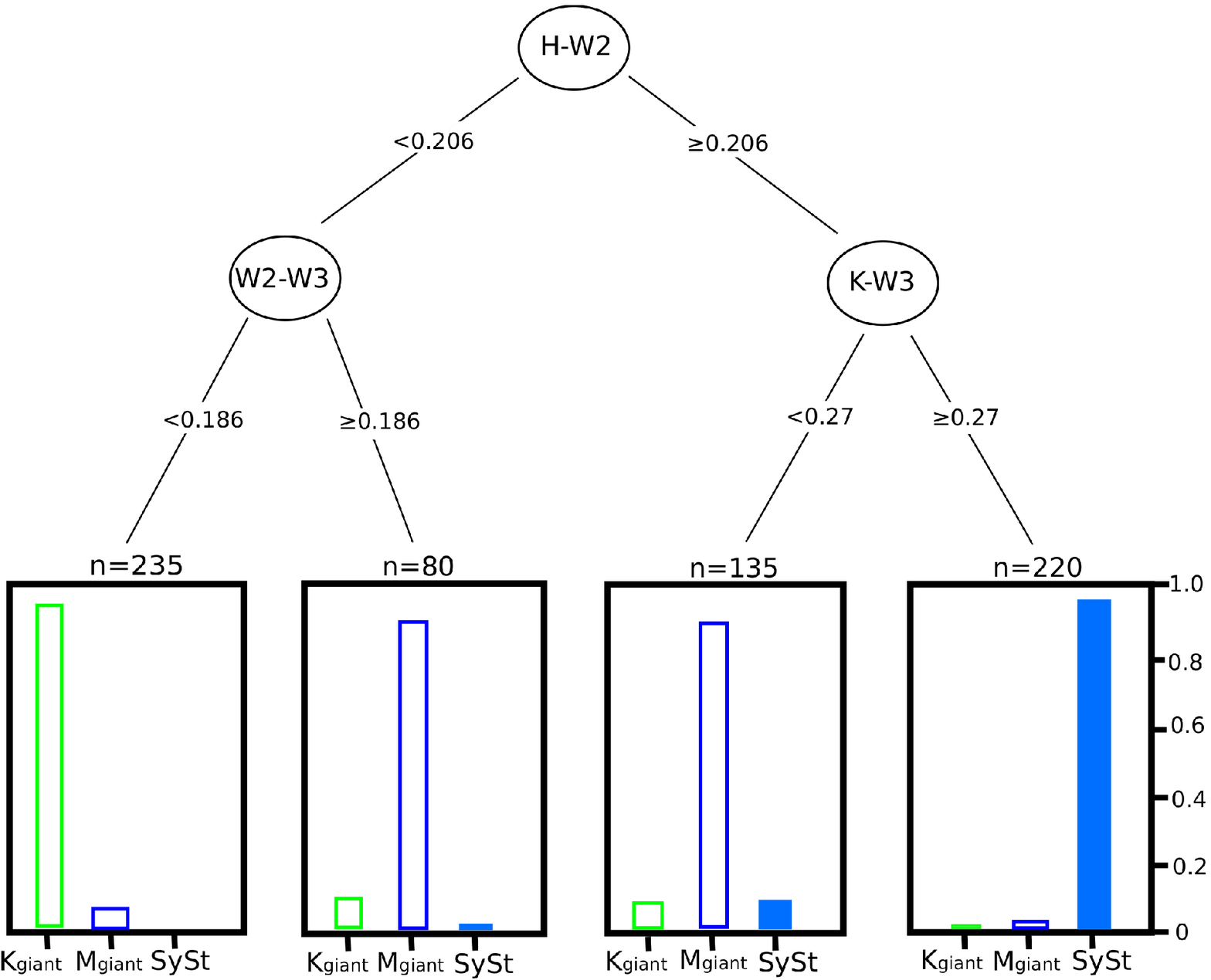}
\caption[]{Classification tree plot from the SySts/K-giants/M-giants training sample.}
\label{figB9}
\end{figure*}

\begin{figure*}
\includegraphics[scale=0.65]{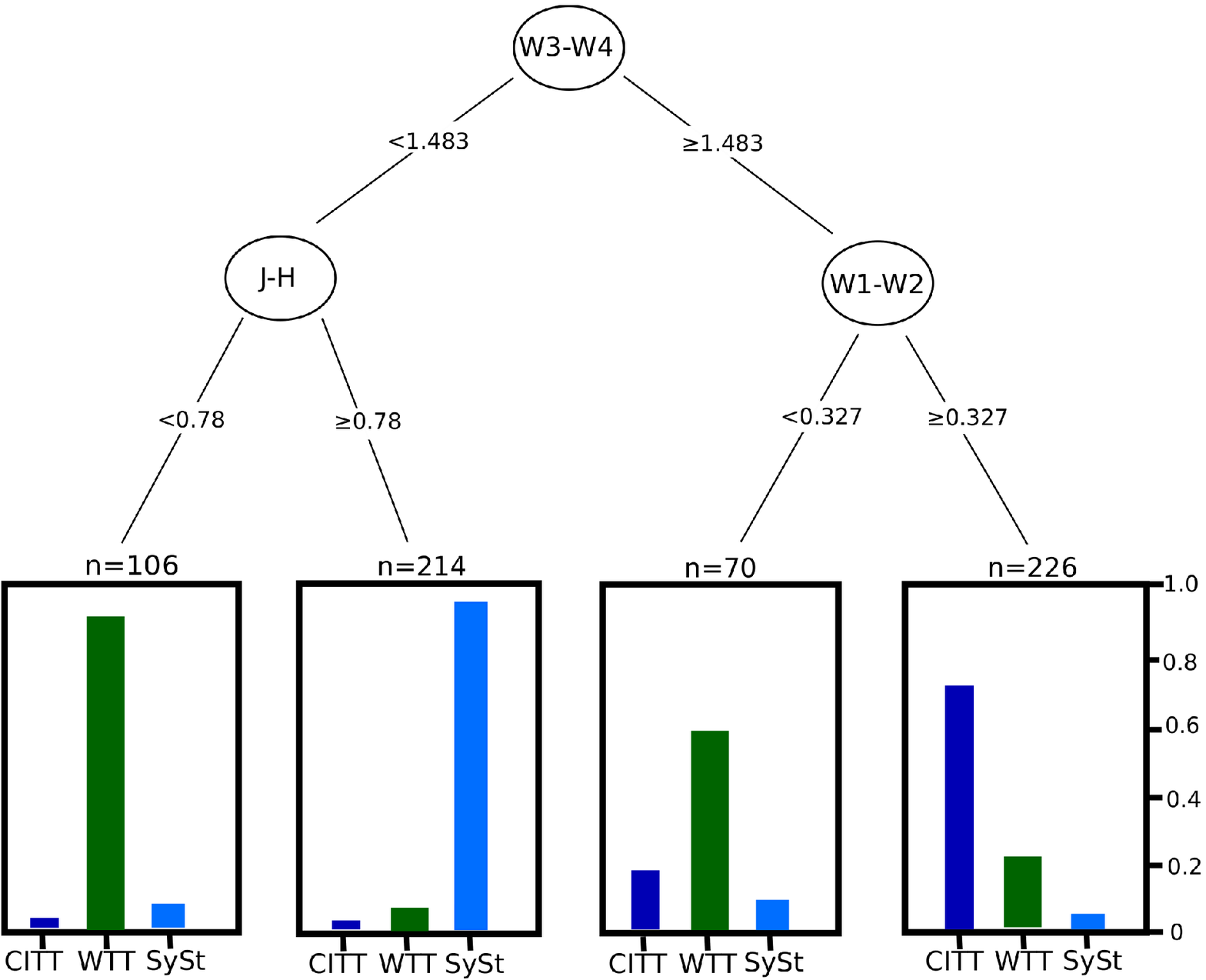}
\caption[]{Classification tree plot from the  SySts/WTT/ClTT training sample.}
\label{figB10}
\end{figure*}

\begin{figure*}
\includegraphics[scale=0.65]{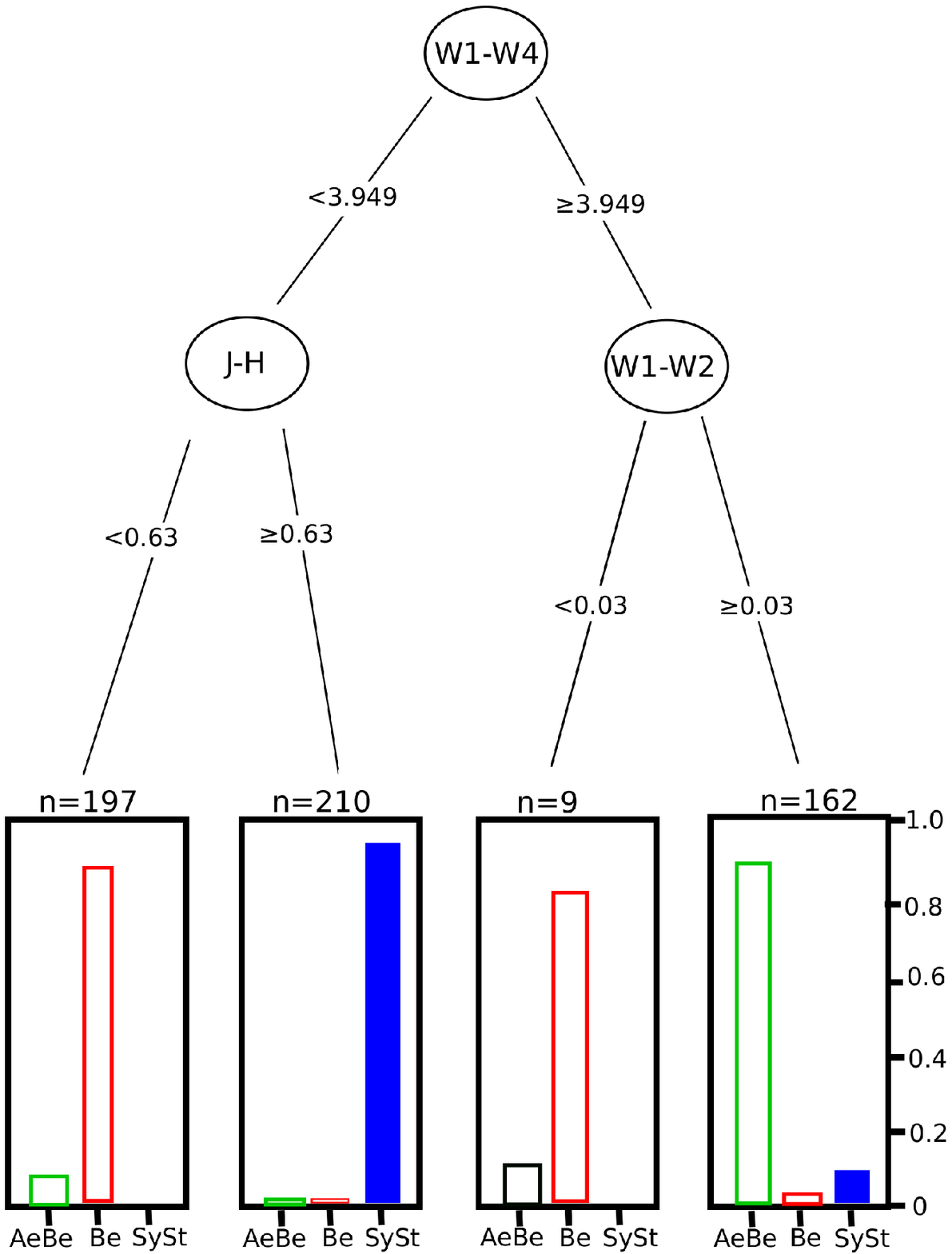}
\caption[]{Classification tree plot from the SySts/Be/AeBe training sample.}
\label{figB11}
\end{figure*}

\clearpage

\newpage

\section{LDA+KNN}

To transform any set of linear discriminant components (LD1 and LD2) obtained from the coefficients in Table~B1 into the range [0,1] 
one should apply the following relations:\\
\\
{ $\bullet$ Data set: SySts - PNe - Be }
\begin{eqnarray}
	Normalized LD1 = \frac{(LD1 + 5.65)}{9.28} \\
    Normalized LD2 = \frac{(LD2 + 8.38)}{11.67},
\end{eqnarray}
\\
{$\bullet$ Data set: SySts - CV - Mira}
\begin{eqnarray}
	Normalized LD1 = \frac{(LD1 + 6.35)}{12.19} \\
    Normalized LD2 = \frac{(LD2 + 3.40)}{8.47},
\end{eqnarray}
\\
{$\bullet$ Data set: SySts - YSO - CV}
\begin{eqnarray}
	Normalized LD1 = \frac{(LD1 + 4.46)}{9.06} \\
    Normalized LD2 = \frac{(LD2 + 3.36)}{8.86},
\end{eqnarray}
\\
{$\bullet$  Data set: SySts - post-AGB - WR}
\begin{eqnarray}
	Normalized LD1 = \frac{(LD1 + 4.87)}{7.49} \\
    Normalized LD2 = \frac{(LD2 + 7.28)}{10.50},
\end{eqnarray}

{$\bullet$ Data set: SySts - M giants - K giants}
\begin{eqnarray}
	Normalized LD1 = \frac{(LD1 + 2.21)}{11.87} \\
    Normalized LD2 = \frac{(LD2 + 4.81)}{10.81},
\end{eqnarray}
\\
{$\bullet$  Data set: SySts - WTT - ClTT}
\begin{eqnarray}
	Normalized LD1 = \frac{(LD1 + 4.77)}{11.53} \\
    Normalized LD2 = \frac{(LD2 + 4.47)}{6.69},
\end{eqnarray}
\\
{$\bullet$ Data set: SySts - Be - AeBe}
\begin{eqnarray}
	Normalized LD1 = \frac{(LD1 + 4.89)}{13.16} \\
    Normalized LD2 = \frac{(LD2 + 2.69)}{7.77},
\end{eqnarray}
\\

The LDA/KNN plots derived from the training samples of the following groups: SySts/PNe/Be, SySts/CV/Mira,  
SySts/CV/YSO, SySts/WR/post-AGB, SySts/K-giants/M-giants, SySts/WTT/ClTT and SySts/Be/AeBe are presented here.\\

\begin{table*}
\centering
\caption{LDA \& KNN modelling} 
\label{table5}
\begin{tabular}{ | c | c |}
\hline
Coefficients spectrum plots & Normalized LD1 vs LD2 KNN plots \\ \hline
\begin{minipage}{.45\textwidth}
\includegraphics[width=\linewidth, height=70mm]{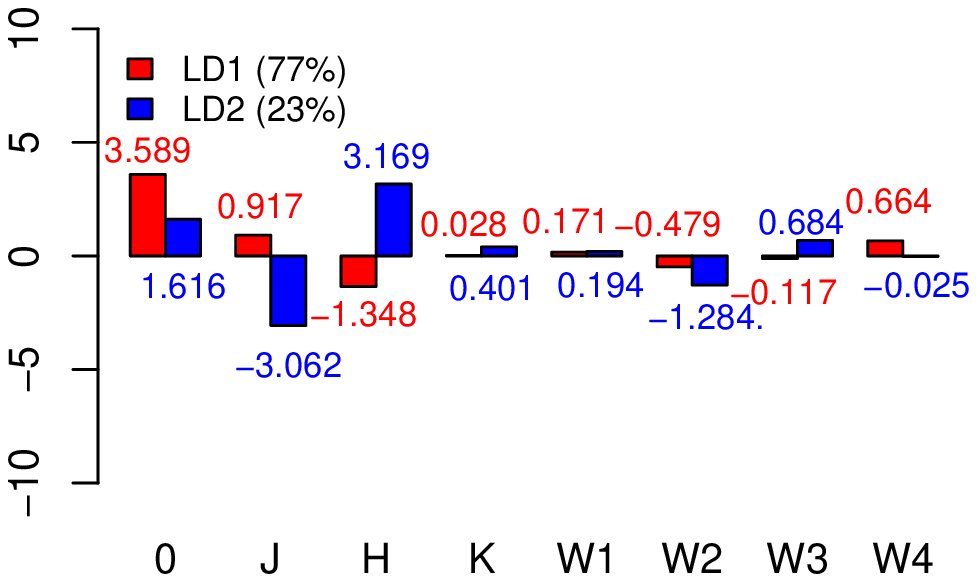}
\end{minipage}
&
\begin{minipage}{.50\textwidth}
\includegraphics[width=\linewidth, height=70mm]{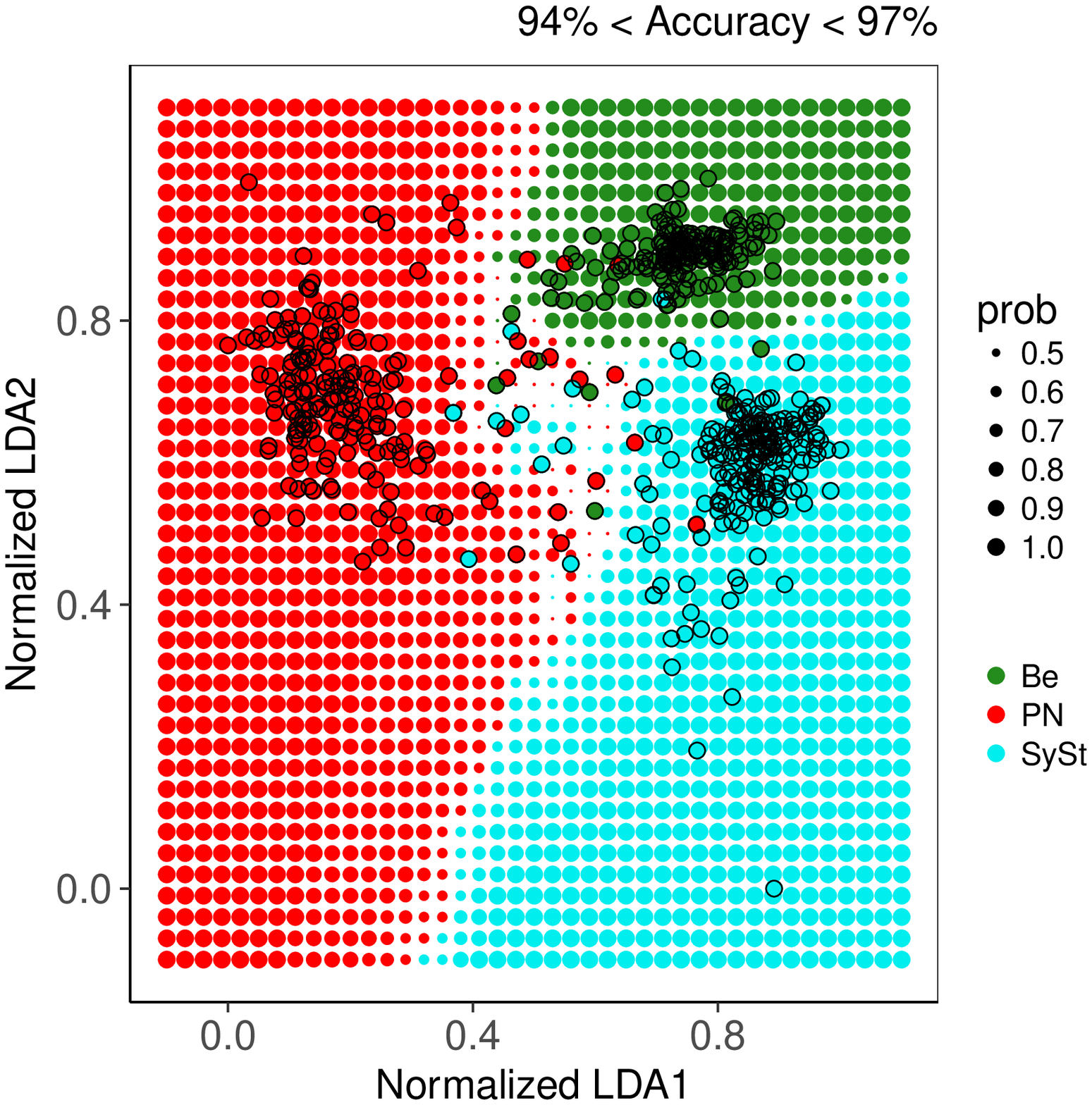}
\end{minipage}
\\ \hline
\begin{minipage}{.45\textwidth}
\includegraphics[width=\linewidth, height=70mm]{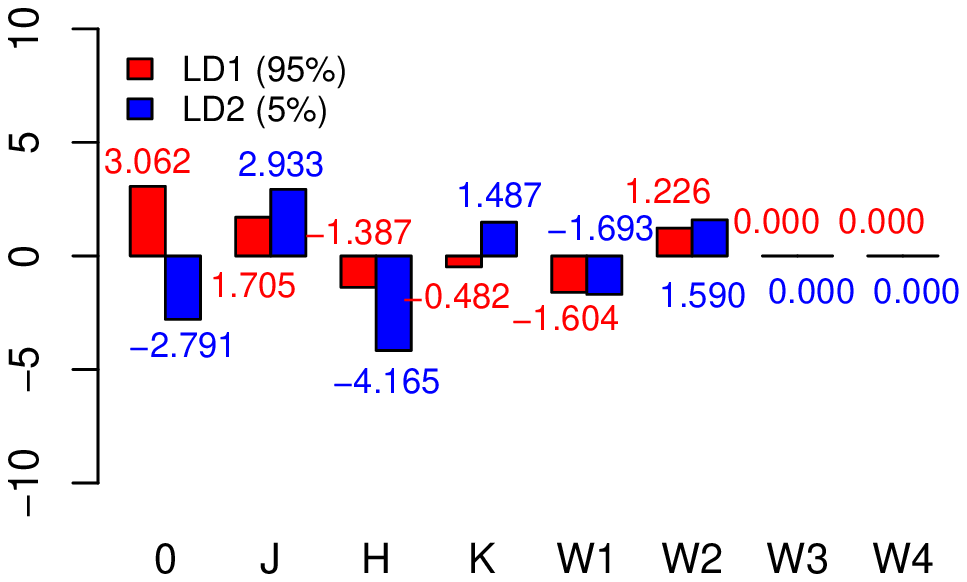}
\end{minipage}
&
\begin{minipage}{.50\textwidth}
\includegraphics[width=\linewidth, height=70mm]{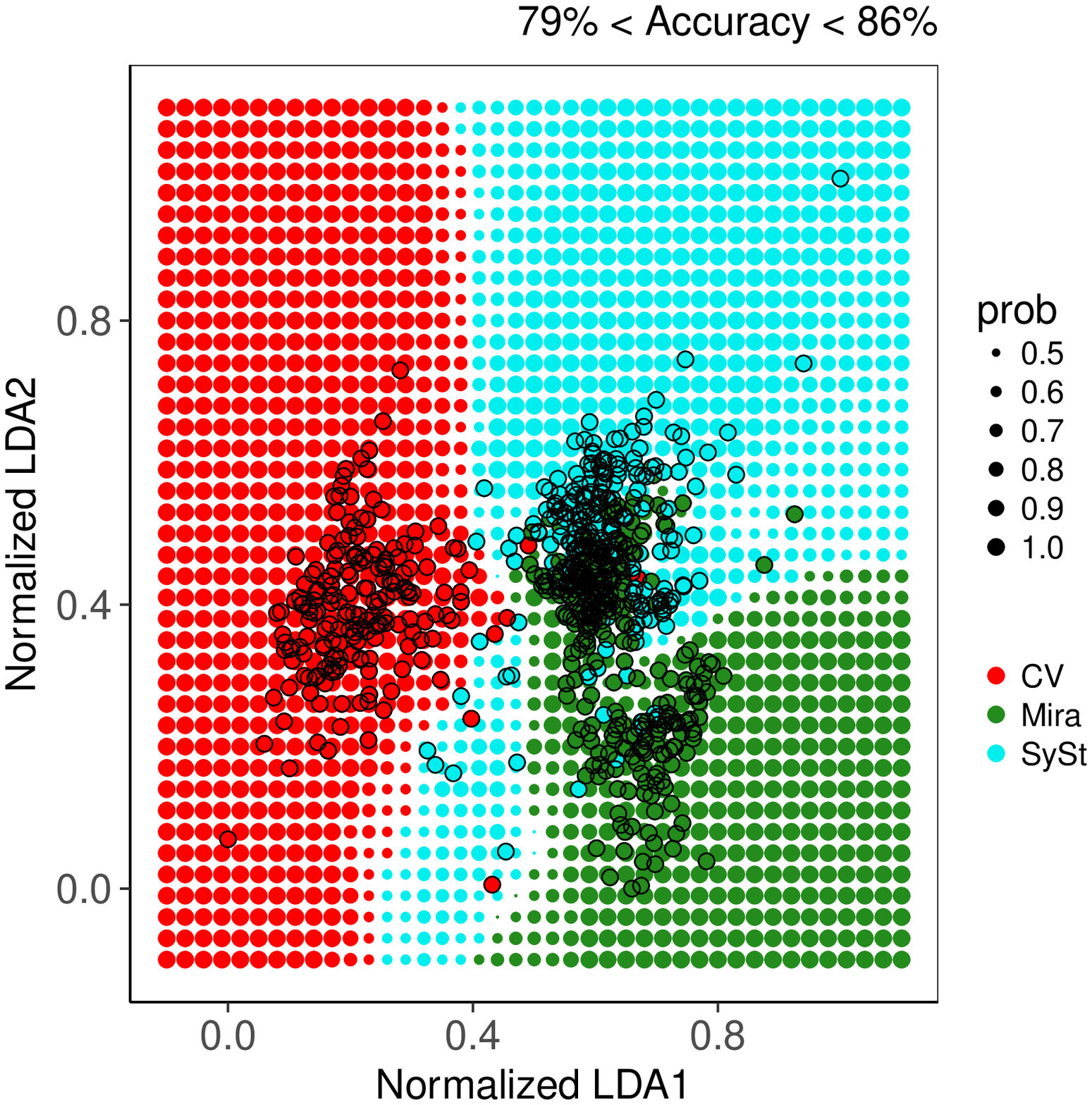}
\end{minipage}
\\ \hline	
\begin{minipage}{.45\textwidth}
\includegraphics[width=\linewidth, height=70mm]{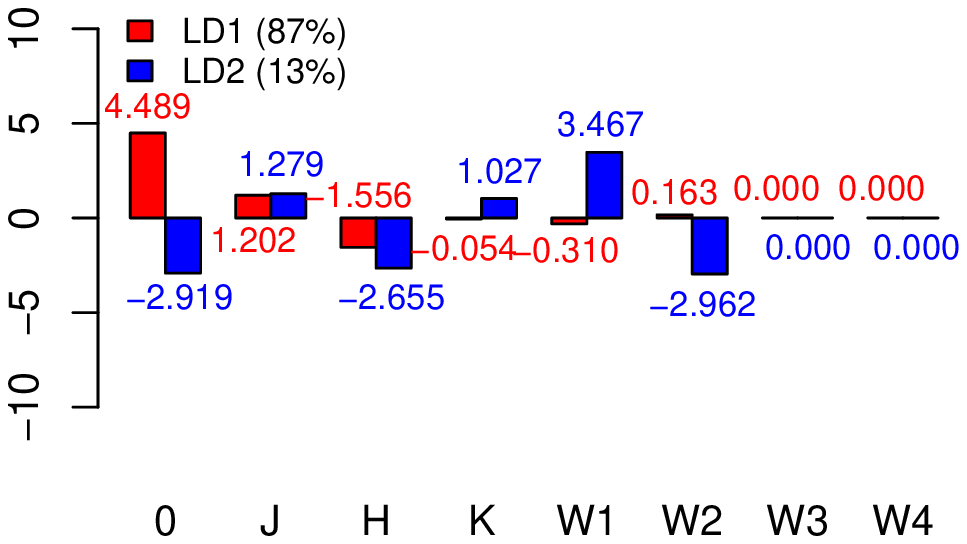}
\end{minipage}
&
\begin{minipage}{.50\textwidth}
\includegraphics[width=\linewidth, height=70mm]{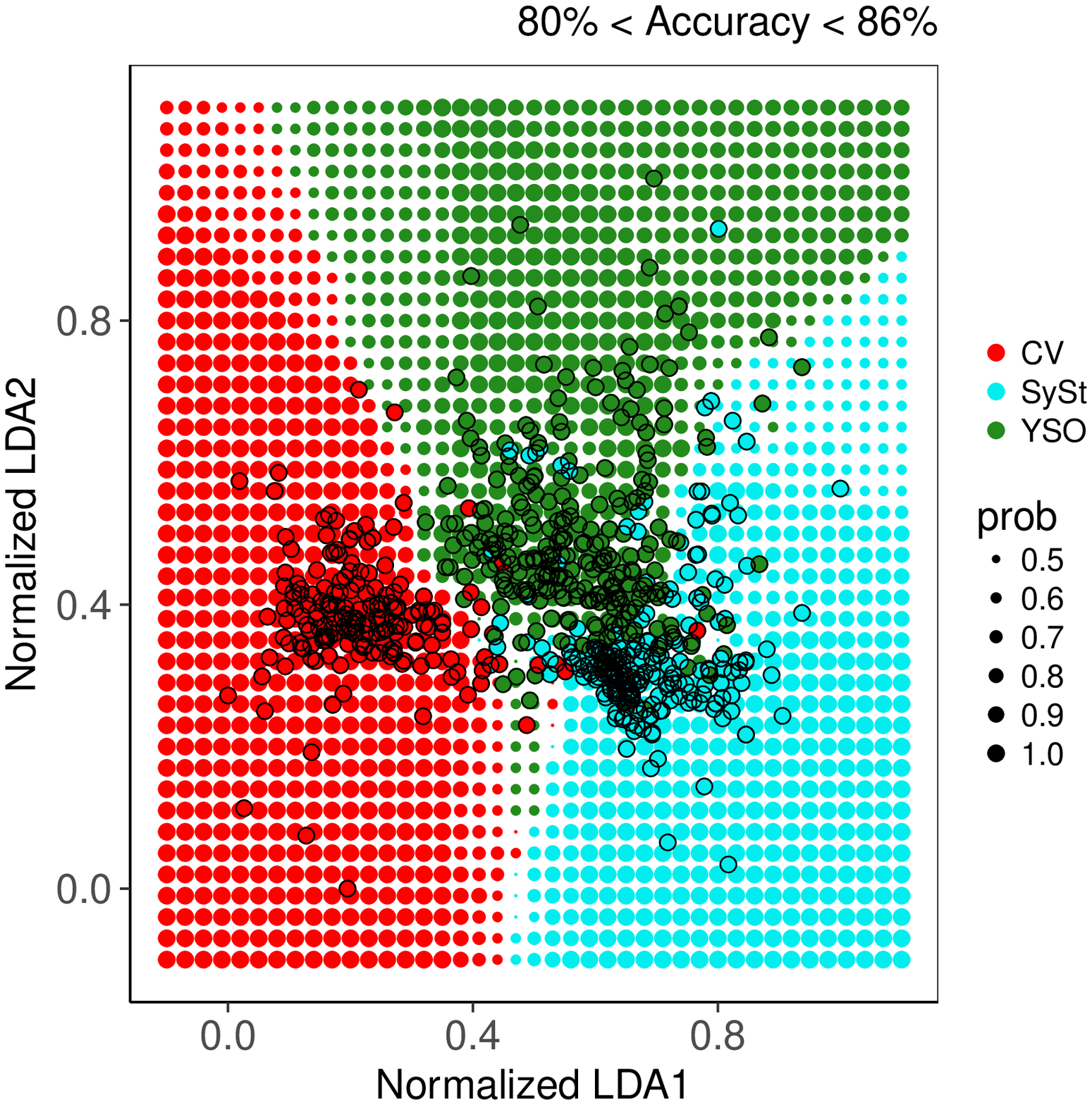}
\end{minipage}
\\ \hline
\end{tabular}
\begin{flushleft}
Left column: Coefficient spectrum plot of the first (red) and second (blue) discriminant components for the seven-dimensional 
space of 2MASS and WISE surveys. "0" variable corresponds to the zero point. The numbers in parenthesis give the percentage of 
discriminabily. Right column: The LDA/KNN plots for different sets of objects. The size of the background circles corresponds 
to the probability of being classified as a specific type. The equations to normalize the LDA components and produce the KNN plots 
are given in Appendix A. 
\end{flushleft}
\end{table*}

\addtocounter{table}{-1}

\begin{table*}
\centering
\caption{LDA \& KNN modelling} 
\begin{tabular}{ | c | c |}
\hline
Coefficients spectrum plots & Normalized LD1 vs LD2 KNN plots \\ \hline
\begin{minipage}{.45\textwidth}
\includegraphics[width=\linewidth, height=70mm]{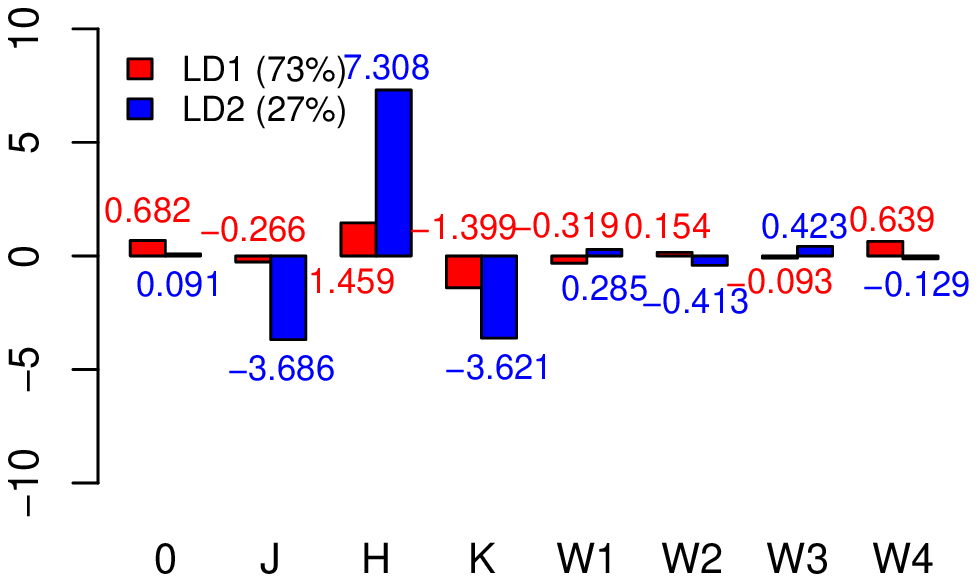}
\end{minipage}
&
\begin{minipage}{.50\textwidth}
\includegraphics[width=\linewidth, height=70mm]{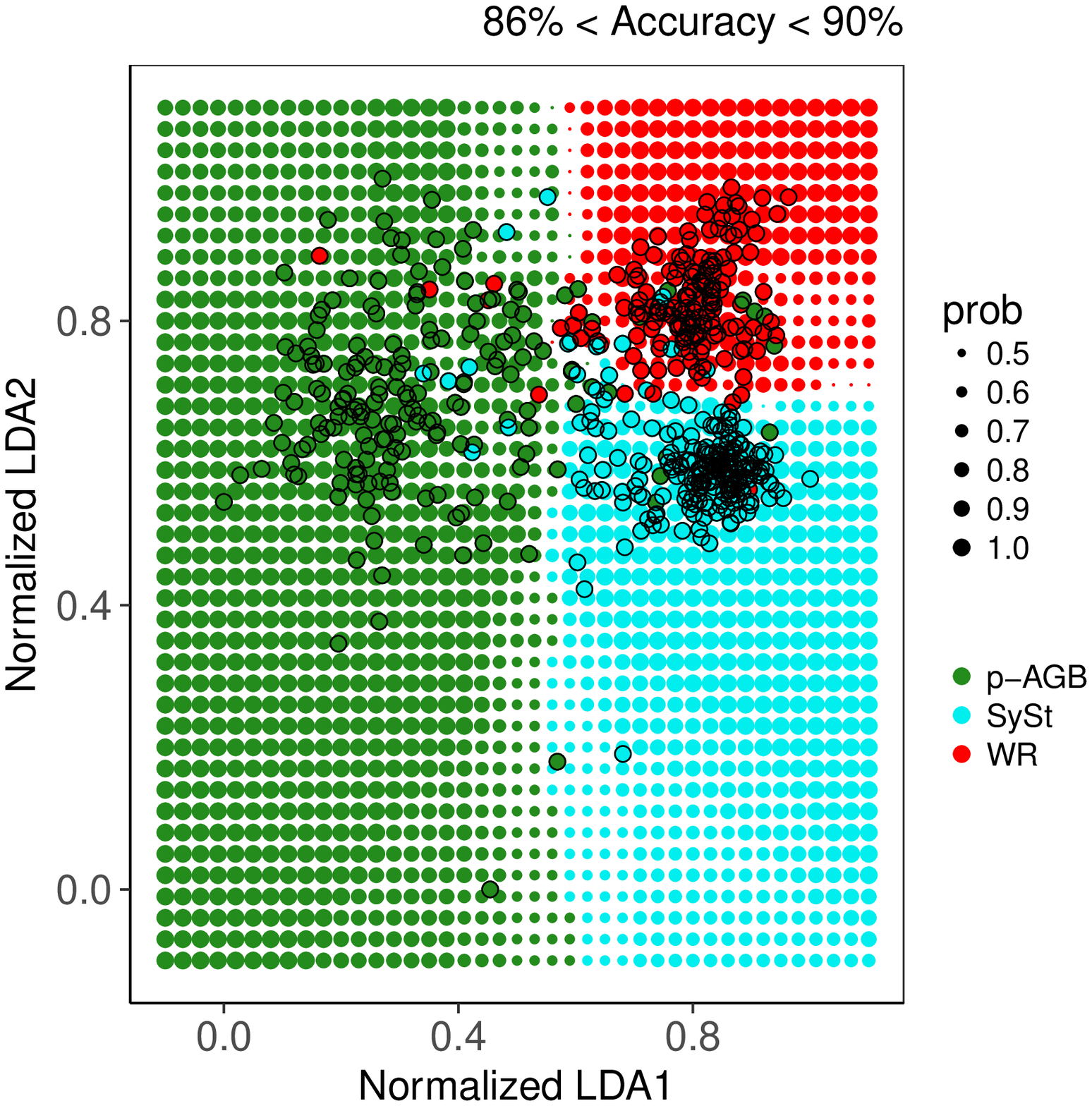}
\end{minipage}
\\ \hline
\begin{minipage}{.45\textwidth}
\includegraphics[width=\linewidth, height=70mm]{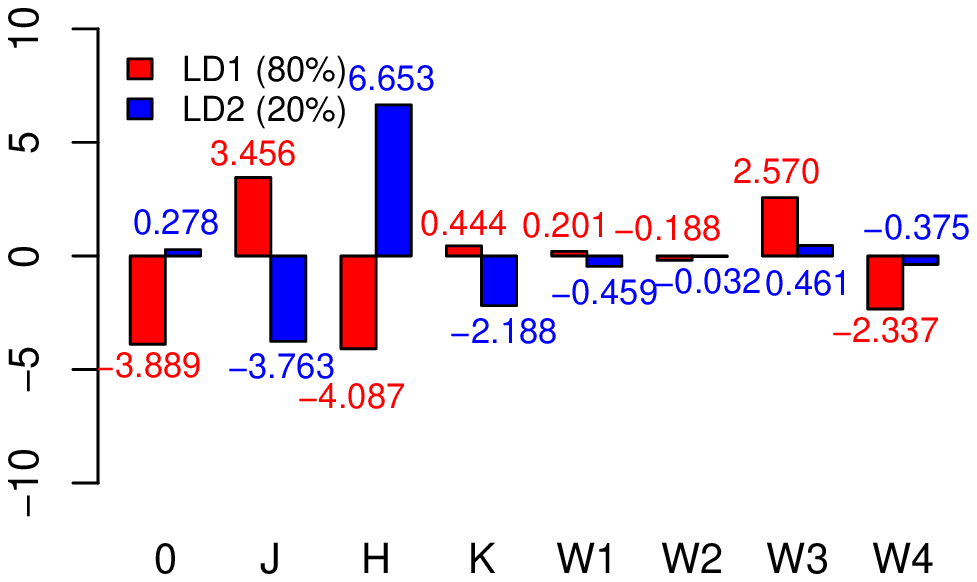}
\end{minipage}
&
\begin{minipage}{.50\textwidth}
\includegraphics[width=\linewidth, height=70mm]{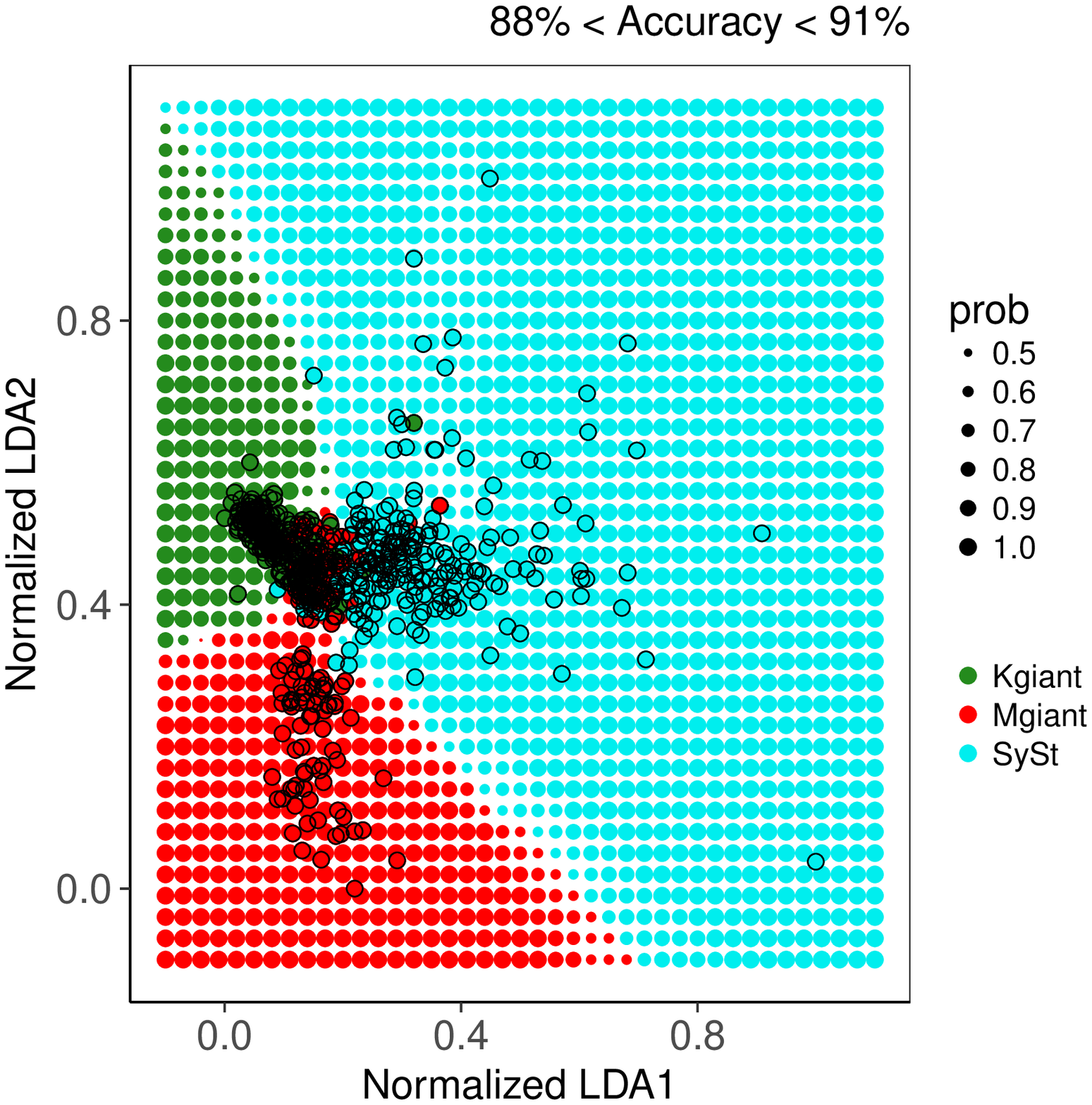}
\end{minipage}
\\ \hline
\begin{minipage}{.45\textwidth}
\includegraphics[width=\linewidth, height=70mm]{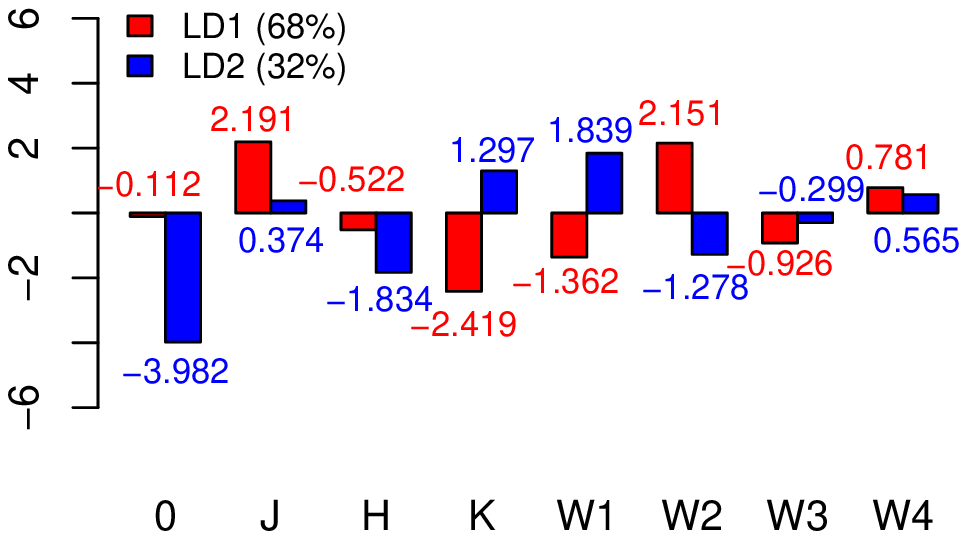}
\end{minipage}
&
\begin{minipage}{.50\textwidth}
\includegraphics[width=\linewidth, height=70mm]{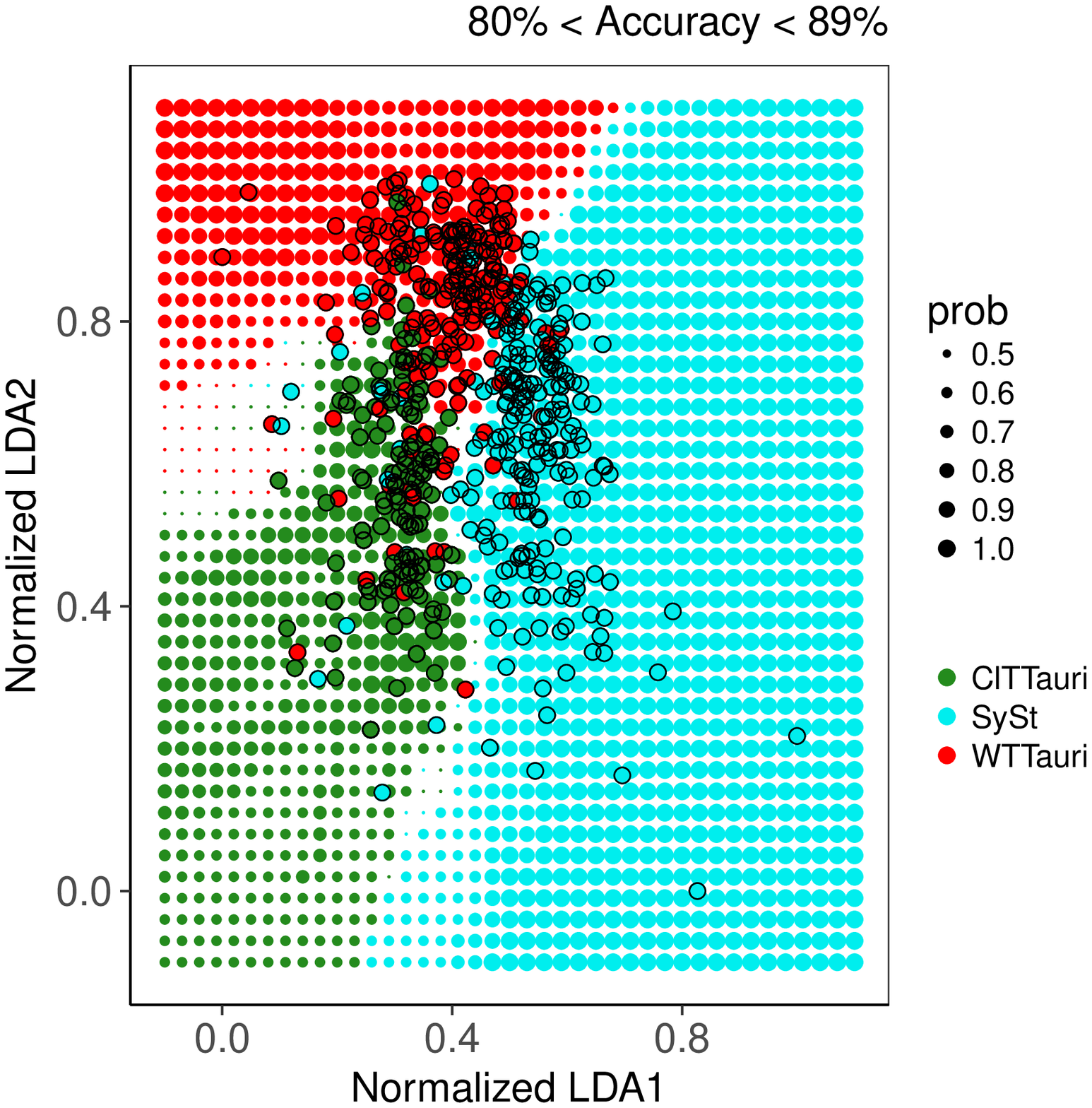}
\end{minipage}
\\ \hline
\end{tabular}
\begin{flushleft}
Left column: Coefficient spectrum plot of the first (red) and second (blue) discriminant components for the seven-dimensional 
space of 2MASS and WISE surveys. "0" variable corresponds to the zero point.  The numbers in parenthesis give the percentage of 
discriminabily. Right column: The LDA/KNN plots for different sets of objects. The size of the background circles corresponds 
to the probability of being classified as a specific type. The equations to normalize the LDA components and produce the KNN plots 
are given above.
\end{flushleft}
\end{table*}

\addtocounter{table}{-1}

\begin{table*}
\centering
\caption{LDA \& KNN modelling} 
\begin{tabular}{ | c | c |}
\hline
Coefficients spectrum plots & Normalized LD1 vs LD2 KNN plots \\ \hline
\begin{minipage}{.45\textwidth}
\includegraphics[width=\linewidth, height=70mm]{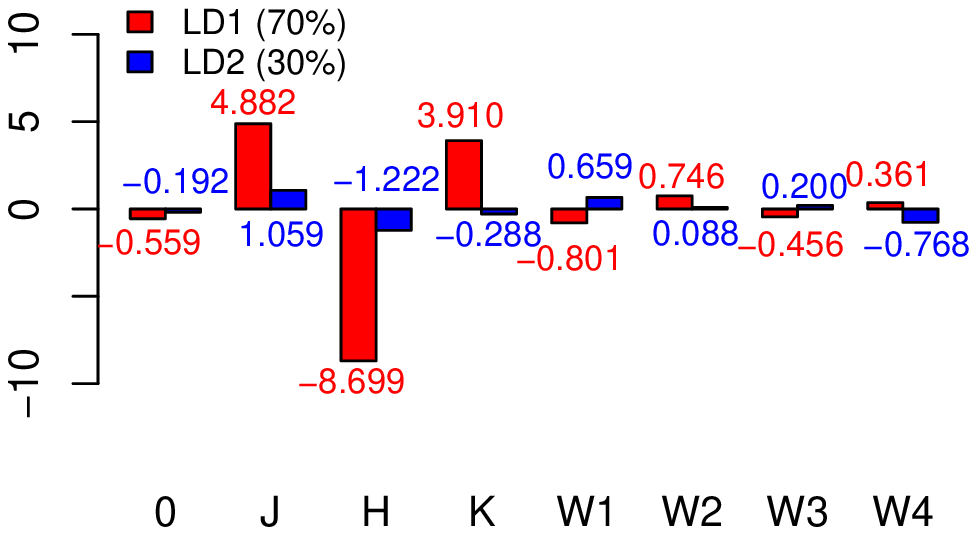}
\end{minipage}
&
\begin{minipage}{.50\textwidth}
\includegraphics[width=\linewidth, height=70mm]{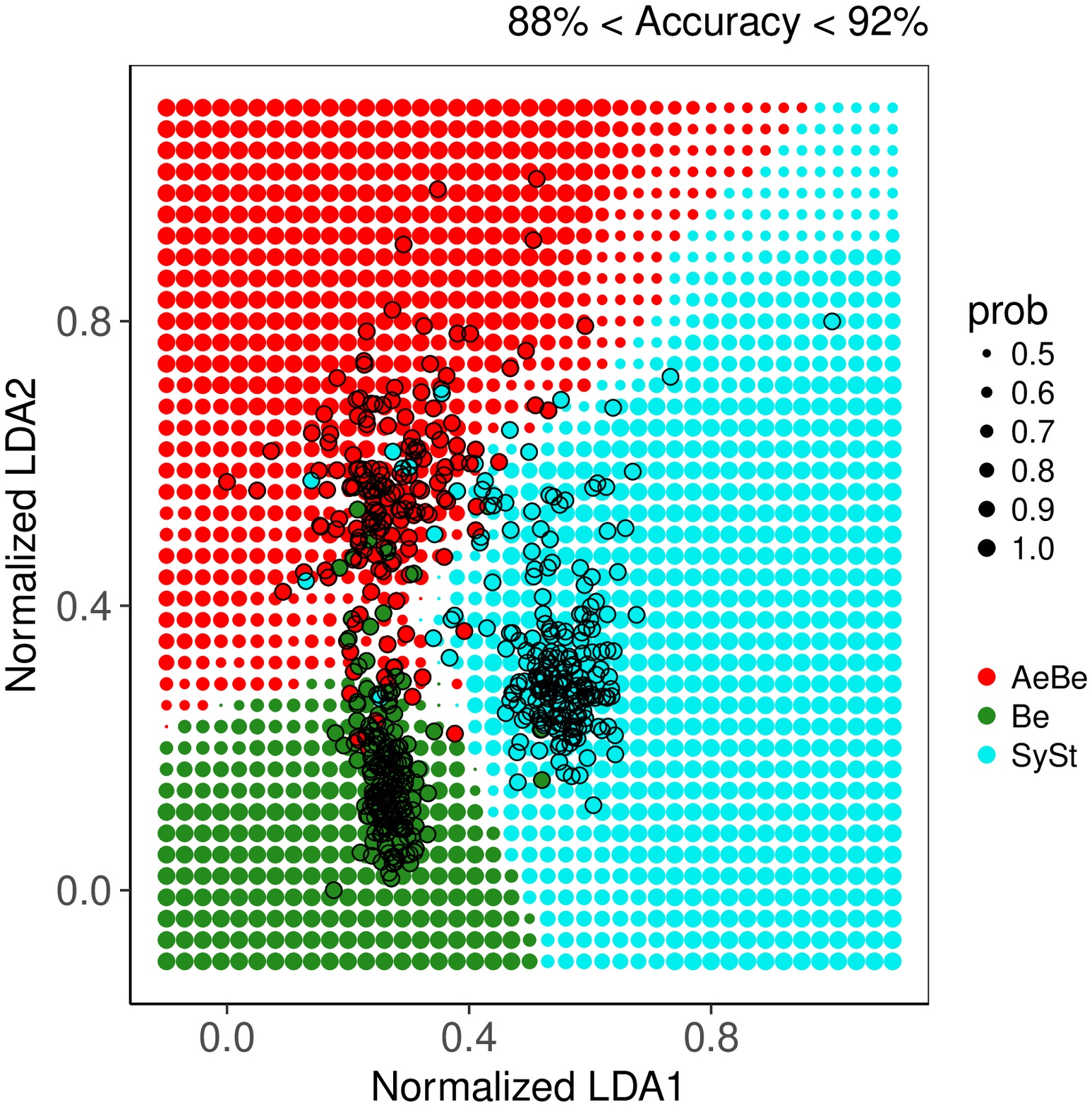}
\end{minipage}
\\ \hline
\end{tabular}
\begin{flushleft}
Left column: Coefficient spectrum plot of the first (red) and second (blue) discriminant components for the seven-dimensional 
space of 2MASS and WISE surveys. "0" variable corresponds to the zero point.  The numbers in parenthesis give the percentage of 
discriminabily. Right column: The LDA/KNN plots for different sets of objects. The size of the background circles corresponds 
to the probability of being classified as a specific type. The equations to normalize the LDA components and produce the KNN plots 
are given in Appendix A.
\end{flushleft}
\end{table*}

\section{Very likely candidate SySts in IPHAS and VPHAS+ catalogues}

The list of 125 sources found in the list of the candidate SySts (paper~I), the IPHAS list of candidate SySts (Corradi et al. 2008) 
and the DR2 VPHAS+ catalogue (Drew et al. 2014) are presented here. The classification of each source based on the classification tree and LD/KNN 
is also provided in the columns 2 to 8 as follows: (a) SySts/PNe/Be, (b) SySts/CV/Mira, (c) SySts/CV/YSO, (d) SySts/WR/post-AGB, 
(e) SySts/K-giants/M-giants, (f) SySts/WTT/ClTT, (g) SySts/Be/AeBe.

\onecolumn
\begin{scriptsize}
\begin{landscape}
\begin{center}
\begin{longtable}{|l|l|l|l|l|l|l|l|l|l|}
\caption{New very likely symbiotic stars found in Paper~I and the IPHAS and VPHAS+ surveys. A further classification of the candidates based on the classification tree analysis is given in columns 2 to 8: (a) SySts/PNe/Be, (b) SySts/CV/Mira, (c) SySts/CV/YSO, (d) SySts/WR/post-AGB, (e) SySts/K-giants/M-giants, (f) SySts/WTT/ClTT, (g) SySts/Be/AeBe.} \label{tab:long} \\

\hline \multicolumn{1}{|c|}{\textbf{Name}} & \multicolumn{1}{c|}{\textbf{(a)}} & \multicolumn{1}{c|}{\textbf{(b)}} & \multicolumn{1}{c|}{\textbf{(c)}} & \multicolumn{1}{c|}{\textbf{(d)}} & \multicolumn{1}{c|}{\textbf{(e)}} & \multicolumn{1}{c|}{\textbf{(f)}} & \multicolumn{1}{c|}{\textbf{(g)}} & \multicolumn{1}{c|}{\textbf{Comments}} \\ \hline 
\endfirsthead

\multicolumn{10}{c}%
{{\bfseries \tablename\ \thetable{} -- continued from previous page}} \\
\hline \multicolumn{1}{|c|}{\textbf{Name}} & \multicolumn{1}{c|}{\textbf{(a)}} & \multicolumn{1}{c|}{\textbf{(b)}} & \multicolumn{1}{c|}{\textbf{(c)}}& \multicolumn{1}{c|}{\textbf{(d)}}& \multicolumn{1}{c|}{\textbf{(e)}}& \multicolumn{1}{c|}{\textbf{f}}& \multicolumn{1}{c|}{\textbf{(g)}}& \multicolumn{1}{c|}{\textbf{Comments}}\\ \hline 
\endhead

\hline \multicolumn{10}{|r|}{{Continued on next page}} \\ \hline
\endfoot

\hline \hline
\endlastfoot
\hline
\hline
{\it Candidates -- Paper~I}\\
\hline
Hen 3-653	    			      &    SySt  &  SySt  &  SySt &  SySt   &  SySt    &  SySt     &  SySt    &      \\
Hen 4-134       			      &    SySt  &  SySt  &  SySt &  SySt   &  SySt    &  SySt     &  SySt    &      \\
Hen 4-137      				      &    SySt  &  SySt  &  SySt &  SySt   &  SySt    &  SySt     &  SySt    &      \\
V748 Cen    			 	      &    SySt  &  SySt  &  SySt &  SySt   &  SySt    &  SySt     &  SySt    &      \\
WRAY 16-294 				      &    SySt  &  SySt  &  SySt &  SySt   &  SySt    &  SySt     &  SySt    &      \\
001.97+02.41 				      &    SySt  &  SySt  &  SySt &  SySt   &  SySt    &  SySt     &  SySt    &      \\
001.33+01.07 				      &    SySt  &  SySt  &  SySt &  SySt   &  SySt    &  SySt     &  SySt    &      \\
001.71+01.14 			          &    SySt  &  SySt  &  SySt &  SySt   &  SySt    &  SySt     &  SySt    &      \\
DASCH J075731.1+201735 		      &    SySt  &  SySt  &  SySt &  SySt   &  Mgiant  &  SySt     &  SySt    &      \\
ASAS J174600-2321.3			      &    SySt  &  SySt  &  SySt &  SySt   &  SySt    &  SySt     &  SySt    &      \\
\hline
{\it dusty} \\
\hline
2MASSJ17145509-393311712	      &    SySt  &  Mira  &  YSO  &  SySt   &  SySt    &  SySt     &  AeBe    &      \\
357.12+01.66             	      &    SySt  &  Mira  &  YSO  &  SySt   &  SySt    &  ClTTauri &  AeBe    &      \\
AS 288   					      &    SySt  &  Mira  &  YSO  &  SySt   &  SySt    &  SySt     &  AeBe    &      \\
\hline
{\it IPHAS}\\
\hline
\hline
{\it stellar} \\
\hline
IPHASJ182906.08-003457.2	      &    SySt  &  SySt  &  SySt &  SySt   &  SySt    &  SySt     &  SySt    &         known SySt	\\  
IPHASJ183501.83+014656.0 	      &    SySt  &  SySt  &  SySt &  SySt   &  SySt    &  SySt     &  SySt    &         known SySt	\\   
DQ Ser                   	      &    SySt  &  SySt  &  SySt &  SySt   &  SySt    &  SySt     &  SySt    &         known SySt	\\  
IPHASJ184446.08+060703.5	      &    SySt  &  SySt  &  SySt &  SySt   &  SySt    &  SySt     &  SySt    &         known SySt	\\  
IPHASJ184733.03+032554.3	      &    SySt  &  SySt  &  SySt &  SySt   &  SySt    &  SySt     &  SySt    &         known SySt	\\  
IPHASJ185039.20+065916.7	      &    SySt  &  SySt  &  SySt &  SySt   &  SySt    &  WTTauri  &  SySt    &     \\ 
IPHASJ185323.58+084955.0	      &    SySt  &  SySt  &  SySt &  SySt   &  SySt    &  SySt     &  SySt    &         known SySt	\\  
IPHASJ190924.64-010910.2          &    SySt  &  SySt  &  SySt &  SySt   &  SySt    &  SySt     &  SySt    &         known SySt	\\   
Ap 3-1                  	      &    SySt  &  SySt  &  SySt &  SySt   &  SySt    &  SySt     &  SySt    &         known SySt	\\  
IPHASJ193436.06+163128.9	      &    SySt  &  SySt  &  SySt &  SySt   &  SySt    &  SySt     &  SySt    &         known SySt	\\  
IPHASJ193501.31+135427.5	      &    SySt  &  SySt  &  SySt &  SySt   &  SySt    &  SySt     &  SySt    &         known SySt	\\  
IPHASJ194120.77+245612.9	      &    SySt  &  SySt  &  SySt &  SySt   &  SySt    &  SySt     &  SySt    &     \\              
\hline
{\it dusty} \\
\hline     
IPHASJ192257.72+113854.8          &    SySt  &  Mira  &  YSO  &  SySt   &  SySt    &  ClTTauri &  AeBe    &     \\     
IPHASJ194907.23+211742.0          &    SySt  &  Mira  &  YSO  &  SySt   &  SySt    &  SySt     &  AeBe    &        young PN (3)\\    
IPHASJ195712.42+301316.1          &    SySt  &  Mira  &  YSO  &  SySt   &  SySt    &  SySt     &  AeBe    &        young PN (3)\\        
IPHASJ201550.96+373004.2	      &    SySt  &  Mira  &  YSO  &  SySt   &  SySt    &  SySt     &  AeBe    &         Be/YSO? (1,2)\\  
IPHASJ202058.52+380949.8 	      &    SySt  &  Mira  &  YSO  &  SySt   &  SySt    &  SySt     &  SySt    &      \\  
IPHASJ202947.93+355926.5 	      &    SySt  &  Mira  &  YSO  &  SySt   &  SySt    &  ClTTauri &  AeBe    &      YSO (3,4) \\  
IPHASJ204713.69+463517.5  	      &    SySt  &  Mira  &  YSO  &  SySt   &  SySt    &  SySt     &  AeBe    &      \\  
IPHASJ215628.47+571445.5 	      &    SySt  &  Mira  &  YSO  &  SySt   &  SySt    &  SySt     &  AeBe    &      \\  
IPHASJ231735.92+634506.4          &    SySt  &  Mira  &  YSO  &  SySt   &  SySt    &  SySt     &  AeBe    &      \\  
IPHASJ203413.39+410157.9          &    SySt  &  Mira  &  YSO  &  SySt   &  SySt    &  ClTTauri &  AeBe    &      \\  
IPHASJ191017.43+065258.1          &    SySt  &  Mira  &  YSO  &  SySt   &  SySt    &  SySt     &  AeBe    &      \\  
\hline
{\it VPHAS+}\\
\hline
\hline
{\it stellar} \\
\hline
VPHASDR2J174455.7-341418.0        &    SySt  &  SySt  &  SySt &  SySt   &  SySt    &  SySt     &  SySt    &         355.39-02.6, known SySt \\
VPHASDR2J174354.4-330845.3        &    SySt  &  SySt  &  SySt &  SySt   &  SySt    &  SySt     &  SySt    &      \\
VPHASDR2J175313.8-301805.8        &    SySt  &  SySt  &  SySt &  SySt   &  Mgiant  &  SySt     &  SySt    &         PN Bl L, known SySt \\
VPHASDR2J175059.8-301247.5 	      &    SySt  &  SySt  &  SySt &  SySt   &  SySt    &  SySt     &  SySt    &         2MASSJ17505978-3012473, ELS \\
VPHASDR2J175320.4-295327.4		  &    SySt  &  SySt  &  SySt &  SySt   &  SySt    &  SySt     &  SySt    &         \\
VPHASDR2J175225.9-294557.0        &    SySt  &  SySt  &  SySt &  SySt   &  SySt    &  SySt     &  SySt    &         PN Bl 3-14, known SySt \\       
VPHASDR2J175231.2-291534.8    	  &    SySt  &  SySt  &  SySt &  SySt   &  SySt    &  SySt     &  SySt    &         000.49-01.45, known SySt \\
VPHASDR2J175346.2-284826.6        &    SySt  &  SySt  &  SySt &  SySt   &  SySt    &  WTTauri  &  SySt    &         \\
VPHASDR2J173007.4-312706.8        &    SySt  &  SySt  &  SySt &  SySt   & SySt     &  SySt     &  SySt    &         \\
VPHASDR2J173123.2-300844.3        &    PN    &  SySt  &  SySt &  SySt   &  SySt    &  WTTauri  &  SySt    &         357.32+01.97, known SySt \\
VPHASDR2J173155.9-301915.6		  &    SySt  &  SySt  &  SySt &  SySt   &  SySt    &  SySt     &  SySt    &         \\ 
VPHASDR2J173435.5-294822.2    	  &    SySt  &  SySt  &  SySt &  SySt   &  SySt    &  SySt     &  SySt    &         357.98+01.57, known SySt \\ 
VPHASDR2J173416.8-292912.0        &    SySt  &  SySt  &  SySt &  SySt   &  SySt    &  SySt     &  SySt    &         PN Th 3-31, known SySt \\
VPHASDR2J173227.9-290509.1        &    SySt  &  SySt  &  SySt &  SySt   &  SySt    &  SySt     &  SySt    &         PN Th 3-29, known SySt\\ 
VPHASDR2J171755.8-300142.6        &    SySt  &  SySt  &  SySt &  SySt   &  SySt    &  SySt     &  SySt    &         PN Sa 3-43, known SySt\\
VPHASDR2J172102.5-292252.8        &    SySt  &  SySt  &  SySt &  SySt   &  SySt    &  SySt     &  SySt    &         PN Th 3-7, known SySt \\
VPHASDR2J172830.6-292124.5        &    SySt  &  SySt  &  SySt &  SySt   &  SySt    &  SySt     &  SySt    &         ELS \\
VPHASDR2J172731.6-290256.4        &    SySt  &  SySt  &  SySt &  SySt   &  SySt    &  SySt     &  SySt    &         PN Th 3-17, known SySt\\
VPHASDR2J173558.5-284954.1        &    SySt  &  SySt  &  SySt &  SySt   &  SySt    &  SySt     &  SySt    &         NSV 22840, known SySt \\
VPHASDR2J174513.7-265242.9        &    SySt  &  SySt  &  SySt &  SySt   &  SySt    &  SySt     &  SySt    &        Carbon star\\
VPHASDR2J174055.7-274748.4		  &    SySt  &  SySt  &  SySt &  SySt   &  SySt    &  SySt     &  SySt    &        Variable star\\
VPHASDR2J173343.4-280721.2        &    SySt  &  SySt  &  SySt &  SySt   &  SySt    &  SySt     &  SySt    &         PN Th 3-30, known SySt\\
VPHASDR2J175648.7-285837.0        &    SySt  &  SySt  &  SySt &  SySt   &  SySt    &  SySt     &  SySt    &         OGLE BLG-LPV 134361, SR-PS \\ 
VPHASDR2J175704.2-285034.8        &    SySt  &  SySt  &  SySt &  SySt   &  SySt    &  SySt     &  SySt    &            \\
VPHASDR2J175645.2-285154.4		  &    SySt  &  SySt  &  SySt &  SySt   &  Mgiant  &  WTTauri  &  SySt    &         \\
VPHASDR2J175828.0-283342.0        &    SySt  &  SySt  &  SySt &  SySt   &  SySt    &  SySt     &  SySt    &         PN H 2-34, known SySt \\
VPHASDR2J175732.5-271825.3   	  &    SySt  &  SySt  &  SySt &  SySt   &  SySt    &  SySt     &  SySt    &         PHR 1757-2718, known SySt \\
VPHASDR2J180913.0-253521.9        &    SySt  &  SySt  &  SySt &  SySt   &  Mgiant  &  SySt     &  SySt    &       \\ 
VPHASDR2J180924.9-253834.8        &    SySt  &  SySt  &  SySt &  SySt   &  SySt    &  WTTauri  &  SySt    &       \\
VPHASDR2J180920.2-253738.5        &    SySt  &  SySt  &  SySt &  SySt   &  SySt    &  SySt     &  SySt    &       \\
VPHASDR2J180923.8-253158.5        &    SySt  &  SySt  &  SySt &  SySt   &  SySt    &  SySt     &  SySt    &       \\
VPHASDR2J180910.5-253003.1        &    SySt  &  SySt  &  SySt &  SySt   &  Mgiant  &  SySt     &  SySt    &       \\
VPHASDR2J180910.6-253023.6        &    SySt  &  SySt  &  SySt &  SySt   &  SySt    &  WTTauri  &  SySt    &       \\
VPHASDR2J180912.0-253053.3        &    SySt  &  SySt  &  SySt &  SySt   &  Mgiant  &  SySt     &  SySt    &       \\
VPHASDR2J180914.2-253827.1        &    SySt  &  SySt  &  SySt &  SySt   &  Mgiant  &  SySt     &  SySt    &       \\
VPHASDR2J180913.6-253106.5        &    SySt  &  SySt  &  SySt &  SySt   &  Mgiant  &  SySt     &  SySt    &       \\ 
VPHASDR2J180915.7-252939.1        &    SySt  &  SySt  &  SySt &  SySt   &  Mgiant  &  SySt     &  SySt    &       \\
VPHASDR2J180910.0-253622.8        &    SySt  &  SySt  &  SySt &  SySt   &  SySt    &  SySt     &  SySt    &       \\
VPHASDR2J181154.5-243536.2        &    SySt  &  SySt  &  SySt &  SySt   &  SySt    &  SySt     &  SySt    &         2MASSJ18115453-2435360, ELS  \\
VPHASDR2J181333.6-245225.0        &    SySt  &  SySt  &  SySt &  SySt   &  SySt    &  SySt     &  SySt    &         [KW2003] 98, ELS \\ 
VPHASDR2J180934.5-245744.2        &    SySt  &  SySt  &  SySt &  SySt   &  SySt    &  SySt     &  SySt    &        \\
VPHASDR2J181123.2-241430.0        &    SySt  &  SySt  &  SySt &  SySt   &  SySt    &  SySt     &  SySt    &         2MASSJ18112322-2414299, ELS \\
VPHASDR2J174512.6-253207.2		  &    SySt  &  SySt  &  SySt &  SySt   &  SySt    &  SySt     &  SySt    &         \\
VPHASDR2J174356.1-250625.6		  &    SySt  &  SySt  &  SySt &  SySt   &  SySt    &  SySt     &  SySt    &         \\ 
VPHASDR2J175527.9-222339.6        &    SySt  &  SySt  &  SySt &  SySt   &  SySt    &  SySt     &  SySt    &         \\
VPHASDR2J181705.6-153203.8        &    SySt  &  SySt  &  SySt &  SySt   &  SySt    &  SySt     &  SySt    &         \\
VPHASDR2J185821.0-071139.5	      &    SySt  &  SySt  &  SySt &  SySt   &  Kgiant  &  WTTauri  &  SySt    &         \\
VPHASDR2J184835.7-064110.4    	  &    SySt  &  SySt  &  SySt &  SySt   &  SySt    &  SySt     &  SySt    &         AS 323, known SySt \\ 
VPHASDR2J191333.7+021813.1	      &    SySt  &  SySt  &  SySt &  SySt   &  SySt    &  SySt     &  SySt    &         V352 Aql, known SySt \\
VPHASDR2J141301.4-653320.1        &    SySt  &  SySt  &  SySt &  SySt   &  SySt    &  SySt     &  SySt    &         WRAY 15-1180, ELS \\
VPHASDR2J160910.9-530245.4        &    SySt  &  SySt  &  SySt &  SySt   &  SySt    &  SySt     &  SySt    &         \\
VPHASDR2J165421.0-404248.0        &    SySt  &  SySt  &  SySt &  SySt   &  SySt    &  WTTauri  &  SySt    &         \\
\hline
{\it dusty} \\
\hline
VPHASDR2J175016.7-305734.6        &    SySt  &  Mira  &  YSO  &  SySt   &  SySt    &  SySt     &  SySt    &     WRAY~16-312, Known SySt     \\
VPHASDR2J175153.5-293053.5        &    SySt  &  Mira  &  YSO  &  SySt   &  SySt    &  SySt     &  AeBe    &     variable star     \\
VPHASDR2J173030.0-304937.2        &    SySt  &  Mira  &  YSO  &  SySt   &  SySt    &  SySt     &  AeBe    &          \\
VPHASDR2J173204.8-302854.6        &    SySt  &  Mira  &  YSO  &  SySt   &  SySt    &  ClTTauri &  AeBe    &          \\
VPHASDR2J173522.2-294519.8        &    SySt  &  Mira  &  YSO  &  SySt   &  SySt    &  SySt     &  AeBe    &     Hen~2-251 Known SySt  \\   
VPHASDR2J175821.9-281452.2        &    SySt  &  Mira  &  YSO  &  SySt   &  SySt    &  SySt     &  SySt    &     PN H~1-45 Known SySt     \\ 
VPHASDR2J180149.5-195828.4        &    SySt  &  Mira  &  YSO  &  SySt   &  SySt    &  SySt     &  SySt    &          \\
VPHASDR2J180803.5-203454.0        &    SySt  &  Mira  &  YSO  &  SySt   &  SySt    &  ClTTauri &  AeBe    &      posible YSO    \\
VPHASDR2J182047.1-173627.3        &    SySt  &  Mira  &  YSO  &  SySt   &  SySt    &  SySt     &  SySt    &          \\
VPHASDR2J182503.1-143031.5        &    SySt  &  Mira  &  YSO  &  SySt   &  SySt    &  ClTTauri &  SySt    &          \\
VPHASDR2J183013.2-135356.7	      &    SySt  &  Mira  &  YSO  &  SySt   &  SySt    &  SySt     &  AeBe    &     Known PN     \\   
VPHASDR2J182831.5-122059.5        &    SySt  &  Mira  &  YSO  &  SySt   &  SySt    &  SySt     &  SySt    &          \\
VPHASDR2J182606.0-122839.3        &    SySt  &  Mira  &  YSO  &  SySt   &  SySt    &  ClTTauri &  SySt    &          \\
VPHASDR2J184024.2-084346.3		  &    SySt  &  Mira  &  YSO  &  SySt   &  SySt    &  SySt     &  SySt    &     PN K~3-9  Known SySt     \\
VPHASDR2J183910.8-085644.4        &    SySt  &  Mira  &  YSO  &  SySt   &  SySt    &  ClTTauri &  AeBe    &          \\
VPHASDR2J183044.6-100757.4        &    SySt  &  Mira  &  YSO  &  SySt   &  SySt    &  SySt     &  SySt    &          \\
VPHASDR2J184303.6-050026.4        &    SySt  &  Mira  &  YSO  &  SySt   &  SySt    &  SySt     &  SySt    &     AGB star   \\
VPHASDR2J184532.1-005029.4        &    SySt  &  Mira  &  YSO  &  SySt   &  SySt    &  SySt     &  AeBe    &          \\
VPHASDR2J184229.2-002144.1        &    SySt  &  Mira  &  YSO  &  SySt   &  SySt    &  SySt     &  AeBe    &          \\
VPHASDR2J101521.0-570706.0        &    SySt  &  Mira  &  YSO  &  SySt   &  SySt    &  SySt     &  AeBe    &          \\
VPHASDR2J124845.2-634948.6        &    SySt  &  Mira  &  YSO  &  SySt   &  SySt    &  SySt     &  SySt    &          \\
VPHASDR2J133405.8-623745.7        &    SySt  &  Mira  &  YSO  &  SySt   &  SySt    &  SySt     &  AeBe    &     AGB star    \\
VPHASDR2J133509.6-614305.8        &    SySt  &  Mira  &  YSO  &  SySt   &  SySt    &  SySt     &  AeBe    &     ELS     \\
VPHASDR2J154125.6-565953.2		  &    SySt  &  Mira  &  YSO  &  SySt   &  SySt    &  ClTTauri &  AeBe    &          \\   
VPHASDR2J152144.3-572220.6        &    SySt  &  Mira  &  YSO  &  SySt   &  SySt    &  SySt     &  AeBe    &          \\
VPHASDR2J160631.4-525616.4        &    SySt  &  Mira  &  YSO  &  SySt   &  SySt    &  SySt     &  SySt    &          \\
VPHASDR2J164646.3-454758.3        &    SySt  &  Mira  &  YSO  &  SySt   &  SySt    &  SySt     &  SySt    &     WR star     \\
VPHASDR2J162446.2-485536.4        &    SySt  &  Mira  &  YSO  &  SySt   &  SySt    &  SySt     &  SySt    &          \\
VPHASDR2J162457.4-484340.2        &    SySt  &  Mira  &  YSO  &  SySt   &  SySt    &  SySt     &  AeBe    &     ELS     \\
VPHASDR2J164300.8-452701.3        &    SySt  &  Mira  &  YSO  &  SySt   &  SySt    &  SySt     &  SySt    &          \\
VPHASDR2J165346.7-434931.0        &    SySt  &  Mira  &  YSO  &  SySt   &  SySt    &  SySt     &  AeBe    &          \\
VPHASDR2J171225.1-412555.4        &    SySt  &  Mira  &  YSO  &  SySt   &  SySt    &  SySt     &  AeBe    &          \\
VPHASDR2J171455.1-393311.7        &    SySt  &  Mira  &  YSO  &  SySt   &  SySt    &  SySt     &  AeBe    &     SySt candidate     \\
VPHASDR2J171527.4-390209.2        &    SySt  &  Mira  &  YSO  &  SySt   &  SySt    &  SySt     &  AeBe    &          \\
VPHASDR2J171513.5-364633.2        &    SySt  &  Mira  &  YSO  &  SySt   &  SySt    &  ClTTauri &  SySt    &          \\
VPHASDR2J171445.0-361838.4        &    SySt  &  Mira  &  YSO  &  SySt   &  SySt    &  ClTTauri &  Syst    &     ELS      \\
VPHASDR2J120916.3-633202.7        &    SySt  &  Mira  &  YSO  &  SySt   &  SySt    &  SySt     &  SySt    &          \\

\end{longtable}
\begin{flushleft}
The classification of some sources as emission line stars (ELS), semi regular pulsating star (SR-PS), asymptotic giant branch stars (AGB),  
Wolf-Rayet stars (WR), planetary nebula (PN) or known/candidate symbiotic stars (SySt) is based on the SIMBAD catalogue, Kohoutek \& Wehmeyer 
(2003) or Paper~I.\\
(1) Rodr\'{i}guez-Flores et al. 2014, (2) Corradi et al. 2010, (3) Viironen et al. (2009b),  (4) Krause et al. (2003)\\
\end{flushleft} 
\end{center}
\end{landscape}
\end{scriptsize}

\newpage

\onecolumn
\begin{scriptsize}
\begin{landscape}
\begin{center}
\begin{longtable}{|l|l|l|l|l|l|l|l|l|l|}
\caption{NNew very likely symbiotic stars found in Paper~I and the IPHAS and VPHAS+ surveys. A further classification of the candidates 
based on the LDA/KNN analysis is given in columns 2 to 8: (a) SySts/PNe/Be, (b) SySts/CV/Mira, (c) SySts/CV/YSO, (d) SySts/WR/post-AGB, 
(e) SySts/K-giants/M-giants, (f) SySts/WTT/ClTT, (g) SySts/Be/AeBe.} \label{tab:long} \\

\hline \multicolumn{1}{|c|}{\textbf{Name}} & \multicolumn{1}{c|}{\textbf{(a)}} & \multicolumn{1}{c|}{\textbf{(b)}} & \multicolumn{1}{c|}{\textbf{(c)}} & \multicolumn{1}{c|}{\textbf{(d)}} & \multicolumn{1}{c|}{\textbf{(e)}} & \multicolumn{1}{c|}{\textbf{(f)}} & \multicolumn{1}{c|}{\textbf{(g)}} & \multicolumn{1}{c|}{\textbf{Comments}} \\ \hline 
\endfirsthead

\multicolumn{10}{c}%
{{\bfseries \tablename\ \thetable{} -- continued from previous page}} \\
\hline \multicolumn{1}{|c|}{\textbf{Name}} & \multicolumn{1}{c|}{\textbf{(a)}} & \multicolumn{1}{c|}{\textbf{(b)}} & \multicolumn{1}{c|}{\textbf{(c)}}& \multicolumn{1}{c|}{\textbf{(d)}}& \multicolumn{1}{c|}{\textbf{(e)}}& \multicolumn{1}{c|}{\textbf{f}}& \multicolumn{1}{c|}{\textbf{(g)}}& \multicolumn{1}{c|}{\textbf{Comments}}\\ \hline 
\endhead

\hline \multicolumn{10}{|r|}{{Continued on next page}} \\ \hline
\endfoot

\hline \hline
\endlastfoot
\hline
\hline
{\it Candidates -- Paper~I}\\
\hline
\hline
{\it stellar} \\
\hline
Hen 3-653    		       	      & SySt  (1.00) & SySt  (0.71) & SySt   (1.00) & SySt  (1.00)   & SySt   (0.85)  & SySt     (1.00) & SySt  (1.00)   &     \\
Hen 4-134       			      & SySt  (1.00) & Mira  (1.00) & SySt   (1.00) & SySt  (1.00)   & Kgiant (0.85)  & SySt     (1.00) & SySt  (0.86)   &     \\
Hen 4-137      				      & SySt  (1.00) & SySt  (0.71) & SySt   (1.00) & SySt  (1.00)   & Mgiant (1.00)  & SySt     (1.00) & SySt  (1.00)   &     \\
V748 Cen    			 	      & SySt  (1.00) & Mira  (0.71) & SySt   (0.86) & SySt  (1.00)   & Mgiant (0.85)  & WTTauri  (0.57) & SySt  (0.86)   &     \\
WRAY 16-294 				      & SySt  (1.00) & SySt  (0.86) & SySt   (0.86) & SySt  (0.57)   & Mgiant (0.43)  & SySt     (1.00) & SySt  (0.86)   &     \\
001.97+02.41 		              & SySt  (1.00) & SySt  (1.00) & SySt   (0.86) & SySt  (1.00)   & SySt   (1.00)  & SySt     (0.71) & SySt  (1.00)   &     \\
001.33+01.07 				      & SySt  (1.00) & SySt  (1.00) & SySt   (1.00) & SySt  (1.00)   & SySt   (1.00)  & SySt     (1.00) & SySt  (1.00)   &     \\
001.71+01.14 			          & SySt  (1.00) & SySt  (1.00) & SySt   (1.00) & SySt  (0.86)   & SySt   (1.00)  & SySt     (1.00) & SySt  (1.00)   &     \\
DASCH J075731.1+201735 		      & SySt  (1.00) & Mira  (0.71) & SySt   (1.00) & SySt  (1.00)   & Mgiant (0.85)  & SySt     (1.00) & SySt  (0.86)   &     \\
ASAS J174600-2321.3 	          & SySt  (1.00) & SySt  (1.00) & SySt   (1.00) & SySt  (0.86)   & SySt   (1.00)  & SySt     (0.71) & SySt  (1.00)   &     \\  
\hline
{\it dusty} \\
\hline
2MASSJ17145509-393311712	      & SySt  (1.00) & SySt  (0.57) & SySt   (0.57) & SySt  (1.00)   & SySt   (1.00)  & SySt     (1.00) & SySt  (0.86)   &     \\
357.12+01.66             	      & SySt  (0.85) & SySt  (0.71) & SySt   (0.86) & SySt  (1.00)   & SySt   (1.00)  & SySt     (1.00) & SySt  (0.86)   &     \\
AS 288   					      & SySt  (0.85) & Mira  (0.57) & YSO    (0.86) & SySt  (0.57)   & SySt   (1.00)  & SySt     (1.00) & SySt  (0.86)   &     \\
\hline
{\it IPHAS}\\
\hline
\hline
{\it stellar} \\
\hline
IPHASJ182906.08-003457.2	      & SySt  (1.00) & SySt  (1.00) & SySt   (0.86) & SySt  (1.00)   & SySt   (1.00)  & SySt     (1.00) & SySt  (1.00)   &        known SySt	\\ 
IPHASJ183501.83+014656.0 	      & SySt  (1.00) & SySt  (1.00) & SySt   (1.00) & SySt  (0.86)   & SySt   (1.00)  & SySt     (1.00) & SySt  (1.00)   &        known SySt	\\   
DQ Ser                   	      & SySt  (1.00) & SySt  (0.86) & SySt   (1.00) & SySt  (1.00)   & SySt   (0.85)  & SySt     (1.00) & SySt  (1.00)   &         known SySt	\\  
IPHASJ184446.08+060703.5	      & SySt  (1.00) & SySt  (1.00) & SySt   (1.00) & SySt  (1.00)   & SySt   (0.57)  & SySt     (1.00) & SySt  (1.00)   &         known SySt	\\  
IPHASJ184733.03+032554.3	      & SySt  (1.00) & SySt  (1.00) & SySt   (1.00) & SySt  (1.00)   & SySt   (1.00)  & SySt     (1.00) & SySt  (1.00)   &         known SySt	\\  
IPHASJ185039.20+065916.7	      & SySt  (1.00) & SySt  (1.00) & YSO    (0.71) & SySt  (1.00)   & SySt   (1.00)  & SySt     (1.00) & SySt  (1.00)   &     \\
IPHASJ185323.58+084955.0	      & SySt  (1.00) & SySt  (1.00) & SySt   (1.00) & SySt  (1.00)   & SySt   (1.00)  & SySt     (1.00) & SySt  (1.00)   &         known SySt	\\  
IPHASJ190924.64-010910.2          & SySt  (1.00) & SySt  (1.00) & SySt   (0.71) & SySt  (0.86)   & SySt   (0.85)  & SySt     (0.71) & SySt  (1.00)   &         known SySt	\\   
Ap 3-1                  	      & SySt  (1.00) & SySt  (1.00) & SySt   (0.71) & SySt  (1.00)   & Mgiant (0.57)  & SySt     (1.00) & SySt  (1.00)   &         known SySt	\\  
IPHASJ193436.06+163128.9	      & SySt  (1.00) & SySt  (1.00) & SySt   (1.00) & SySt  (0.86)   & SySt   (1.00)  & SySt     (0.86) & SySt  (1.00)   &         known SySt	\\  
IPHASJ193501.31+135427.5	      & SySt  (1.00) & SySt  (1.00) & SySt   (0.71) & SySt  (1.00)   & SySt   (1.00)  & SySt     (1.00) & SySt  (1.00)   &        known SySt	\\  
IPHASJ194120.77+245612.9	      & SySt  (1.00) & SySt  (1.00) & SySt   (0.71) & SySt  (1.00)   & SySt   (1.00)  & SySt     (1.00) & SySt  (1.00)   &    \\
\hline
{\it dusty} \\
\hline 
IPHASJ192257.72+113854.8 	      & SySt  (0.57) & SySt  (1.00) & YSO    (0.85) & SySt  (0.57)   & SySt   (1.00)  & SySt     (1.00) & SySt  (1.00)   &    \\     
IPHASJ194907.23+211742.0          & SySt  (0.71) & SySt  (0.86) & YSO    (1.00) & WR    (0.57)   & SySt   (1.00)  & SySt     (0.86) & SySt  (1.00)   &    young PN (3) \\  
IPHASJ195712.42+301316.1          & SySt  (0.71) & Mira  (0.86) & YSO    (1.00) & WR    (1.00)   & SySt   (0.86)  & ClTTauri (0.75) & AeBE  (0.71)   &    young PN (3) \\  
IPHASJ201550.96+373004.2	      & SySt  (0.71) & Mira  (0.86) & YSO    (1.00) & WR    (0.86)   & SySt   (1.00)  & SySt     (0.85) & SySt  (0.71)   &        Be/YSO? (1,2)\\
IPHASJ202058.52+380949.8 	      & SySt  (0.86) & Mira  (0.86) & YSO    (1.00) & WR    (1.00)   & SySt   (0.86)  & SySt     (1.00) & AeBe  (0.43)   &     \\
IPHASJ202947.93+355926.5 	      & SySt  (0.71) & Mira  (0.86) & YSO    (0.86) & SySt  (0.71)   & SySt   (1.00)  & SySt     (0.63) & SySt  (0.57)   &     YSO (3,4) \\
IPHASJ204713.69+463517.5  	      & SySt  (0.57) & Mira  (0.71) & YSO    (0.71) & WR    (0.57)   & SySt   (0.85)  & ClTTauri (0.86) & AeBe  (0.71)   &     \\
IPHASJ215628.47+571445.5 	      & SySt  (0.43) & SySt  (0.71) & YSO    (1.00) & WR    (1.00)   & SySt   (1.00)  & ClTTauri (0.71) & Be    (0.57)   &     \\
IPHASJ231735.92+634506.4          & SySt  (0.71) & Mira  (1.00) & YSO    (1.00) & WR    (0.57)   & SySt   (1.00)  & SySt     (0.86) & SySt  (0.57)   &     \\
IPHASJ203413.39+410157.9          & SySt  (0.86) & SySt  (1.00) & YSO    (0.86) & SySt  (1.00)   & SySt   (1.00)  & SySt     (1.00) & SySt  (1.00)   &      \\  
IPHASJ191017.43+065258.1          & SySt  (0.86) & Mira  (0.71) & YSO    (0.86) & SySt  (0.57)   & SySt   (1.00)  & SySt     (0.57) & SySt  (0.57)   &      \\  
\hline
{\it VPHAS+}\\
\hline
\hline
{\it stellar} \\
\hline
VPHASDR2J174455.7-341418.0        & SySt  (1.00) & SySt  (1.00) & SySt   (1.00) & SySt  (1.00)   & SySt   (1.00)  & SySt     (1.00) & SySt  (1.00)   &        355.39-02.6, known SySt \\      
VPHASDR2J174354.4-330845.3        & SySt  (1.00) & SySt  (1.00) & SySt   (0.71) & SySt  (1.00)   & SySt   (1.00)  & SySt     (1.00) & SySt  (1.00)   &     \\
VPHASDR2J175313.8-301805.8        & SySt  (1.00) & SySt  (1.00) & SySt   (1.00) & SySt  (0.86)   & SySt   (1.00)  & SySt     (0.57) & SySt  (1.00)   &        PN Bl L, known SySt \\
VPHASDR2J175059.8-301247.5        & SySt  (1.00) & SySt  (1.00) & SySt   (0.86) & SySt  (1.00)   & SySt   (1.00)  & SySt     (1.00) & SySt  (1.00)   &        2MASSJ17505978-3012473, ELS \\       
VPHASDR2J175320.4-295327.4	      & SySt  (1.00) & SySt  (1.00) & SySt   (1.00) & SySt  (1.00)   & SySt   (1.00)  & SySt     (1.00) & SySt  (1.00)   &        \\
VPHASDR2J175225.9-294557.0        & SySt  (1.00) & SySt  (1.00) & SySt   (1.00) & SySt  (1.00)   & SySt   (1.00)  & SySt     (1.00) & SySt  (1.00)   &       PN Bl 3-14, known SySt \\
VPHASDR2J175231.2-291534.8        & SySt  (1.00) & SySt  (0.86) & SySt   (1.00) & SySt  (1.00)   & SySt   (1.00)  & SySt     (1.00) & SySt  (1.00)   &        000.49-01.45, known SySt \\    
VPHASDR2J175346.2-284826.6        & SySt  (1.00) & SySt  (1.00) & SySt   (1.00) & SySt  (1.00)   & SySt   (1.00)  & SySt     (0.86) & SySt  (1.00)   &        \\
VPHASDR2J173007.4-312706.8        & SySt  (1.00) & SySt  (1.00) & SySt   (0.86) & SySt  (1.00)   & SySt   (1.00)  & SySt     (1.00) & SySt  (1.00)   &        \\ 
VPHASDR2J173123.2-300844.3        & SySt  (0.85) & SySt  (1.00) & SySt   (0.85) & SySt  (1.00)   & SySt   (1.00)  & SySt     (0.57) & SySt  (1.00)   &        357.32+01.97, known SySt \\
VPHASDR2J173155.9-301915.6		  & SySt  (1.00) & SySt  (1.00) & SySt   (1.00) & SySt  (1.00)   & SySt   (1.00)  & SySt     (1.00) & SySt  (1.00)   &        \\       
VPHASDR2J173435.5-294822.2        & SySt  (1.00) & SySt  (1.00) & SySt   (0.86) & SySt  (1.00)   & SySt   (1.00)  & SySt     (1.00) & SySt  (1.00)   &        357.98+01.57, known SySt \\ 
VPHASDR2J173416.8-292912.0        & SySt  (1.00) & SySt  (1.00) & SySt   (0.86) & SySt  (0.86)   & SySt   (1.00)  & SySt     (0.71) & SySt  (1.00)   &        PN Th 3-31, known SySt \\ 
VPHASDR2J173227.9-290509.1        & SySt  (1.00) & SySt  (0.86) & SySt   (1.00) & SySt  (1.00)   & SySt   (1.00)  & SySt     (1.00) & SySt  (1.00)   &        PN Th 3-29, known SySt\\ 
VPHASDR2J171755.8-300142.6        & SySt  (1.00) & SySt  (1.00) & SySt   (1.00) & SySt  (1.00)   & SySt   (1.00)  & SySt     (1.00) & SySt  (0.86)   &        PN Sa 3-43, known SySt  \\    
VPHASDR2J172102.5-292252.8        & SySt  (1.00) & SySt  (1.00) & SySt   (1.00) & SySt  (1.00)   & SySt   (1.00)  & SySt     (1.00) & SySt  (0.86)   &        PN Th 3-7, known SySt \\
VPHASDR2J172830.6-292124.5        & SySt  (1.00) & SySt  (0.86) & SySt   (1.00) & SySt  (1.00)   & SySt   (0.71)  & SySt     (1.00) & SySt  (1.00)   &        ELS \\
VPHASDR2J172731.6-290256.4        & SySt  (1.00) & SySt  (1.00) & SySt   (1.00) & SySt  (0.86)   & SySt   (1.00)  & SySt     (1.00) & SySt  (1.00)   &        PN Th 3-17, known SySt\\
VPHASDR2J173558.5-284954.1        & SySt  (1.00) & SySt  (1.00) & SySt   (1.00) & SySt  (1.00)   & SySt   (1.00)  & SySt     (0.71) & SySt  (1.00)   &        NSV 22840, known SySt \\
VPHASDR2J174513.7-265242.9        & SySt  (1.00) & SySt  (1.00) & SySt   (1.00) & SySt  (1.00)   & SySt   (1.00)  & SySt     (1.00) & SySt  (1.00)   &        Carbon star\\
VPHASDR2J174055.7-274748.4        & SySt  (1.00) & SySt  (1.00) & SySt   (1.00) & SySt  (1.00)   & SySt   (1.00)  & SySt     (1.00) & SySt  (1.00)   &        Variable star\\
VPHASDR2J173343.4-280721.2        & SySt  (1.00) & SySt  (1.00) & SySt   (1.00) & SySt  (1.00)   & SySt   (1.00)  & SySt     (1.00) & SySt  (1.00)   &        PN Th 3-30, known SySt\\
VPHASDR2J175648.7-285837.0        & SySt  (1.00) & SySt  (1.00) & SySt   (1.00) & SySt  (1.00)   & SySt   (1.00)  & SySt     (0.86) & SySt  (1.00)   &        OGLE BLG-LPV 134361, SR-PS \\
VPHASDR2J175704.2-285034.8        & SySt  (1.00) & SySt  (1.00) & SySt   (1.00) & SySt  (1.00)   & SySt   (1.00)  & SySt     (1.00) & SySt  (1.00)   &        \\
VPHASDR2J175645.2-285154.4        & SySt  (1.00) & SySt  (1.00) & YSO    (0.57) & SySt  (1.00)   & SySt   (1.00)  & WTTauri  (1.00) & SySt  (1.00)   &        \\
VPHASDR2J175828.0-283342.0        & SySt  (1.00) & SySt  (1.00) & SySt   (1.00) & SySt  (1.00)   & SySt   (1.00)  & SySt     (1.00) & SySt  (1.00)   &        PN H 2-34, known SySt \\ 
VPHASDR2J175732.5-271825.3        & SySt  (1.00) & SySt  (1.00) & SySt   (1.00) & SySt  (1.00)   & SySt   (1.00)  & SySt     (1.00) & SySt  (1.00)   &        PHR 1757-2718, known SySt \\   
VPHASDR2J180913.0-253521.9        & SySt  (1.00) & SySt  (1.00) & SySt   (1.00) & SySt  (1.00)   & SySt   (1.00)  & SySt     (0.71) & SySt  (1.00)   &      \\ 
VPHASDR2J180924.9-253834.8        & SySt  (1.00) & SySt  (1.00) & SySt   (0.71) & SySt  (1.00)   & SySt   (1.00)  & SySt     (0.57) & SySt  (1.00)   &      \\
VPHASDR2J180920.2-253738.5        & SySt  (1.00) & SySt  (1.00) & SySt   (1.00) & SySt  (1.00)   & SySt   (1.00)  & SySt     (0.71) & SySt  (1.00)   &        \\ 
VPHASDR2J180923.8-253158.5        & SySt  (1.00) & SySt  (1.00) & SySt   (1.00) & SySt  (1.00)   & SySt   (1.00)  & SySt     (0.71) & SySt  (1.00)   &        \\
VPHASDR2J180910.5-253003.1        & SySt  (1.00) & SySt  (1.00) & CV     (0.57) & SySt  (1.00)   & SySt   (1.00)  & WTTauri  (0.86) & SySt  (0.86)   &        \\
VPHASDR2J180910.6-253023.6        & SySt  (0.86) & SySt  (1.00) & YSO    (0.57) & SySt  (1.00)   & SySt   (1.00)  & WTTauri  (1.00) & SySt  (1.00)   &      \\
VPHASDR2J180912.0-253053.3        & SySt  (1.00) & SySt  (1.00) & SySt   (0.86) & SySt  (1.00)   & SySt   (1.00)  & SySt     (1.00) & SySt  (1.00)   &      \\       
VPHASDR2J180914.2-253827.1        & SySt  (1.00) & SySt  (1.00) & SySt   (1.00) & SySt  (1.00)   & SySt   (1.00)  & SySt     (0.85) & SySt  (1.00)   &      \\      
VPHASDR2J180913.6-253106.5        & SySt  (1.00) & SySt  (1.00) & SySt   (1.00) & SySt  (1.00)   & SySt   (0.86)  & SySt     (1.00) & SySt  (0.86)   &        \\  
VPHASDR2J180915.7-252939.1        & SySt  (1.00) & SySt  (1.00) & SySt   (0.86) & SySt  (1.00)   & SySt   (1.00)  & SySt     (0.71) & SySt  (1.00)   &      \\
VPHASDR2J180910.0-253622.8        & SySt  (1.00) & SySt  (1.00) & SySt   (0.57) & SySt  (1.00)   & SySt   (1.00)  & WTTauri  (1.00) & SySt  (0.86)   &        \\  
VPHASDR2J181154.5-243536.2        & SySt  (1.00) & SySt  (1.00) & SySt   (0.71) & SySt  (1.00)   & SySt   (1.00)  & SySt     (1.00) & SySt  (1.00)   &        2MASSJ18115453-2435360, ELS  \\
VPHASDR2J181333.6-245225.0        & SySt  (1.00) & SySt  (1.00) & SySt   (0.86) & SySt  (1.00)   & SySt   (1.00)  & SySt     (0.57) & SySt  (1.00)   &        [KW2003] 98, ELS \\       
VPHASDR2J180934.5-245744.2        & SySt  (1.00) & SySt  (1.00) & SySt   (1.00) & SySt  (1.00)   & SySt   (1.00)  & SySt     (1.00) & SySt  (1.00)   &       \\
VPHASDR2J181123.2-241430.0        & SySt  (1.00) & SySt  (1.00) & SySt   (1.00) & SySt  (1.00)   & SySt   (1.00)  & SySt     (1.00) & SySt  (1.00)   &        2MASSJ18112322-2414299, ELS \\      
VPHASDR2J174512.6-253207.2		  & SySt  (1.00) & SySt  (1.00) & SySt   (0.86) & SySt  (1.00)   & SySt   (1.00)  & SySt     (0.86) & SySt  (1.00)   &        \\       
VPHASDR2J174356.1-250625.6		  & SySt  (1.00) & SySt  (1.00) & SySt   (0.86) & SySt  (1.00)   & SySt   (1.00)  & SySt     (1.00) & SySt  (1.00)   &        \\ 
VPHASDR2J175527.9-222339.6        & SySt  (1.00) & SySt  (1.00) & SySt   (1.00) & SySt  (1.00)   & SySt   (1.00)  & SySt     (1.00) & SySt  (1.00)   &        \\   
VPHASDR2J181705.6-153203.8        & SySt  (1.00) & SySt  (1.00) & SySt   (1.00) & SySt  (1.00)   & SySt   (1.00)  & SySt     (1.00) & SySt  (1.00)   &        \\ 
VPHASDR2J185821.0-071139.5	      & SySt  (1.00) & SySt  (0.86) & SySt   (0.43) & SySt  (1.00)   & SySt   (1.00)  & WTTauri  (0.86) & SySt  (1.00)   &        \\
VPHASDR2J184835.7-064110.4        & SySt  (1.00) & SySt  (0.86) & SySt   (1.00) & SySt  (1.00)   & Mgiant (0.42)  & SySt     (1.00) & SySt  (1.00)   &        AS 323, known SySt \\      
VPHASDR2J191333.7+021813.1        & SySt  (1.00) & SySt  (1.00) & SySt   (1.00) & SySt  (1.00)   & Kgiant (0.71)  & SySt     (1.00) & SySt  (1.00)   &        V352 Aql, known SySt \\
VPHASDR2J141301.4-653320.1        & SySt  (1.00) & SySt  (1.00) & SySt   (1.00) & SySt  (1.00)   & SySt   (0.71)  & SySt     (1.00) & SySt  (1.00)   &        WRAY 15-1180, ELS \\     
VPHASDR2J160910.9-530245.4        & SySt  (1.00) & SySt  (1.00) & SySt   (1.00) & SySt  (1.00)   & SySt   (1.00)  & SySt     (1.00) & SySt  (1.00)   &        \\ 
VPHASDR2J165421.0-404248.0        & SySt  (0.71) & SySt  (1.00) & CV     (0.57) & SySt  (1.00)   & SySt   (1.00)  & WTTauri  (0.86) & SySt  (1.00)   &        \\   
\hline
{\it dusty} \\
\hline
VPHASDR2J175016.7-305734.6        & SySt  (1.00) & SySt  (1.00) & YSO    (0.86) & SySt  (1.00)   & SySt   (1.00)  & SySt     (1.00) & SySt  (1.00)   &     WRAY~16-312, Known SySt     \\
VPHASDR2J175153.6-293053.5		  & SySt  (0.86) & SySt  (0.75) & SySt   (0.57) & SySt  (0.86)   & SySt   (1.00)  & SySt     (0.86) & SySt  (0.71)   &    variable star     \\
VPHASDR2J173030.0-304937.2		  & SySt  (0.86) & Mira  (0.57) & SySt   (0.57) & SySt  (0.86)   & SySt   (1.00)  & SySt     (1.00) & SySt  (1.00)   &         \\
VPHASDR2J173204.8-302854.6	      & SySt  (0.86) & SySt  (0.71) & SySt   (0.85) & SySt  (1.00)   & SySt   (1.00)  & SySt     (1.00) & SySt  (0.86)   &          \\
VPHASDR2J173522.2-294519.8        & SySt  (0.86) & SySt  (1.00) & SySt   (0.71) & SySt  (0.57)   & SySt   (1.00)  & SySt     (1.00) & SySt  (0.57)   &     Hen~2-251 Known SySt  \\   
VPHASDR2J175821.9-281452.2		  & SySt  (1.00) & SySt  (0.86) & SySt   (0.86) & SySt  (1.00)   & SySt   (1.00)  & SySt     (1.00) & SySt  (1.00)   &     PN H~1-45 Known SySt     \\ 
VPHASDR2J180149.5-195828.4        & SySt  (0.86) & SySt  (0.86) & YSO    (0.86) & SySt  (0.86)   & SySt   (1.00)  & SySt     (0.86) & SySt  (0.86)   &          \\
VPHASDR2J180803.5-203454.0        & SySt  (0.57) & Mira  (0.86) & YSO    (1.00) & SySt  (0.71)   & SySt   (1.00)  & SySt     (1.00) & SySt  (0.86)   &      posible YSO    \\
VPHASDR2J182047.1-173627.3        & SySt  (1.00) & SySt  (1.00) & SySt   (1.00) & SySt  (1.00)   & SySt   (1.00)  & SySt     (1.00) & SySt  (0.86)   &          \\
VPHASDR2J182503.1-143031.5	      & SySt  (0.86) & SySt  (0.71) & SySt   (0.86) & SySt  (1.00)   & SySt   (1.00)  & SySt     (1.00) & SySt  (1.00)   &         \\
VPHASDR2J183013.2-135356.7        & SySt  (1.00) & SySt  (1.00) & SySt   (1.00) & SySt  (1.00)   & SySt   (1.00)  & SySt     (1.00) & SySt  (0.71)   &    Known PN     \\   
VPHASDR2J182831.5-122059.5        & SySt  (1.00) & Mira  (0.86) & YSO    (0.57) & SySt  (1.00)   & SySt   (1.00)  & SySt     (1.00) & SySt  (0.71)   &          \\
VPHASDR2J182606.0-122839.3		  & SySt  (0.86) & SySt  (0.71) & SySt   (1.00) & SySt  (0.86)   & SySt   (1.00)  & SySt     (1.00) & SySt  (1.00)   &         \\
VPHASDR2J184024.2-084346.3		  & SySt  (1.00) & Mira  (0.86) & YSO    (0.86) & SySt  (1.00)   & SySt   (1.00)  & ClTTauri (0.86) & SySt  (0.86)   &     PN K~3-9  Known SySt     \\
VPHASDR2J183910.8-085644.4		  & SySt  (0.86) & Mira  (0.57) & YSO    (1.00) & WR    (0.86)   & SySt   (1.00)  & SySt     (1.00) & SySt  (0.57)   &           \\
VPHASDR2J183044.6-100757.4        & SySt  (0.86) & SySt  (0.57) & SySt   (1.00) & SySt  (0.57)   & SySt   (1.00)  & SySt     (1.00) & SySt  (0.86)   &          \\
VPHASDR2J184303.6-050026.4        & SySt  (0.71) & Mira  (0.86) & YSO    (1.00) & SySt  (1.00)   & SySt   (1.00)  & SySt     (0.86) & SySt  (0.71)   &     AGB star   \\
VPHASDR2J184532.1-005029.4        & SySt  (1.00) & Mira  (0.71) & YSO    (0.86) & SySt  (1.00)   & SySt   (1.00)  & SySt     (1.00) & SySt  (0.86)   &          \\
VPHASDR2J184229.2-002144.1        & SySt  (1.00) & SySt  (0.71) & YSO    (1.00) & SySt  (1.00)   & SySt   (1.00)  & SySt     (1.00) & SySt  (1.00)   &          \\
VPHASDR2J101521.0-570706.0		  & SySt  (0.71) & Mira  (0.86) & YSO    (1.00) & WR    (0.86)   & SySt   (0.86)  & SySt     (0.71) & AeBe  (0.86)   &         \\
VPHASDR2J124845.2-634948.6        & SySt  (1.00) & Mira  (1.00) & YSO    (0.71) & WR    (0.71)   & SySt   (1.00)  & SySt     (1.00) & AeBe  (0.57)   &          \\
VPHASDR2J133405.8-623745.7        & SySt  (0.71) & Mira  (0.86) & YSO    (0.86) & SySt  (0.71)   & SySt   (1.00)  & ClTTauri (0.86) & AeBe  (0.86)   &     AGB star    \\
VPHASDR2J133509.6-614305.8        & SySt  (0.43) & Mira  (0.71) & YSO    (0.86) & WR    (1.00)   & SySt   (1.00)  & ClTTauri (0.86) & AeBe  (0.57)   &      ELS     \\
VPHASDR2J154125.6-565953.2		  & SySt  (0.86) & SySt  (0.57) & YSO    (1.00) & SySt  (1.00)   & SySt   (1.00)  & SySt     (1.00) & SySt  (1.00)   &          \\   
VPHASDR2J152144.3-572220.6		  & SySt  (0.86) & Mira  (0.57) & SySt   (0.57) & SySt  (0.86)   & SySt   (1.00)  & SySt     (1.00) & SySt  (0.71)   &         \\
VPHASDR2J160631.4-525616.4        & SySt  (0.71) & Mira  (0.86) & YSO    (0.86) & WR    (0.86)   & SySt   (1.00)  & SySt     (1.00) & AeBe  (0.71)   &          \\
VPHASDR2J164646.3-454758.2		  & SySt  (1.00) & SySt  (0.86) & SySt   (1.00) & SySt  (1.00)   & SySt   (1.00)  & SySt     (1.00) & SySt  (1.00)   &      WR star     \\
VPHASDR2J162446.2-485536.4        & SySt  (1.00) & Mira  (0.71) & YSO    (0.71) & SySt  (1.00)   & SySt   (1.00)  & SySt     (1.00) & AeBe  (0.57)   &          \\
VPHASDR2J162457.4-484340.2        & SySt  (1.00) & Mira  (1.00) & YSO    (0.71) & WR    (0.86)   & SySt   (1.00)  & SySt     (1.00) & AeBe  (0.86)   &     ELS     \\
VPHASDR2J164300.8-452701.3        & SySt  (1.00) & SySt  (1.00) & SySt   (0.86) & SySt  (1.00)   & SySt   (1.00)  & SySt     (1.00) & SySt  (1.00)   &          \\
VPHASDR2J165346.7-434931.0        & SySt  (0.71) & Mira  (0.86) & YSO    (0.86) & SySt  (1.00)   & SySt   (1.00)  & SySt     (1.00) & SySt  (0.86)   &          \\
VPHASDR2J171225.1-412555.4        & SySt  (0.86) & Mira  (0.86) & YSO    (0.86) & SySt  (1.00)   & SySt   (1.00)  & SySt     (0.86) & SySt  (0.86)   &          \\
VPHASDR2J171455.1-393311.7        & SySt  (1.00) & SySt  (1.00) & YSO    (0.57) & SySt  (1.00)   & SySt   (1.00)  & SySt     (0.86) & SySt  (1.00)   &     SySt candidate     \\
VPHASDR2J171527.4-390209.2        & SySt  (1.00) & SySt  (0.86) & YSO    (0.71) & SySt  (1.00)   & SySt   (1.00)  & SySt     (1.00) & AeBe  (1.00)   &          \\
VPHASDR2J171513.5-364633.2        & SySt  (0.86) & SySt  (0.71) & YSO    (0.86) & SySt  (1.00)   & SySt   (1.00)  & SySt     (0.86) & SySt  (0.71)   &          \\
VPHASDR2J171445.0-361838.4        & SySt  (0.71) & Mira  (0.86) & SySt   (1.00) & SySt  (1.00)   & SySt   (1.00)  & ClTTauri (0.71) & AeBe  (1.00)   &     ELS      \\
VPHASDR2J120916.3-633202.7        & SySt  (0.86) & SySt  (1.00) & YSO    (0.86) & SySt  (1.00)   & SySt   (1.00)  & SySt     (0.71) & SySt  (1.00)   &          \\

\end{longtable}
\begin{flushleft}
The classification of some sources as emission line stars (ELS), semi regular pulsating star (SR-PS), asymptotic giant branch stars (AGB),  
Wolf-Rayet stars (WR), planetary nebula (PN) or known/candidate symbiotic stars (SySt) is based on the SIMBAD catalogue, Kohoutek \& Wehmeyer 
(2003) or Paper~I.\\
(1) Rodr\'{i}guez-Flores et al. 2014, (2) Corradi et al. 2010, (3) Viironen et al. (2009b),  (4) Krause et al. (2003)
\end{flushleft} 
\end{center}
\end{landscape}
\end{scriptsize}

\end{document}